\documentclass[apjl,iop]{emulateapj}
\usepackage{comment}
\usepackage{ifthen}
\newcommand{\forloop}[5][1]%
{%
\setcounter{#2}{#3}%
\ifthenelse{#4}%
	{%
	#5%
	\addtocounter{#2}{#1}%
	\forloop[#1]{#2}{\value{#2}}{#4}{#5}%
	}%
	{%
	}%
}% 

\newcommand{\ctbd}[1]{}

\newcommand{\lc}{light curve}
\newcommand{\lcs}{light curves}
\newcommand{\Lc}{Light curve}

\newcommand{\masy}{\ensuremath{\rm mas\,yr^{-1}}}
\newcommand{\kms}{\ensuremath{\rm km\,s^{-1}}}
\newcommand{\ms}{\ensuremath{\rm m\,s^{-1}}}

\newcommand{\gcmc}{\ensuremath{\rm g\,cm^{-3}}}

\newcommand{\vsini}{\ensuremath{v \sin{i}}}
\newcommand{\feh}{\ensuremath{\rm [Fe/H]}}

\newcommand{\vmac}{\ensuremath{v_{\rm mac}}}
\newcommand{\vmic}{\ensuremath{v_{\rm mic}}}
\newcommand{\rsun}{\ensuremath{R_\sun}}
\newcommand{\msun}{\ensuremath{M_\sun}}
\newcommand{\lsun}{\ensuremath{L_\sun}}

\newcommand{\rstar}{\ensuremath{R_\star}}
\newcommand{\mstar}{\ensuremath{M_\star}}
\newcommand{\lstar}{\ensuremath{L_\star}}

\newcommand{\teffstar}{\ensuremath{T_{\rm eff\star}}}
\newcommand{\rhostar}{\ensuremath{\rho_\star}}
\newcommand{\loggstar}{\ensuremath{\log{g_{\star}}}}

\newcommand{\rpl}{\ensuremath{R_{p}}}
\newcommand{\mpl}{\ensuremath{M_{p}}}

\newcommand{\rhopl}{\ensuremath{\rho_{p}}}

\newcommand{\arstar}{\ensuremath{a/\rstar}}
\newcommand{\zrstar}{\ensuremath{\zeta/\rstar}}

\newcommand{\rjup}{\ensuremath{R_{\rm J}}}
\newcommand{\mjup}{\ensuremath{M_{\rm J}}}

\newcommand{\reffigl}[1]{Figure~\ref{fig:#1}}
\newcommand{\refsecl}[1]{\mbox{Section \ref{sec:#1}}}

\newcommand{\reftabl}[1]{Table~\ref{tab:#1}}

\newcommand{\loopand}{\ifnum\value{planetcounter}=2 and \else\fi}
\newcommand{\loopcomma}{\ifnum\value{planetcounter}<2 ,\else. \fi}
\newcommand{\loopcommanoperiod}{\ifnum\value{planetcounter}<2 ,\else \space\fi}
\newcommand{\loopcommanospace}{\ifnum\value{planetcounter}<2 ,\else \fi}

\newcommand{\hatcurhtrxxxxxA}{HTR342-006}                       % Original HTR name of target
\newcommand{\hatcurfieldxxxxxA}{\ensuremath{string}}            % HTR field
\newcommand{\hatcurCCraxxxxxA}{\ensuremath{21^{\mathrm h}03^{\mathrm m}37.44{\mathrm s}}}                     % Right Ascension
\newcommand{\hatcurCCdecxxxxxA}{\ensuremath{+11{\arcdeg}59{\arcmin}21.9{\arcsec}}}                    % Declination
\newcommand{\hatcurCCmagxxxxxA}{13.145}                         % apparent V-band magnitude
\newcommand{\hatcurCCtwomassxxxxxA}{2MASS~21033731+1159218}     % 2MASS identifier
\newcommand{\hatcurCCgscxxxxxA}{GSC~1111-00383}                 % GSC(1.2) identifier
\newcommand{\hatcurCCtassmvxxxxxA}{\ensuremath{13.145\pm0.029}} % APASS V-band magnitude
\newcommand{\hatcurCCtassmvshortxxxxxA}{\ensuremath{13.1}}      % APASS V-band magnitude
\newcommand{\hatcurCCtassmBxxxxxA}{\ensuremath{13.818\pm0.021}} % APASS B-band magnitude
\newcommand{\hatcurCCtassmBshortxxxxxA}{\ensuremath{13.8}}      % APASS B-band magnitude
\newcommand{\hatcurCCtassmIxxxxxA}{\ensuremath{12.46\pm0.10}}   % TASS I-band magnitude
\newcommand{\hatcurCCtassmIshortxxxxxA}{\ensuremath{12.5}}      % TASS I-band magnitude
\newcommand{\hatcurCCtassmgxxxxxA}{\ensuremath{13.445\pm0.016}} % APASS g-band magnitude
\newcommand{\hatcurCCtassmgshortxxxxxA}{\ensuremath{13.4}}      % APASS g-band magnitude
\newcommand{\hatcurCCtassmrxxxxxA}{\ensuremath{12.948\pm0.033}} % APASS r-band magnitude
\newcommand{\hatcurCCtassmrshortxxxxxA}{\ensuremath{12.9}}      % APASS r-band magnitude
\newcommand{\hatcurCCtassmixxxxxA}{\ensuremath{12.784\pm0.097}} % APASS i-band magnitude
\newcommand{\hatcurCCtassmishortxxxxxA}{\ensuremath{12.8}}      % APASS i-band magnitude
\newcommand{\hatcurCCtwomassJmagxxxxxA}{\ensuremath{11.892\pm0.026}} % 2MASS ORIG MAG
\newcommand{\hatcurCCtwomassHmagxxxxxA}{\ensuremath{11.604\pm0.022}} % 2MASS ORIG MAG
\newcommand{\hatcurCCtwomassKmagxxxxxA}{\ensuremath{11.528\pm0.025}} % 2MASS ORIG MAG
\newcommand{\hatcurCCcitJmagxxxxxA}{\ensuremath{11.909\pm0.026}} % 2MASS CIT MAG
\newcommand{\hatcurCCcitHmagxxxxxA}{\ensuremath{11.599\pm0.022}} % 2MASS CIT MAG
\newcommand{\hatcurCCcitKmagxxxxxA}{\ensuremath{11.552\pm0.025}} % 2MASS CIT MAG
\newcommand{\hatcurCCbbJmagxxxxxA}{\ensuremath{11.958\pm0.028}} % 2MASS BB MAG
\newcommand{\hatcurCCbbHmagxxxxxA}{\ensuremath{11.620\pm0.023}} % 2MASS BB MAG
\newcommand{\hatcurCCbbKmagxxxxxA}{\ensuremath{11.572\pm0.025}} % 2MASS BB MAG
\newcommand{\hatcurCCesoJmagxxxxxA}{\ensuremath{11.960\pm0.029}} % 2MASS ESO MAG
\newcommand{\hatcurCCesoHmagxxxxxA}{\ensuremath{11.616\pm0.026}} % 2MASS ESO MAG
\newcommand{\hatcurCCesoKmagxxxxxA}{\ensuremath{11.571\pm0.026}} % 2MASS ESO MAG
\newcommand{\hatcurCCesoJHmagxxxxxA}{\ensuremath{0.344\pm0.037}} % 2MASS ESO JH COLOR
\newcommand{\hatcurCCesoJKmagxxxxxA}{\ensuremath{0.390\pm0.039}} % 2MASS ESO JK COLOR
\newcommand{\hatcurCCesoHKmagxxxxxA}{\ensuremath{0.045\pm0.037}} % 2MASS ESO HK COLOR
\newcommand{\hatcurLCdipxxxxxA}{\ensuremath{13.5}}              % BLS detected dip (mmag)
\newcommand{\hatcurLCrprstarxxxxxA}{\ensuremath{0.1045\pm0.0024}} % Rp/R*
\newcommand{\hatcurLCbsqxxxxxA}{\ensuremath{0.215_{-0.079}^{+0.070}}} % impact parameter square
\newcommand{\hatcurLCimpxxxxxA}{\ensuremath{0.464_{-0.094}^{+0.070}}} % impact parameter
\newcommand{\hatcurLCzetaxxxxxA}{\ensuremath{12.438\pm0.080}}   % zeta/R*
\newcommand{\hatcurLCdurxxxxxA}{\ensuremath{0.1819\pm0.0022}}   % transit duration (days)
\newcommand{\hatcurLCdurshortxxxxxA}{\ensuremath{0.1819}}       % transit duration (days)
\newcommand{\hatcurLCdurhrxxxxxA}{\ensuremath{4.366\pm0.052}}   % transit duration (hours)
\newcommand{\hatcurLCdurhrshortxxxxxA}{\ensuremath{4.366}}      % transit duration (hours)
\newcommand{\hatcurLCqxxxxxA}{\ensuremath{0.06980\pm0.00083}}   % fractional transit duration (days)
\newcommand{\hatcurLCqshortxxxxxA}{\ensuremath{0.070}}          % fractional transit duration (days)
\newcommand{\hatcurLCingdurxxxxxA}{\ensuremath{0.0215\pm0.0023}} % ingress/egress duration (days)
\newcommand{\hatcurLCPxxxxxA}{\ensuremath{2.6054552\pm0.0000031}} % period (days)
\newcommand{\hatcurLCPprecxxxxxA}{\ensuremath{2.6054552}}       % period (days)
\newcommand{\hatcurLCPshortxxxxxA}{\ensuremath{2.6055}}         % period (days)
\newcommand{\hatcurLCTxxxxxA}{\ensuremath{2456409.33263\pm0.00046}} % epoch (BJD)
\newcommand{\hatcurLCTAxxxxxA}{\ensuremath{2455090.9723\pm0.0015}} % TA (BJD)
\newcommand{\hatcurLCTBxxxxxA}{\ensuremath{2456568.26541\pm0.00052}} % TB (BJD)
\newcommand{\hatcurLChatnetmxxxxxA}{\ensuremath{12.94498\pm0.00020}} % HATNet OOT level
\newcommand{\hatcurLCiblendxxxxxA}{\ensuremath{1\pm0}}          % HATNet iblend factor
\newcommand{\hatcurLCrhoxxxxxA}{\ensuremath{0.266\pm0.035}}     % stellar density no isochrone constraint (cgs)
\newcommand{\hatcurSMEiteffxxxxxA}{\ensuremath{5943\pm51}}      % Ini SME, stellar effective temperature
\newcommand{\hatcurSMEizfehxxxxxA}{\ensuremath{0.170\pm0.080}}  % Ini SME, stellar metallicity
\newcommand{\hatcurSMEizfehshortxxxxxA}{\ensuremath{0.17}}      % Ini SME, stellar metallicity
\newcommand{\hatcurSMEiloggxxxxxA}{\ensuremath{4.18\pm0.10}}    % Ini SME, stellar surface gravity
\newcommand{\hatcurSMEivsinxxxxxA}{\ensuremath{6.90\pm0.50}}    % Ini SME, stellar rotational velocity
\newcommand{\hatcurSMEivmacxxxxxA}{\ensuremath{0.0}}            % Ini SME, stellar macroturbulence
\newcommand{\hatcurSMEivmicxxxxxA}{\ensuremath{0.0}}            % Ini SME, stellar microturbulence
\newcommand{\hatcurSMEiiteffxxxxxA}{\ensuremath{5835\pm51}}     % Final SME, stellar effective temperature
\newcommand{\hatcurSMEiizfehxxxxxA}{\ensuremath{0.100\pm0.080}} % Final SME, stellar metallicity
\newcommand{\hatcurSMEiizfehshortxxxxxA}{\ensuremath{0.10}}     % Final SME, stellar metallicity
\newcommand{\hatcurSMEiiloggxxxxxA}{\ensuremath{3.992\pm0.038}} % Final SME, stellar surface gravity
\newcommand{\hatcurSMEiivsinxxxxxA}{\ensuremath{7.10\pm0.50}}   % Final SME, stellar rotational velocity
\newcommand{\hatcurLBizxxxxxA}{\ensuremath{0.1949}}             % Limb darkening parameters, Gamma1, z-band
\newcommand{\hatcurLBiizxxxxxA}{\ensuremath{0.3379}}            % Limb darkening parameters, Gamma2, z-band
\newcommand{\hatcurLBiixxxxxA}{\ensuremath{0.2544}}             % Limb darkening parameters, Gamma1, i-band
\newcommand{\hatcurLBiiixxxxxA}{\ensuremath{0.3414}}            % Limb darkening parameters, Gamma2, i-band
\newcommand{\hatcurLBiIxxxxxA}{\ensuremath{0.2336}}             % Limb darkening parameters, Gamma1, I-band
\newcommand{\hatcurLBiiIxxxxxA}{\ensuremath{0.3416}}            % Limb darkening parameters, Gamma2, I-band
\newcommand{\hatcurLBigxxxxxA}{\ensuremath{0.5412}}             % Limb darkening parameters, Gamma1, g-band
\newcommand{\hatcurLBiigxxxxxA}{\ensuremath{0.2456}}            % Limb darkening parameters, Gamma2, g-band
\newcommand{\hatcurLBirxxxxxA}{\ensuremath{0.3439}}             % Limb darkening parameters, Gamma1, r-band
\newcommand{\hatcurLBiirxxxxxA}{\ensuremath{0.3359}}            % Limb darkening parameters, Gamma2, r-band
\newcommand{\hatcurLBiRxxxxxA}{\ensuremath{0.3190}}             % Limb darkening parameters, Gamma1, R-band
\newcommand{\hatcurLBiiRxxxxxA}{\ensuremath{0.3385}}            % Limb darkening parameters, Gamma2, R-band
\newcommand{\hatcurLBikepxxxxxA}{\ensuremath{0.1000}}           % Limb darkening parameters, Gamma1, Kep-band
\newcommand{\hatcurLBiikepxxxxxA}{\ensuremath{0.1000}}          % Limb darkening parameters, Gamma2, Kep-band
\newcommand{\hatcurISOmxxxxxA}{\ensuremath{1.212\pm0.050}}      % stellar mass
\newcommand{\hatcurISOmshortxxxxxA}{\ensuremath{1.21}}          % stellar mass
\newcommand{\hatcurISOmlongxxxxxA}{\ensuremath{1.212\pm0.050}}  % stellar mass
\newcommand{\hatcurISOrxxxxxA}{\ensuremath{1.860\pm0.096}}      % stellar radius
\newcommand{\hatcurISOrshortxxxxxA}{\ensuremath{1.86}}          % stellar radius
\newcommand{\hatcurISOrlongxxxxxA}{\ensuremath{1.860\pm0.096}}  % stellar radius
\newcommand{\hatcurISOrhoxxxxxA}{\ensuremath{0.266\pm0.036}}    % stellar density (cgs)
\newcommand{\hatcurISOrholongxxxxxA}{\ensuremath{0.266\pm0.036}} % stellar density (cgs)
\newcommand{\hatcurISOloggxxxxxA}{\ensuremath{3.983\pm0.035}}   % stellar surface gravity from isochrones
\newcommand{\hatcurISOlumxxxxxA}{\ensuremath{3.59\pm0.40}}      % stellar luminosity
\newcommand{\hatcurISOlumshortxxxxxA}{\ensuremath{3.59}}        % stellar luminosity
\newcommand{\hatcurISOmvxxxxxA}{\ensuremath{3.43\pm0.13}}       % stellar absolute magnitude
\newcommand{\hatcurISOvixxxxxA}{\ensuremath{0.679\pm0.015}}     % stellar V-I index
\newcommand{\hatcurISOagexxxxxA}{\ensuremath{5.46\pm0.61}}      % stellar age
\newcommand{\hatcurISOsigmaxxxxxA}{\ensuremath{0.000200\pm0.000041}} % system mass-correction sigma parameter
\newcommand{\hatcurISOMJxxxxxA}{\ensuremath{2.31\pm0.12}}       % stellar absolute J magnitude
\newcommand{\hatcurISOMHxxxxxA}{\ensuremath{1.99\pm0.12}}       % stellar absolute H magnitude
\newcommand{\hatcurISOMKxxxxxA}{\ensuremath{1.93\pm0.12}}       % stellar absolute K magnitude
\newcommand{\hatcurISOJKxxxxxA}{\ensuremath{0.380\pm0.010}}     % J-K color index from isochrones.
\newcommand{\hatcurISOspecxxxxxA}{G}                            % stellar spectral type
\newcommand{\hatcurRVKxxxxxA}{\ensuremath{68\pm11}}             % RV semi-amplitude [m/s]
\newcommand{\hatcurRVrkxxxxxA}{\ensuremath{0\pm0}}              % sqrt(e)*cos(omega)
\newcommand{\hatcurRVrhxxxxxA}{\ensuremath{0\pm0}}              % sqrt(e)*sin(omega)
\newcommand{\hatcurRVkxxxxxA}{\ensuremath{0\pm0}}               % e*cos(omega)
\newcommand{\hatcurRVhxxxxxA}{\ensuremath{0\pm0}}               % e*sin(omega)
\newcommand{\hatcurRVtronexxxxxA}{\ensuremath{0\pm0}}           % RV linear trend tr1 factor
\newcommand{\hatcurRVtrtwoxxxxxA}{\ensuremath{0\pm0}}           % RV linear trend tr2 factor
\newcommand{\hatcurRVgammaxxxxxA}{\ensuremath{6.4\pm8.4}}       % RV gamma velocity, relative scale
\newcommand{\hatcurRVjitterxxxxxA}{\ensuremath{26.0\pm7.1}}     % RV jitter (m/s)
\newcommand{\hatcurRVjittertwosiglimxxxxxA}{\ensuremath{<39.6}} % RV jitter (m/s) 95 percent confidence upper limit
\newcommand{\hatcurRVfitrmsxxxxxA}{\ensuremath{.1fym}}          % 
\newcommand{\hatcurRVeccenxxxxxA}{\ensuremath{0\pm0}}           % eccentricity
\newcommand{\hatcurRVeccentwosiglimxxxxxA}{\ensuremath{<0.000}} % eccentricity
\newcommand{\hatcurRVomegaxxxxxA}{\ensuremath{0\pm0}}           % argument of pericenter
\newcommand{\hatcurPPixxxxxA}{\ensuremath{84.2\pm1.3}}          % orbital inclination
\newcommand{\hatcurPPgxxxxxA}{\ensuremath{3.63\pm0.74}}         % planetary surface gravity (m/s^2)
\newcommand{\hatcurPPloggxxxxxA}{\ensuremath{2.560\pm0.090}}    % planetary surface gravity (log cgs)
\newcommand{\hatcurPParxxxxxA}{\ensuremath{4.57\pm0.20}}        % relative orbital radius (a/R*)
\newcommand{\hatcurPParelxxxxxA}{\ensuremath{0.03951\pm0.00054}} % semimajor axis (AU)
\newcommand{\hatcurPPrhoxxxxxA}{\ensuremath{0.096\pm0.025}}     % planetary density (cgs)
\newcommand{\hatcurPPmxxxxxA}{\ensuremath{0.527\pm0.083}}       % planetary mass (M_jup)
\newcommand{\hatcurPPmshortxxxxxA}{\ensuremath{0.53}}           % planetary mass (M_jup)
\newcommand{\hatcurPPmlongxxxxxA}{\ensuremath{0.527\pm0.083}}   % planetary mass (M_jup)
\newcommand{\hatcurPPmexxxxxA}{\ensuremath{167\pm26}}           % planetary mass (M_earth)
\newcommand{\hatcurPPmeshortxxxxxA}{\ensuremath{167.5}}         % planetary mass (M_earth)
\newcommand{\hatcurPPmelongxxxxxA}{\ensuremath{167\pm26}}       % planetary mass (M_earth)
\newcommand{\hatcurPPrxxxxxA}{\ensuremath{1.89\pm0.13}}         % planetary radius (R_jup)
\newcommand{\hatcurPPrshortxxxxxA}{\ensuremath{1.89}}           % planetary radius (R_jup)
\newcommand{\hatcurPPrlongxxxxxA}{\ensuremath{1.89\pm0.13}}     % planetary radius (R_jup)
\newcommand{\hatcurPPrexxxxxA}{\ensuremath{21.2\pm1.5}}         % planetary radius (R_earth)
\newcommand{\hatcurPPreshortxxxxxA}{\ensuremath{21.2}}          % planetary radius (R_earth)
\newcommand{\hatcurPPrelongxxxxxA}{\ensuremath{21.2\pm1.5}}     % planetary radius (R_earth)
\newcommand{\hatcurPPmrcorrxxxxxA}{\ensuremath{0.10}}           % mass/radius correlation
\newcommand{\hatcurPPteffxxxxxA}{\ensuremath{1930\pm45}}        % planetary temperature (K)
\newcommand{\hatcurPPthetaxxxxxA}{\ensuremath{0.0180\pm0.0031}} % Safranov number
\newcommand{\hatcurPPfluxperixxxxxA}{\ensuremath{3.13\pm0.29}}  % flux @ periastron (CGS)
\newcommand{\hatcurPPfluxperidimxxxxxA}{\ensuremath{9}}         % flux @ periastron (CGS) units.
\newcommand{\hatcurPPfluxapxxxxxA}{\ensuremath{3.13\pm0.29}}    % flux @ apastron (CGS)
\newcommand{\hatcurPPfluxapdimxxxxxA}{\ensuremath{9}}           % flux @ apastron (CGS) units.
\newcommand{\hatcurPPfluxavgxxxxxA}{\ensuremath{3.13\pm0.29}}   % flux on average (CGS)
\newcommand{\hatcurPPfluxavgdimxxxxxA}{\ensuremath{9}}          % flux average (CGS) units.
\newcommand{\hatcurPPfluxavglogxxxxxA}{\ensuremath{9.495\pm0.041}} % log10 flux on average (CGS)
\newcommand{\hatcurXsecphasexxxxxA}{\ensuremath{0\pm0}}         % Phase of secondary eclipse
\newcommand{\hatcurXsecondaryxxxxxA}{\ensuremath{2456410.63536\pm0.00046}} % Secondary eclipse epoch
\newcommand{\hatcurXsecdurxxxxxA}{\ensuremath{0.1819\pm0.0022}} % sec eclipse duration (days)
\newcommand{\hatcurXsecingdurxxxxxA}{\ensuremath{0.0215\pm0.0023}} % sec I/E duration (days)
\newcommand{\hatcurPPphiconjxxxxxA}{\ensuremath{0\pm0}}         % phase diff between conjunction and periastron
\newcommand{\hatcurPPperixxxxxA}{\ensuremath{2456408.68127\pm0.00046}} % time of periastron passage.
\newcommand{\hatcurPPaequivxxxxxA}{\ensuremath{0.02090\pm0.00098}} % equivalent semi-major axis
\newcommand{\hatcurPPtcircxxxxxA}{\ensuremath{4.7_{-1.4}^{+2.1}}} % circularization timescale
\newcommand{\hatcurPPtinfallxxxxxA}{\ensuremath{115_{-27}^{+37}}} % infall timescale
\newcommand{\hatcurXdistxxxxxA}{\ensuremath{848\pm46}}          % distance (pc), no reddenning correction
\newcommand{\hatcurXAvxxxxxA}{\ensuremath{0.090\pm0.052}}       % Av (mag)
\newcommand{\hatcurXdistredxxxxxA}{\ensuremath{841\pm45}}       % distance with Av correction (pc)
\newcommand{\hatcurXEBVxxxxxA}{\ensuremath{0.029\pm0.017}}      % E(B-V) (mag)
\newcommand{\hatcurXmvisoredxxxxxA}{\ensuremath{13.145\pm0.028}} % Expected m_v with reddening (mag)
\newcommand{\hatcurXmiisoredxxxxxA}{\ensuremath{12.419\pm0.015}} % Expected m_i with reddening (mag)
\newcommand{\hatcurXmjisoredxxxxxA}{\ensuremath{11.961\pm0.014}} % Expected m_j with reddening (mag)
\newcommand{\hatcurXmhisoredxxxxxA}{\ensuremath{11.626\pm0.015}} % Expected m_h with reddening (mag)
\newcommand{\hatcurXmkisoredxxxxxA}{\ensuremath{11.563\pm0.017}} % Expected m_k with reddening (mag)
\newcommand{\hatcurXviisoredxxxxxA}{\ensuremath{0.727\pm0.021}} % Expected V-I with reddening (mag)
\newcommand{\hatcurXvkisoredxxxxxA}{\ensuremath{1.583\pm0.034}} % Expected V-K with reddening (mag)
\newcommand{\hatcurXjhisoredxxxxxA}{\ensuremath{0.3360\pm0.0066}} % Expected J-H with reddening (mag)
\newcommand{\hatcurXjkisoredxxxxxA}{\ensuremath{0.3990\pm0.0078}} % Expected J-K with reddening (mag)
\newcommand{\hatcurCCpmraxxxxxA}{\ensuremath{5.5\pm1.9}}        % proper motion, in RA
\newcommand{\hatcurCCpmdecxxxxxA}{\ensuremath{-4.0\pm1.9}}      % proper motion, in DEC
\newcommand{\hatcurCCpmxxxxxA}{\ensuremath{6.8\pm2.7}}          % proper motion

\newcommand{\hatcurhtrxxxxxB}{HTR101-005}                       % Original HTR name of target
\newcommand{\hatcurfieldxxxxxB}{\ensuremath{string}}            % HTR field
\newcommand{\hatcurCCraxxxxxB}{\ensuremath{10^{\mathrm h}02^{\mathrm m}17.52{\mathrm s}}}                     % Right Ascension
\newcommand{\hatcurCCdecxxxxxB}{\ensuremath{+53{\arcdeg}57{\arcmin}03.1{\arcsec}}}                    % Declination
\newcommand{\hatcurCCmagxxxxxB}{12.993}                         % apparent V-band magnitude
\newcommand{\hatcurCCtwomassxxxxxB}{2MASS~10021743+5357031}     % 2MASS identifier
\newcommand{\hatcurCCgscxxxxxB}{GSC~3814-00307}                 % GSC(1.2) identifier
\newcommand{\hatcurCCtassmvxxxxxB}{\ensuremath{12.993\pm0.052}} % APASS V-band magnitude
\newcommand{\hatcurCCtassmvshortxxxxxB}{\ensuremath{13.0}}      % APASS V-band magnitude
\newcommand{\hatcurCCtassmBxxxxxB}{\ensuremath{13.552\pm0.027}} % APASS B-band magnitude
\newcommand{\hatcurCCtassmBshortxxxxxB}{\ensuremath{13.6}}      % APASS B-band magnitude
\newcommand{\hatcurCCtassmIxxxxxB}{\ensuremath{12.339\pm0.084}} % TASS I-band magnitude
\newcommand{\hatcurCCtassmIshortxxxxxB}{\ensuremath{12.3}}      % TASS I-band magnitude
\newcommand{\hatcurCCtassmgxxxxxB}{\ensuremath{13.209\pm0.021}} % APASS g-band magnitude
\newcommand{\hatcurCCtassmgshortxxxxxB}{\ensuremath{13.2}}      % APASS g-band magnitude
\newcommand{\hatcurCCtassmrxxxxxB}{\ensuremath{12.859\pm0.064}} % APASS r-band magnitude
\newcommand{\hatcurCCtassmrshortxxxxxB}{\ensuremath{12.9}}      % APASS r-band magnitude
\newcommand{\hatcurCCtassmixxxxxB}{\ensuremath{12.771\pm0.064}} % APASS i-band magnitude
\newcommand{\hatcurCCtassmishortxxxxxB}{\ensuremath{12.8}}      % APASS i-band magnitude
\newcommand{\hatcurCCtwomassJmagxxxxxB}{\ensuremath{12.001\pm0.022}} % 2MASS ORIG MAG
\newcommand{\hatcurCCtwomassHmagxxxxxB}{\ensuremath{11.735\pm0.022}} % 2MASS ORIG MAG
\newcommand{\hatcurCCtwomassKmagxxxxxB}{\ensuremath{11.675\pm0.022}} % 2MASS ORIG MAG
\newcommand{\hatcurCCcitJmagxxxxxB}{\ensuremath{12.020\pm0.023}} % 2MASS CIT MAG
\newcommand{\hatcurCCcitHmagxxxxxB}{\ensuremath{11.730\pm0.023}} % 2MASS CIT MAG
\newcommand{\hatcurCCcitKmagxxxxxB}{\ensuremath{11.699\pm0.022}} % 2MASS CIT MAG
\newcommand{\hatcurCCbbJmagxxxxxB}{\ensuremath{12.065\pm0.023}} % 2MASS BB MAG
\newcommand{\hatcurCCbbHmagxxxxxB}{\ensuremath{11.751\pm0.023}} % 2MASS BB MAG
\newcommand{\hatcurCCbbKmagxxxxxB}{\ensuremath{11.719\pm0.022}} % 2MASS BB MAG
\newcommand{\hatcurCCesoJmagxxxxxB}{\ensuremath{12.068\pm0.024}} % 2MASS ESO MAG
\newcommand{\hatcurCCesoHmagxxxxxB}{\ensuremath{11.746\pm0.024}} % 2MASS ESO MAG
\newcommand{\hatcurCCesoKmagxxxxxB}{\ensuremath{11.718\pm0.023}} % 2MASS ESO MAG
\newcommand{\hatcurCCesoJHmagxxxxxB}{\ensuremath{0.3220\pm0.0070}} % 2MASS ESO JH COLOR
\newcommand{\hatcurCCesoJKmagxxxxxB}{\ensuremath{0.3490\pm0.0090}} % 2MASS ESO JK COLOR
\newcommand{\hatcurCCesoHKmagxxxxxB}{\ensuremath{0.0270\pm0.0080}} % 2MASS ESO HK COLOR
\newcommand{\hatcurLCdipxxxxxB}{\ensuremath{0.0}}               % BLS detected dip (mmag)
\newcommand{\hatcurLCrprstarxxxxxB}{\ensuremath{0.0872\pm0.0024}} % Rp/R*
\newcommand{\hatcurLCbsqxxxxxB}{\ensuremath{0.110_{-0.081}^{+0.106}}} % impact parameter square
\newcommand{\hatcurLCimpxxxxxB}{\ensuremath{0.33_{-0.16}^{+0.13}}} % impact parameter
\newcommand{\hatcurLCzetaxxxxxB}{\ensuremath{11.226\pm0.096}}   % zeta/R*
\newcommand{\hatcurLCdurxxxxxB}{\ensuremath{0.1958\pm0.0028}}   % transit duration (days)
\newcommand{\hatcurLCdurshortxxxxxB}{\ensuremath{0.1958}}       % transit duration (days)
\newcommand{\hatcurLCdurhrxxxxxB}{\ensuremath{4.699\pm0.066}}   % transit duration (hours)
\newcommand{\hatcurLCdurhrshortxxxxxB}{\ensuremath{4.699}}      % transit duration (hours)
\newcommand{\hatcurLCqxxxxxB}{\ensuremath{0.06590\pm0.00093}}   % fractional transit duration (days)
\newcommand{\hatcurLCqshortxxxxxB}{\ensuremath{0.066}}          % fractional transit duration (days)
\newcommand{\hatcurLCingdurxxxxxB}{\ensuremath{0.0174\pm0.0025}} % ingress/egress duration (days)
\newcommand{\hatcurLCPxxxxxB}{\ensuremath{2.9720860\pm0.0000057}} % period (days)
\newcommand{\hatcurLCPprecxxxxxB}{\ensuremath{2.9720860}}       % period (days)
\newcommand{\hatcurLCPshortxxxxxB}{\ensuremath{2.9721}}         % period (days)
\newcommand{\hatcurLCTxxxxxB}{\ensuremath{2457258.79907\pm0.00072}} % epoch (BJD)
\newcommand{\hatcurLCTAxxxxxB}{\ensuremath{2455609.2914\pm0.0031}} % TA (BJD)
\newcommand{\hatcurLCTBxxxxxB}{\ensuremath{2457365.79418\pm0.00079}} % TB (BJD)
\newcommand{\hatcurLChatnetmxxxxxB}{\ensuremath{13.29962\pm0.00013}} % HATNet OOT level
\newcommand{\hatcurLCiblendxxxxxB}{\ensuremath{1\pm0}}          % HATNet iblend factor
\newcommand{\hatcurLCrhoxxxxxB}{\ensuremath{0.269\pm0.040}}     % stellar density no isochrone constraint (cgs)
\newcommand{\hatcurSMEiteffxxxxxB}{\ensuremath{6002\pm50}}      % Ini SME, stellar effective temperature
\newcommand{\hatcurSMEizfehxxxxxB}{\ensuremath{0.035\pm0.080}}  % Ini SME, stellar metallicity
\newcommand{\hatcurSMEizfehshortxxxxxB}{\ensuremath{0.04}}      % Ini SME, stellar metallicity
\newcommand{\hatcurSMEiloggxxxxxB}{\ensuremath{3.96\pm0.10}}    % Ini SME, stellar surface gravity
\newcommand{\hatcurSMEivsinxxxxxB}{\ensuremath{7.57\pm0.50}}    % Ini SME, stellar rotational velocity
\newcommand{\hatcurSMEivmacxxxxxB}{\ensuremath{0.0}}            % Ini SME, stellar macroturbulence
\newcommand{\hatcurSMEivmicxxxxxB}{\ensuremath{0.0}}            % Ini SME, stellar microturbulence
\newcommand{\hatcurLBizxxxxxB}{\ensuremath{0.1703}}             % Limb darkening parameters, Gamma1, z-band
\newcommand{\hatcurLBiizxxxxxB}{\ensuremath{0.3491}}            % Limb darkening parameters, Gamma2, z-band
\newcommand{\hatcurLBiixxxxxB}{\ensuremath{0.2249}}             % Limb darkening parameters, Gamma1, i-band
\newcommand{\hatcurLBiiixxxxxB}{\ensuremath{0.3551}}            % Limb darkening parameters, Gamma2, i-band
\newcommand{\hatcurLBiIxxxxxB}{\ensuremath{0.2054}}             % Limb darkening parameters, Gamma1, I-band
\newcommand{\hatcurLBiiIxxxxxB}{\ensuremath{0.3547}}            % Limb darkening parameters, Gamma2, I-band
\newcommand{\hatcurLBigxxxxxB}{\ensuremath{0.4936}}             % Limb darkening parameters, Gamma1, g-band
\newcommand{\hatcurLBiigxxxxxB}{\ensuremath{0.2793}}            % Limb darkening parameters, Gamma2, g-band
\newcommand{\hatcurLBirxxxxxB}{\ensuremath{0.3077}}             % Limb darkening parameters, Gamma1, r-band
\newcommand{\hatcurLBiirxxxxxB}{\ensuremath{0.3559}}            % Limb darkening parameters, Gamma2, r-band
\newcommand{\hatcurLBiRxxxxxB}{\ensuremath{0.2845}}             % Limb darkening parameters, Gamma1, R-band
\newcommand{\hatcurLBiiRxxxxxB}{\ensuremath{0.3570}}            % Limb darkening parameters, Gamma2, R-band
\newcommand{\hatcurLBikepxxxxxB}{\ensuremath{0.1000}}           % Limb darkening parameters, Gamma1, Kep-band
\newcommand{\hatcurLBiikepxxxxxB}{\ensuremath{0.1000}}          % Limb darkening parameters, Gamma2, Kep-band
\newcommand{\hatcurISOmxxxxxB}{\ensuremath{1.255_{-0.054}^{+0.107}}} % stellar mass
\newcommand{\hatcurISOmshortxxxxxB}{\ensuremath{1.25}}          % stellar mass
\newcommand{\hatcurISOmlongxxxxxB}{\ensuremath{1.255_{-0.054}^{+0.107}}} % stellar mass
\newcommand{\hatcurISOrxxxxxB}{\ensuremath{1.881_{-0.095}^{+0.151}}} % stellar radius
\newcommand{\hatcurISOrshortxxxxxB}{\ensuremath{1.88}}          % stellar radius
\newcommand{\hatcurISOrlongxxxxxB}{\ensuremath{1.881_{-0.095}^{+0.151}}} % stellar radius
\newcommand{\hatcurISOrhoxxxxxB}{\ensuremath{0.269_{-0.048}^{+0.035}}} % stellar density (cgs)
\newcommand{\hatcurISOrholongxxxxxB}{\ensuremath{0.269_{-0.048}^{+0.035}}} % stellar density (cgs)
\newcommand{\hatcurISOloggxxxxxB}{\ensuremath{3.993\pm0.045}}   % stellar surface gravity from isochrones
\newcommand{\hatcurISOlumxxxxxB}{\ensuremath{4.12_{-0.46}^{+0.71}}} % stellar luminosity
\newcommand{\hatcurISOlumshortxxxxxB}{\ensuremath{4.12}}        % stellar luminosity
\newcommand{\hatcurISOmvxxxxxB}{\ensuremath{3.26\pm0.15}}       % stellar absolute magnitude
\newcommand{\hatcurISOvixxxxxB}{\ensuremath{0.629\pm0.014}}     % stellar V-I index
\newcommand{\hatcurISOagexxxxxB}{\ensuremath{4.66_{-1.12}^{+0.52}}} % stellar age
\newcommand{\hatcurISOsigmaxxxxxB}{\ensuremath{0.000300\pm0.000040}} % system mass-correction sigma parameter
\newcommand{\hatcurISOMJxxxxxB}{\ensuremath{2.22\pm0.14}}       % stellar absolute J magnitude
\newcommand{\hatcurISOMHxxxxxB}{\ensuremath{1.92\pm0.14}}       % stellar absolute H magnitude
\newcommand{\hatcurISOMKxxxxxB}{\ensuremath{1.86\pm0.14}}       % stellar absolute K magnitude
\newcommand{\hatcurISOJKxxxxxB}{\ensuremath{0.360\pm0.010}}     % J-K color index from isochrones.
\newcommand{\hatcurISOspecxxxxxB}{F}                            % stellar spectral type
\newcommand{\hatcurRVKxxxxxB}{\ensuremath{93.5\pm5.7}}          % RV semi-amplitude [m/s]
\newcommand{\hatcurRVrkxxxxxB}{\ensuremath{0\pm0}}              % sqrt(e)*cos(omega)
\newcommand{\hatcurRVrhxxxxxB}{\ensuremath{0\pm0}}              % sqrt(e)*sin(omega)
\newcommand{\hatcurRVkxxxxxB}{\ensuremath{0\pm0}}               % e*cos(omega)
\newcommand{\hatcurRVhxxxxxB}{\ensuremath{0\pm0}}               % e*sin(omega)
\newcommand{\hatcurRVtronexxxxxB}{\ensuremath{0\pm0}}           % RV linear trend tr1 factor
\newcommand{\hatcurRVtrtwoxxxxxB}{\ensuremath{0\pm0}}           % RV linear trend tr2 factor
\newcommand{\hatcurRVgammaAxxxxxB}{\ensuremath{-9\pm18}}        % RV gamma velocity, relative scale
\newcommand{\hatcurRVjitterAxxxxxB}{\ensuremath{15\pm14}}       % RV jitter (m/s)
\newcommand{\hatcurRVjittertwosiglimAxxxxxB}{\ensuremath{<43.2}} % RV jitter (m/s) 95 percent confidence upper limit
\newcommand{\hatcurRVfitrmsAxxxxxB}{\ensuremath{0.0}}           % RVfitrms
\newcommand{\hatcurRVgammaBxxxxxB}{\ensuremath{132\pm19}}       % RV gamma velocity, relative scale
\newcommand{\hatcurRVjitterBxxxxxB}{\ensuremath{0.1\pm9.1}}     % RV jitter (m/s)
\newcommand{\hatcurRVjittertwosiglimBxxxxxB}{\ensuremath{<22.1}} % RV jitter (m/s) 95 percent confidence upper limit
\newcommand{\hatcurRVfitrmsBxxxxxB}{\ensuremath{0.0}}           % RVfitrms
\newcommand{\hatcurRVjitterCxxxxxB}{\ensuremath{0.4\pm6.8}}     % RV jitter (m/s)
\newcommand{\hatcurRVjittertwosiglimCxxxxxB}{\ensuremath{<16.3}} % RV jitter (m/s) 95 percent confidence upper limit
\newcommand{\hatcurRVeccenxxxxxB}{\ensuremath{0\pm0}}           % eccentricity
\newcommand{\hatcurRVeccentwosiglimxxxxxB}{\ensuremath{<0.000}} % eccentricity
\newcommand{\hatcurRVomegaxxxxxB}{\ensuremath{0\pm0}}           % argument of pericenter
\newcommand{\hatcurPPixxxxxB}{\ensuremath{86.2\pm1.8}}          % orbital inclination
\newcommand{\hatcurPPgxxxxxB}{\ensuremath{7.65_{-1.30}^{+0.96}}} % planetary surface gravity (m/s^2)
\newcommand{\hatcurPPloggxxxxxB}{\ensuremath{2.884_{-0.081}^{+0.051}}} % planetary surface gravity (log cgs)
\newcommand{\hatcurPParxxxxxB}{\ensuremath{5.01_{-0.32}^{+0.21}}} % relative orbital radius (a/R*)
\newcommand{\hatcurPParelxxxxxB}{\ensuremath{0.04363_{-0.00064}^{+0.00121}}} % semimajor axis (AU)
\newcommand{\hatcurPPrhoxxxxxB}{\ensuremath{0.242_{-0.061}^{+0.045}}} % planetary density (cgs)
\newcommand{\hatcurPPmxxxxxB}{\ensuremath{0.783\pm0.057}}       % planetary mass (M_jup)
\newcommand{\hatcurPPmshortxxxxxB}{\ensuremath{0.78}}           % planetary mass (M_jup)
\newcommand{\hatcurPPmlongxxxxxB}{\ensuremath{0.783\pm0.057}}   % planetary mass (M_jup)
\newcommand{\hatcurPPmexxxxxB}{\ensuremath{249\pm18}}           % planetary mass (M_earth)
\newcommand{\hatcurPPmeshortxxxxxB}{\ensuremath{248.8}}         % planetary mass (M_earth)
\newcommand{\hatcurPPmelongxxxxxB}{\ensuremath{249\pm18}}       % planetary mass (M_earth)
\newcommand{\hatcurPPrxxxxxB}{\ensuremath{1.59_{-0.10}^{+0.16}}} % planetary radius (R_jup)
\newcommand{\hatcurPPrshortxxxxxB}{\ensuremath{1.59}}           % planetary radius (R_jup)
\newcommand{\hatcurPPrlongxxxxxB}{\ensuremath{1.59_{-0.10}^{+0.16}}} % planetary radius (R_jup)
\newcommand{\hatcurPPrexxxxxB}{\ensuremath{17.9_{-1.1}^{+1.8}}} % planetary radius (R_earth)
\newcommand{\hatcurPPreshortxxxxxB}{\ensuremath{17.9}}          % planetary radius (R_earth)
\newcommand{\hatcurPPrelongxxxxxB}{\ensuremath{17.9_{-1.1}^{+1.8}}} % planetary radius (R_earth)
\newcommand{\hatcurPPmrcorrxxxxxB}{\ensuremath{0.29}}           % mass/radius correlation
\newcommand{\hatcurPPteffxxxxxB}{\ensuremath{1896_{-42}^{+66}}} % planetary temperature (K)
\newcommand{\hatcurPPthetaxxxxxB}{\ensuremath{0.0336\pm0.0034}} % Safranov number
\newcommand{\hatcurPPfluxperixxxxxB}{\ensuremath{2.92_{-0.25}^{+0.43}}} % flux @ periastron (CGS)
\newcommand{\hatcurPPfluxperidimxxxxxB}{\ensuremath{9}}         % flux @ periastron (CGS) units.
\newcommand{\hatcurPPfluxapxxxxxB}{\ensuremath{2.92_{-0.25}^{+0.43}}} % flux @ apastron (CGS)
\newcommand{\hatcurPPfluxapdimxxxxxB}{\ensuremath{9}}           % flux @ apastron (CGS) units.
\newcommand{\hatcurPPfluxavgxxxxxB}{\ensuremath{2.92_{-0.25}^{+0.43}}} % flux on average (CGS)
\newcommand{\hatcurPPfluxavgdimxxxxxB}{\ensuremath{9}}          % flux average (CGS) units.
\newcommand{\hatcurPPfluxavglogxxxxxB}{\ensuremath{9.465_{-0.039}^{+0.060}}} % log10 flux on average (CGS)
\newcommand{\hatcurXsecphasexxxxxB}{\ensuremath{0\pm0}}         % Phase of secondary eclipse
\newcommand{\hatcurXsecondaryxxxxxB}{\ensuremath{2457260.28511\pm0.00072}} % Secondary eclipse epoch
\newcommand{\hatcurXsecdurxxxxxB}{\ensuremath{0.1958\pm0.0028}} % sec eclipse duration (days)
\newcommand{\hatcurXsecingdurxxxxxB}{\ensuremath{0.0174\pm0.0025}} % sec I/E duration (days)
\newcommand{\hatcurPPphiconjxxxxxB}{\ensuremath{0\pm0}}         % phase diff between conjunction and periastron
\newcommand{\hatcurPPperixxxxxB}{\ensuremath{2457258.05605\pm0.00072}} % time of periastron passage.
\newcommand{\hatcurPPaequivxxxxxB}{\ensuremath{0.02160_{-0.00140}^{+0.00100}}} % equivalent semi-major axis
\newcommand{\hatcurPPtcircxxxxxB}{\ensuremath{31\pm10}}         % circularization timescale
\newcommand{\hatcurPPtinfallxxxxxB}{\ensuremath{143\pm35}}      % infall timescale
\newcommand{\hatcurXdistxxxxxB}{\ensuremath{935_{-49}^{+75}}}   % distance (pc), no reddenning correction
\newcommand{\hatcurXAvxxxxxB}{\ensuremath{0.0000\pm0.0062}}     % Av (mag)
\newcommand{\hatcurXdistredxxxxxB}{\ensuremath{927_{-49}^{+75}}} % distance with Av correction (pc)
\newcommand{\hatcurXEBVxxxxxB}{\ensuremath{0.0000\pm0.0020}}    % E(B-V) (mag)
\newcommand{\hatcurXmvisoredxxxxxB}{\ensuremath{13.097\pm0.029}} % Expected m_v with reddening (mag)
\newcommand{\hatcurXmiisoredxxxxxB}{\ensuremath{12.468\pm0.018}} % Expected m_i with reddening (mag)
\newcommand{\hatcurXmjisoredxxxxxB}{\ensuremath{12.053\pm0.014}} % Expected m_j with reddening (mag)
\newcommand{\hatcurXmhisoredxxxxxB}{\ensuremath{11.751\pm0.014}} % Expected m_h with reddening (mag)
\newcommand{\hatcurXmkisoredxxxxxB}{\ensuremath{11.697\pm0.014}} % Expected m_k with reddening (mag)
\newcommand{\hatcurXviisoredxxxxxB}{\ensuremath{0.629\pm0.014}} % Expected V-I with reddening (mag)
\newcommand{\hatcurXvkisoredxxxxxB}{\ensuremath{1.400\pm0.031}} % Expected V-K with reddening (mag)
\newcommand{\hatcurXjhisoredxxxxxB}{\ensuremath{0.3030\pm0.0081}} % Expected J-H with reddening (mag)
\newcommand{\hatcurXjkisoredxxxxxB}{\ensuremath{0.3560\pm0.0090}} % Expected J-K with reddening (mag)
\newcommand{\hatcurCCpmraxxxxxB}{\ensuremath{-9.2\pm1.8}}       % proper motion, in RA
\newcommand{\hatcurCCpmdecxxxxxB}{\ensuremath{-11.4\pm2.4}}     % proper motion, in DEC
\newcommand{\hatcurCCpmxxxxxB}{\ensuremath{14.6\pm3.0}}         % proper motion

\newcommand{\hatcurCCbbHmag}[1]{\ifnum#1=65 %
\hatcurCCbbHmagxxxxxA
\else
\ifnum#1=66 %
\hatcurCCbbHmagxxxxxB
\else
??????\fi
\fi
}
\newcommand{\hatcurCCbbJmag}[1]{\ifnum#1=65 %
\hatcurCCbbJmagxxxxxA
\else
\ifnum#1=66 %
\hatcurCCbbJmagxxxxxB
\else
??????\fi
\fi
}
\newcommand{\hatcurCCbbKmag}[1]{\ifnum#1=65 %
\hatcurCCbbKmagxxxxxA
\else
\ifnum#1=66 %
\hatcurCCbbKmagxxxxxB
\else
??????\fi
\fi
}
\newcommand{\hatcurCCcitHmag}[1]{\ifnum#1=65 %
\hatcurCCcitHmagxxxxxA
\else
\ifnum#1=66 %
\hatcurCCcitHmagxxxxxB
\else
??????\fi
\fi
}
\newcommand{\hatcurCCcitJmag}[1]{\ifnum#1=65 %
\hatcurCCcitJmagxxxxxA
\else
\ifnum#1=66 %
\hatcurCCcitJmagxxxxxB
\else
??????\fi
\fi
}
\newcommand{\hatcurCCcitKmag}[1]{\ifnum#1=65 %
\hatcurCCcitKmagxxxxxA
\else
\ifnum#1=66 %
\hatcurCCcitKmagxxxxxB
\else
??????\fi
\fi
}
\newcommand{\hatcurCCdec}[1]{\ifnum#1=65 %
\hatcurCCdecxxxxxA
\else
\ifnum#1=66 %
\hatcurCCdecxxxxxB
\else
??????\fi
\fi
}
\newcommand{\hatcurCCesoHKmag}[1]{\ifnum#1=65 %
\hatcurCCesoHKmagxxxxxA
\else
\ifnum#1=66 %
\hatcurCCesoHKmagxxxxxB
\else
??????\fi
\fi
}
\newcommand{\hatcurCCesoHmag}[1]{\ifnum#1=65 %
\hatcurCCesoHmagxxxxxA
\else
\ifnum#1=66 %
\hatcurCCesoHmagxxxxxB
\else
??????\fi
\fi
}
\newcommand{\hatcurCCesoJHmag}[1]{\ifnum#1=65 %
\hatcurCCesoJHmagxxxxxA
\else
\ifnum#1=66 %
\hatcurCCesoJHmagxxxxxB
\else
??????\fi
\fi
}
\newcommand{\hatcurCCesoJKmag}[1]{\ifnum#1=65 %
\hatcurCCesoJKmagxxxxxA
\else
\ifnum#1=66 %
\hatcurCCesoJKmagxxxxxB
\else
??????\fi
\fi
}
\newcommand{\hatcurCCesoJmag}[1]{\ifnum#1=65 %
\hatcurCCesoJmagxxxxxA
\else
\ifnum#1=66 %
\hatcurCCesoJmagxxxxxB
\else
??????\fi
\fi
}
\newcommand{\hatcurCCesoKmag}[1]{\ifnum#1=65 %
\hatcurCCesoKmagxxxxxA
\else
\ifnum#1=66 %
\hatcurCCesoKmagxxxxxB
\else
??????\fi
\fi
}
\newcommand{\hatcurCCgsc}[1]{\ifnum#1=65 %
\hatcurCCgscxxxxxA
\else
\ifnum#1=66 %
\hatcurCCgscxxxxxB
\else
??????\fi
\fi
}
\newcommand{\hatcurCCmag}[1]{\ifnum#1=65 %
\hatcurCCmagxxxxxA
\else
\ifnum#1=66 %
\hatcurCCmagxxxxxB
\else
??????\fi
\fi
}
\newcommand{\hatcurCCpm}[1]{\ifnum#1=65 %
\hatcurCCpmxxxxxA
\else
\ifnum#1=66 %
\hatcurCCpmxxxxxB
\else
??????\fi
\fi
}
\newcommand{\hatcurCCpmdec}[1]{\ifnum#1=65 %
\hatcurCCpmdecxxxxxA
\else
\ifnum#1=66 %
\hatcurCCpmdecxxxxxB
\else
??????\fi
\fi
}
\newcommand{\hatcurCCpmra}[1]{\ifnum#1=65 %
\hatcurCCpmraxxxxxA
\else
\ifnum#1=66 %
\hatcurCCpmraxxxxxB
\else
??????\fi
\fi
}
\newcommand{\hatcurCCra}[1]{\ifnum#1=65 %
\hatcurCCraxxxxxA
\else
\ifnum#1=66 %
\hatcurCCraxxxxxB
\else
??????\fi
\fi
}
\newcommand{\hatcurCCtassmB}[1]{\ifnum#1=65 %
\hatcurCCtassmBxxxxxA
\else
\ifnum#1=66 %
\hatcurCCtassmBxxxxxB
\else
??????\fi
\fi
}
\newcommand{\hatcurCCtassmBshort}[1]{\ifnum#1=65 %
\hatcurCCtassmBshortxxxxxA
\else
\ifnum#1=66 %
\hatcurCCtassmBshortxxxxxB
\else
??????\fi
\fi
}
\newcommand{\hatcurCCtassmg}[1]{\ifnum#1=65 %
\hatcurCCtassmgxxxxxA
\else
\ifnum#1=66 %
\hatcurCCtassmgxxxxxB
\else
??????\fi
\fi
}
\newcommand{\hatcurCCtassmgshort}[1]{\ifnum#1=65 %
\hatcurCCtassmgshortxxxxxA
\else
\ifnum#1=66 %
\hatcurCCtassmgshortxxxxxB
\else
??????\fi
\fi
}
\newcommand{\hatcurCCtassmi}[1]{\ifnum#1=65 %
\hatcurCCtassmixxxxxA
\else
\ifnum#1=66 %
\hatcurCCtassmixxxxxB
\else
??????\fi
\fi
}
\newcommand{\hatcurCCtassmI}[1]{\ifnum#1=65 %
\hatcurCCtassmIxxxxxA
\else
\ifnum#1=66 %
\hatcurCCtassmIxxxxxB
\else
??????\fi
\fi
}
\newcommand{\hatcurCCtassmishort}[1]{\ifnum#1=65 %
\hatcurCCtassmishortxxxxxA
\else
\ifnum#1=66 %
\hatcurCCtassmishortxxxxxB
\else
??????\fi
\fi
}
\newcommand{\hatcurCCtassmIshort}[1]{\ifnum#1=65 %
\hatcurCCtassmIshortxxxxxA
\else
\ifnum#1=66 %
\hatcurCCtassmIshortxxxxxB
\else
??????\fi
\fi
}
\newcommand{\hatcurCCtassmr}[1]{\ifnum#1=65 %
\hatcurCCtassmrxxxxxA
\else
\ifnum#1=66 %
\hatcurCCtassmrxxxxxB
\else
??????\fi
\fi
}
\newcommand{\hatcurCCtassmrshort}[1]{\ifnum#1=65 %
\hatcurCCtassmrshortxxxxxA
\else
\ifnum#1=66 %
\hatcurCCtassmrshortxxxxxB
\else
??????\fi
\fi
}
\newcommand{\hatcurCCtassmv}[1]{\ifnum#1=65 %
\hatcurCCtassmvxxxxxA
\else
\ifnum#1=66 %
\hatcurCCtassmvxxxxxB
\else
??????\fi
\fi
}
\newcommand{\hatcurCCtassmvshort}[1]{\ifnum#1=65 %
\hatcurCCtassmvshortxxxxxA
\else
\ifnum#1=66 %
\hatcurCCtassmvshortxxxxxB
\else
??????\fi
\fi
}
\newcommand{\hatcurCCtwomass}[1]{\ifnum#1=65 %
\hatcurCCtwomassxxxxxA
\else
\ifnum#1=66 %
\hatcurCCtwomassxxxxxB
\else
??????\fi
\fi
}
\newcommand{\hatcurCCtwomassHmag}[1]{\ifnum#1=65 %
\hatcurCCtwomassHmagxxxxxA
\else
\ifnum#1=66 %
\hatcurCCtwomassHmagxxxxxB
\else
??????\fi
\fi
}
\newcommand{\hatcurCCtwomassJmag}[1]{\ifnum#1=65 %
\hatcurCCtwomassJmagxxxxxA
\else
\ifnum#1=66 %
\hatcurCCtwomassJmagxxxxxB
\else
??????\fi
\fi
}
\newcommand{\hatcurCCtwomassKmag}[1]{\ifnum#1=65 %
\hatcurCCtwomassKmagxxxxxA
\else
\ifnum#1=66 %
\hatcurCCtwomassKmagxxxxxB
\else
??????\fi
\fi
}
\newcommand{\hatcurfield}[1]{\ifnum#1=65 %
\hatcurfieldxxxxxA
\else
\ifnum#1=66 %
\hatcurfieldxxxxxB
\else
??????\fi
\fi
}
\newcommand{\hatcurhtr}[1]{\ifnum#1=65 %
\hatcurhtrxxxxxA
\else
\ifnum#1=66 %
\hatcurhtrxxxxxB
\else
??????\fi
\fi
}
\newcommand{\hatcurISOage}[1]{\ifnum#1=65 %
\hatcurISOagexxxxxA
\else
\ifnum#1=66 %
\hatcurISOagexxxxxB
\else
??????\fi
\fi
}
\newcommand{\hatcurISOJK}[1]{\ifnum#1=65 %
\hatcurISOJKxxxxxA
\else
\ifnum#1=66 %
\hatcurISOJKxxxxxB
\else
??????\fi
\fi
}
\newcommand{\hatcurISOlogg}[1]{\ifnum#1=65 %
\hatcurISOloggxxxxxA
\else
\ifnum#1=66 %
\hatcurISOloggxxxxxB
\else
??????\fi
\fi
}
\newcommand{\hatcurISOlum}[1]{\ifnum#1=65 %
\hatcurISOlumxxxxxA
\else
\ifnum#1=66 %
\hatcurISOlumxxxxxB
\else
??????\fi
\fi
}
\newcommand{\hatcurISOlumshort}[1]{\ifnum#1=65 %
\hatcurISOlumshortxxxxxA
\else
\ifnum#1=66 %
\hatcurISOlumshortxxxxxB
\else
??????\fi
\fi
}
\newcommand{\hatcurISOm}[1]{\ifnum#1=65 %
\hatcurISOmxxxxxA
\else
\ifnum#1=66 %
\hatcurISOmxxxxxB
\else
??????\fi
\fi
}
\newcommand{\hatcurISOMH}[1]{\ifnum#1=65 %
\hatcurISOMHxxxxxA
\else
\ifnum#1=66 %
\hatcurISOMHxxxxxB
\else
??????\fi
\fi
}
\newcommand{\hatcurISOMJ}[1]{\ifnum#1=65 %
\hatcurISOMJxxxxxA
\else
\ifnum#1=66 %
\hatcurISOMJxxxxxB
\else
??????\fi
\fi
}
\newcommand{\hatcurISOMK}[1]{\ifnum#1=65 %
\hatcurISOMKxxxxxA
\else
\ifnum#1=66 %
\hatcurISOMKxxxxxB
\else
??????\fi
\fi
}
\newcommand{\hatcurISOmlong}[1]{\ifnum#1=65 %
\hatcurISOmlongxxxxxA
\else
\ifnum#1=66 %
\hatcurISOmlongxxxxxB
\else
??????\fi
\fi
}
\newcommand{\hatcurISOmshort}[1]{\ifnum#1=65 %
\hatcurISOmshortxxxxxA
\else
\ifnum#1=66 %
\hatcurISOmshortxxxxxB
\else
??????\fi
\fi
}
\newcommand{\hatcurISOmv}[1]{\ifnum#1=65 %
\hatcurISOmvxxxxxA
\else
\ifnum#1=66 %
\hatcurISOmvxxxxxB
\else
??????\fi
\fi
}
\newcommand{\hatcurISOr}[1]{\ifnum#1=65 %
\hatcurISOrxxxxxA
\else
\ifnum#1=66 %
\hatcurISOrxxxxxB
\else
??????\fi
\fi
}
\newcommand{\hatcurISOrho}[1]{\ifnum#1=65 %
\hatcurISOrhoxxxxxA
\else
\ifnum#1=66 %
\hatcurISOrhoxxxxxB
\else
??????\fi
\fi
}
\newcommand{\hatcurISOrholong}[1]{\ifnum#1=65 %
\hatcurISOrholongxxxxxA
\else
\ifnum#1=66 %
\hatcurISOrholongxxxxxB
\else
??????\fi
\fi
}
\newcommand{\hatcurISOrlong}[1]{\ifnum#1=65 %
\hatcurISOrlongxxxxxA
\else
\ifnum#1=66 %
\hatcurISOrlongxxxxxB
\else
??????\fi
\fi
}
\newcommand{\hatcurISOrshort}[1]{\ifnum#1=65 %
\hatcurISOrshortxxxxxA
\else
\ifnum#1=66 %
\hatcurISOrshortxxxxxB
\else
??????\fi
\fi
}
\newcommand{\hatcurISOsigma}[1]{\ifnum#1=65 %
\hatcurISOsigmaxxxxxA
\else
\ifnum#1=66 %
\hatcurISOsigmaxxxxxB
\else
??????\fi
\fi
}
\newcommand{\hatcurISOspec}[1]{\ifnum#1=65 %
\hatcurISOspecxxxxxA
\else
\ifnum#1=66 %
\hatcurISOspecxxxxxB
\else
??????\fi
\fi
}
\newcommand{\hatcurISOvi}[1]{\ifnum#1=65 %
\hatcurISOvixxxxxA
\else
\ifnum#1=66 %
\hatcurISOvixxxxxB
\else
??????\fi
\fi
}
\newcommand{\hatcurLBig}[1]{\ifnum#1=65 %
\hatcurLBigxxxxxA
\else
\ifnum#1=66 %
\hatcurLBigxxxxxB
\else
??????\fi
\fi
}
\newcommand{\hatcurLBii}[1]{\ifnum#1=65 %
\hatcurLBiixxxxxA
\else
\ifnum#1=66 %
\hatcurLBiixxxxxB
\else
??????\fi
\fi
}
\newcommand{\hatcurLBiI}[1]{\ifnum#1=65 %
\hatcurLBiIxxxxxA
\else
\ifnum#1=66 %
\hatcurLBiIxxxxxB
\else
??????\fi
\fi
}
\newcommand{\hatcurLBiig}[1]{\ifnum#1=65 %
\hatcurLBiigxxxxxA
\else
\ifnum#1=66 %
\hatcurLBiigxxxxxB
\else
??????\fi
\fi
}
\newcommand{\hatcurLBiii}[1]{\ifnum#1=65 %
\hatcurLBiiixxxxxA
\else
\ifnum#1=66 %
\hatcurLBiiixxxxxB
\else
??????\fi
\fi
}
\newcommand{\hatcurLBiiI}[1]{\ifnum#1=65 %
\hatcurLBiiIxxxxxA
\else
\ifnum#1=66 %
\hatcurLBiiIxxxxxB
\else
??????\fi
\fi
}
\newcommand{\hatcurLBiikep}[1]{\ifnum#1=65 %
\hatcurLBiikepxxxxxA
\else
\ifnum#1=66 %
\hatcurLBiikepxxxxxB
\else
??????\fi
\fi
}
\newcommand{\hatcurLBiir}[1]{\ifnum#1=65 %
\hatcurLBiirxxxxxA
\else
\ifnum#1=66 %
\hatcurLBiirxxxxxB
\else
??????\fi
\fi
}
\newcommand{\hatcurLBiiR}[1]{\ifnum#1=65 %
\hatcurLBiiRxxxxxA
\else
\ifnum#1=66 %
\hatcurLBiiRxxxxxB
\else
??????\fi
\fi
}
\newcommand{\hatcurLBiiz}[1]{\ifnum#1=65 %
\hatcurLBiizxxxxxA
\else
\ifnum#1=66 %
\hatcurLBiizxxxxxB
\else
??????\fi
\fi
}
\newcommand{\hatcurLBikep}[1]{\ifnum#1=65 %
\hatcurLBikepxxxxxA
\else
\ifnum#1=66 %
\hatcurLBikepxxxxxB
\else
??????\fi
\fi
}
\newcommand{\hatcurLBir}[1]{\ifnum#1=65 %
\hatcurLBirxxxxxA
\else
\ifnum#1=66 %
\hatcurLBirxxxxxB
\else
??????\fi
\fi
}
\newcommand{\hatcurLBiR}[1]{\ifnum#1=65 %
\hatcurLBiRxxxxxA
\else
\ifnum#1=66 %
\hatcurLBiRxxxxxB
\else
??????\fi
\fi
}
\newcommand{\hatcurLBiz}[1]{\ifnum#1=65 %
\hatcurLBizxxxxxA
\else
\ifnum#1=66 %
\hatcurLBizxxxxxB
\else
??????\fi
\fi
}
\newcommand{\hatcurLCbsq}[1]{\ifnum#1=65 %
\hatcurLCbsqxxxxxA
\else
\ifnum#1=66 %
\hatcurLCbsqxxxxxB
\else
??????\fi
\fi
}
\newcommand{\hatcurLCdip}[1]{\ifnum#1=65 %
\hatcurLCdipxxxxxA
\else
\ifnum#1=66 %
\hatcurLCdipxxxxxB
\else
??????\fi
\fi
}
\newcommand{\hatcurLCdur}[1]{\ifnum#1=65 %
\hatcurLCdurxxxxxA
\else
\ifnum#1=66 %
\hatcurLCdurxxxxxB
\else
??????\fi
\fi
}
\newcommand{\hatcurLCdurhr}[1]{\ifnum#1=65 %
\hatcurLCdurhrxxxxxA
\else
\ifnum#1=66 %
\hatcurLCdurhrxxxxxB
\else
??????\fi
\fi
}
\newcommand{\hatcurLCdurhrshort}[1]{\ifnum#1=65 %
\hatcurLCdurhrshortxxxxxA
\else
\ifnum#1=66 %
\hatcurLCdurhrshortxxxxxB
\else
??????\fi
\fi
}
\newcommand{\hatcurLCdurshort}[1]{\ifnum#1=65 %
\hatcurLCdurshortxxxxxA
\else
\ifnum#1=66 %
\hatcurLCdurshortxxxxxB
\else
??????\fi
\fi
}
\newcommand{\hatcurLChatnetm}[1]{\ifnum#1=65 %
\hatcurLChatnetmxxxxxA
\else
\ifnum#1=66 %
\hatcurLChatnetmxxxxxB
\else
??????\fi
\fi
}
\newcommand{\hatcurLCiblend}[1]{\ifnum#1=65 %
\hatcurLCiblendxxxxxA
\else
\ifnum#1=66 %
\hatcurLCiblendxxxxxB
\else
??????\fi
\fi
}
\newcommand{\hatcurLCimp}[1]{\ifnum#1=65 %
\hatcurLCimpxxxxxA
\else
\ifnum#1=66 %
\hatcurLCimpxxxxxB
\else
??????\fi
\fi
}
\newcommand{\hatcurLCingdur}[1]{\ifnum#1=65 %
\hatcurLCingdurxxxxxA
\else
\ifnum#1=66 %
\hatcurLCingdurxxxxxB
\else
??????\fi
\fi
}
\newcommand{\hatcurLCP}[1]{\ifnum#1=65 %
\hatcurLCPxxxxxA
\else
\ifnum#1=66 %
\hatcurLCPxxxxxB
\else
??????\fi
\fi
}
\newcommand{\hatcurLCPprec}[1]{\ifnum#1=65 %
\hatcurLCPprecxxxxxA
\else
\ifnum#1=66 %
\hatcurLCPprecxxxxxB
\else
??????\fi
\fi
}
\newcommand{\hatcurLCPshort}[1]{\ifnum#1=65 %
\hatcurLCPshortxxxxxA
\else
\ifnum#1=66 %
\hatcurLCPshortxxxxxB
\else
??????\fi
\fi
}
\newcommand{\hatcurLCq}[1]{\ifnum#1=65 %
\hatcurLCqxxxxxA
\else
\ifnum#1=66 %
\hatcurLCqxxxxxB
\else
??????\fi
\fi
}
\newcommand{\hatcurLCqshort}[1]{\ifnum#1=65 %
\hatcurLCqshortxxxxxA
\else
\ifnum#1=66 %
\hatcurLCqshortxxxxxB
\else
??????\fi
\fi
}
\newcommand{\hatcurLCrho}[1]{\ifnum#1=65 %
\hatcurLCrhoxxxxxA
\else
\ifnum#1=66 %
\hatcurLCrhoxxxxxB
\else
??????\fi
\fi
}
\newcommand{\hatcurLCrprstar}[1]{\ifnum#1=65 %
\hatcurLCrprstarxxxxxA
\else
\ifnum#1=66 %
\hatcurLCrprstarxxxxxB
\else
??????\fi
\fi
}
\newcommand{\hatcurLCT}[1]{\ifnum#1=65 %
\hatcurLCTxxxxxA
\else
\ifnum#1=66 %
\hatcurLCTxxxxxB
\else
??????\fi
\fi
}
\newcommand{\hatcurLCTA}[1]{\ifnum#1=65 %
\hatcurLCTAxxxxxA
\else
\ifnum#1=66 %
\hatcurLCTAxxxxxB
\else
??????\fi
\fi
}
\newcommand{\hatcurLCTB}[1]{\ifnum#1=65 %
\hatcurLCTBxxxxxA
\else
\ifnum#1=66 %
\hatcurLCTBxxxxxB
\else
??????\fi
\fi
}
\newcommand{\hatcurLCzeta}[1]{\ifnum#1=65 %
\hatcurLCzetaxxxxxA
\else
\ifnum#1=66 %
\hatcurLCzetaxxxxxB
\else
??????\fi
\fi
}
\newcommand{\hatcurPPaequiv}[1]{\ifnum#1=65 %
\hatcurPPaequivxxxxxA
\else
\ifnum#1=66 %
\hatcurPPaequivxxxxxB
\else
??????\fi
\fi
}
\newcommand{\hatcurPPar}[1]{\ifnum#1=65 %
\hatcurPParxxxxxA
\else
\ifnum#1=66 %
\hatcurPParxxxxxB
\else
??????\fi
\fi
}
\newcommand{\hatcurPParel}[1]{\ifnum#1=65 %
\hatcurPParelxxxxxA
\else
\ifnum#1=66 %
\hatcurPParelxxxxxB
\else
??????\fi
\fi
}
\newcommand{\hatcurPPfluxap}[1]{\ifnum#1=65 %
\hatcurPPfluxapxxxxxA
\else
\ifnum#1=66 %
\hatcurPPfluxapxxxxxB
\else
??????\fi
\fi
}
\newcommand{\hatcurPPfluxapdim}[1]{\ifnum#1=65 %
\hatcurPPfluxapdimxxxxxA
\else
\ifnum#1=66 %
\hatcurPPfluxapdimxxxxxB
\else
??????\fi
\fi
}
\newcommand{\hatcurPPfluxavg}[1]{\ifnum#1=65 %
\hatcurPPfluxavgxxxxxA
\else
\ifnum#1=66 %
\hatcurPPfluxavgxxxxxB
\else
??????\fi
\fi
}
\newcommand{\hatcurPPfluxavgdim}[1]{\ifnum#1=65 %
\hatcurPPfluxavgdimxxxxxA
\else
\ifnum#1=66 %
\hatcurPPfluxavgdimxxxxxB
\else
??????\fi
\fi
}
\newcommand{\hatcurPPfluxavglog}[1]{\ifnum#1=65 %
\hatcurPPfluxavglogxxxxxA
\else
\ifnum#1=66 %
\hatcurPPfluxavglogxxxxxB
\else
??????\fi
\fi
}
\newcommand{\hatcurPPfluxperi}[1]{\ifnum#1=65 %
\hatcurPPfluxperixxxxxA
\else
\ifnum#1=66 %
\hatcurPPfluxperixxxxxB
\else
??????\fi
\fi
}
\newcommand{\hatcurPPfluxperidim}[1]{\ifnum#1=65 %
\hatcurPPfluxperidimxxxxxA
\else
\ifnum#1=66 %
\hatcurPPfluxperidimxxxxxB
\else
??????\fi
\fi
}
\newcommand{\hatcurPPg}[1]{\ifnum#1=65 %
\hatcurPPgxxxxxA
\else
\ifnum#1=66 %
\hatcurPPgxxxxxB
\else
??????\fi
\fi
}
\newcommand{\hatcurPPi}[1]{\ifnum#1=65 %
\hatcurPPixxxxxA
\else
\ifnum#1=66 %
\hatcurPPixxxxxB
\else
??????\fi
\fi
}
\newcommand{\hatcurPPlogg}[1]{\ifnum#1=65 %
\hatcurPPloggxxxxxA
\else
\ifnum#1=66 %
\hatcurPPloggxxxxxB
\else
??????\fi
\fi
}
\newcommand{\hatcurPPm}[1]{\ifnum#1=65 %
\hatcurPPmxxxxxA
\else
\ifnum#1=66 %
\hatcurPPmxxxxxB
\else
??????\fi
\fi
}
\newcommand{\hatcurPPme}[1]{\ifnum#1=65 %
\hatcurPPmexxxxxA
\else
\ifnum#1=66 %
\hatcurPPmexxxxxB
\else
??????\fi
\fi
}
\newcommand{\hatcurPPmelong}[1]{\ifnum#1=65 %
\hatcurPPmelongxxxxxA
\else
\ifnum#1=66 %
\hatcurPPmelongxxxxxB
\else
??????\fi
\fi
}
\newcommand{\hatcurPPmeshort}[1]{\ifnum#1=65 %
\hatcurPPmeshortxxxxxA
\else
\ifnum#1=66 %
\hatcurPPmeshortxxxxxB
\else
??????\fi
\fi
}
\newcommand{\hatcurPPmlong}[1]{\ifnum#1=65 %
\hatcurPPmlongxxxxxA
\else
\ifnum#1=66 %
\hatcurPPmlongxxxxxB
\else
??????\fi
\fi
}
\newcommand{\hatcurPPmrcorr}[1]{\ifnum#1=65 %
\hatcurPPmrcorrxxxxxA
\else
\ifnum#1=66 %
\hatcurPPmrcorrxxxxxB
\else
??????\fi
\fi
}
\newcommand{\hatcurPPmshort}[1]{\ifnum#1=65 %
\hatcurPPmshortxxxxxA
\else
\ifnum#1=66 %
\hatcurPPmshortxxxxxB
\else
??????\fi
\fi
}
\newcommand{\hatcurPPperi}[1]{\ifnum#1=65 %
\hatcurPPperixxxxxA
\else
\ifnum#1=66 %
\hatcurPPperixxxxxB
\else
??????\fi
\fi
}
\newcommand{\hatcurPPphiconj}[1]{\ifnum#1=65 %
\hatcurPPphiconjxxxxxA
\else
\ifnum#1=66 %
\hatcurPPphiconjxxxxxB
\else
??????\fi
\fi
}
\newcommand{\hatcurPPr}[1]{\ifnum#1=65 %
\hatcurPPrxxxxxA
\else
\ifnum#1=66 %
\hatcurPPrxxxxxB
\else
??????\fi
\fi
}
\newcommand{\hatcurPPre}[1]{\ifnum#1=65 %
\hatcurPPrexxxxxA
\else
\ifnum#1=66 %
\hatcurPPrexxxxxB
\else
??????\fi
\fi
}
\newcommand{\hatcurPPrelong}[1]{\ifnum#1=65 %
\hatcurPPrelongxxxxxA
\else
\ifnum#1=66 %
\hatcurPPrelongxxxxxB
\else
??????\fi
\fi
}
\newcommand{\hatcurPPreshort}[1]{\ifnum#1=65 %
\hatcurPPreshortxxxxxA
\else
\ifnum#1=66 %
\hatcurPPreshortxxxxxB
\else
??????\fi
\fi
}
\newcommand{\hatcurPPrho}[1]{\ifnum#1=65 %
\hatcurPPrhoxxxxxA
\else
\ifnum#1=66 %
\hatcurPPrhoxxxxxB
\else
??????\fi
\fi
}
\newcommand{\hatcurPPrlong}[1]{\ifnum#1=65 %
\hatcurPPrlongxxxxxA
\else
\ifnum#1=66 %
\hatcurPPrlongxxxxxB
\else
??????\fi
\fi
}
\newcommand{\hatcurPPrshort}[1]{\ifnum#1=65 %
\hatcurPPrshortxxxxxA
\else
\ifnum#1=66 %
\hatcurPPrshortxxxxxB
\else
??????\fi
\fi
}
\newcommand{\hatcurPPtcirc}[1]{\ifnum#1=65 %
\hatcurPPtcircxxxxxA
\else
\ifnum#1=66 %
\hatcurPPtcircxxxxxB
\else
??????\fi
\fi
}
\newcommand{\hatcurPPteff}[1]{\ifnum#1=65 %
\hatcurPPteffxxxxxA
\else
\ifnum#1=66 %
\hatcurPPteffxxxxxB
\else
??????\fi
\fi
}
\newcommand{\hatcurPPtheta}[1]{\ifnum#1=65 %
\hatcurPPthetaxxxxxA
\else
\ifnum#1=66 %
\hatcurPPthetaxxxxxB
\else
??????\fi
\fi
}
\newcommand{\hatcurPPtinfall}[1]{\ifnum#1=65 %
\hatcurPPtinfallxxxxxA
\else
\ifnum#1=66 %
\hatcurPPtinfallxxxxxB
\else
??????\fi
\fi
}
\newcommand{\hatcurRVeccen}[1]{\ifnum#1=65 %
\hatcurRVeccenxxxxxA
\else
\ifnum#1=66 %
\hatcurRVeccenxxxxxB
\else
??????\fi
\fi
}
\newcommand{\hatcurRVeccentwosiglim}[1]{\ifnum#1=65 %
\hatcurRVeccentwosiglimxxxxxA
\else
\ifnum#1=66 %
\hatcurRVeccentwosiglimxxxxxB
\else
??????\fi
\fi
}
\newcommand{\hatcurRVfitrms}[1]{\ifnum#1=65 %
\hatcurRVfitrmsxxxxxA
\else
??????\fi
}
\newcommand{\hatcurRVfitrmsA}[1]{\ifnum#1=66 %
\hatcurRVfitrmsAxxxxxB
\else
??????\fi
}
\newcommand{\hatcurRVfitrmsB}[1]{\ifnum#1=66 %
\hatcurRVfitrmsBxxxxxB
\else
??????\fi
}
\newcommand{\hatcurRVgamma}[1]{\ifnum#1=65 %
\hatcurRVgammaxxxxxA
\else
??????\fi
}
\newcommand{\hatcurRVgammaA}[1]{\ifnum#1=66 %
\hatcurRVgammaAxxxxxB
\else
??????\fi
}
\newcommand{\hatcurRVgammaB}[1]{\ifnum#1=66 %
\hatcurRVgammaBxxxxxB
\else
??????\fi
}
\newcommand{\hatcurRVh}[1]{\ifnum#1=65 %
\hatcurRVhxxxxxA
\else
\ifnum#1=66 %
\hatcurRVhxxxxxB
\else
??????\fi
\fi
}
\newcommand{\hatcurRVjitter}[1]{\ifnum#1=65 %
\hatcurRVjitterxxxxxA
\else
??????\fi
}
\newcommand{\hatcurRVjitterA}[1]{\ifnum#1=66 %
\hatcurRVjitterAxxxxxB
\else
??????\fi
}
\newcommand{\hatcurRVjitterB}[1]{\ifnum#1=66 %
\hatcurRVjitterBxxxxxB
\else
??????\fi
}
\newcommand{\hatcurRVjitterC}[1]{\ifnum#1=66 %
\hatcurRVjitterCxxxxxB
\else
??????\fi
}
\newcommand{\hatcurRVjittertwosiglim}[1]{\ifnum#1=65 %
\hatcurRVjittertwosiglimxxxxxA
\else
??????\fi
}
\newcommand{\hatcurRVjittertwosiglimA}[1]{\ifnum#1=66 %
\hatcurRVjittertwosiglimAxxxxxB
\else
??????\fi
}
\newcommand{\hatcurRVjittertwosiglimB}[1]{\ifnum#1=66 %
\hatcurRVjittertwosiglimBxxxxxB
\else
??????\fi
}
\newcommand{\hatcurRVjittertwosiglimC}[1]{\ifnum#1=66 %
\hatcurRVjittertwosiglimCxxxxxB
\else
??????\fi
}
\newcommand{\hatcurRVk}[1]{\ifnum#1=65 %
\hatcurRVkxxxxxA
\else
\ifnum#1=66 %
\hatcurRVkxxxxxB
\else
??????\fi
\fi
}
\newcommand{\hatcurRVK}[1]{\ifnum#1=65 %
\hatcurRVKxxxxxA
\else
\ifnum#1=66 %
\hatcurRVKxxxxxB
\else
??????\fi
\fi
}
\newcommand{\hatcurRVomega}[1]{\ifnum#1=65 %
\hatcurRVomegaxxxxxA
\else
\ifnum#1=66 %
\hatcurRVomegaxxxxxB
\else
??????\fi
\fi
}
\newcommand{\hatcurRVrh}[1]{\ifnum#1=65 %
\hatcurRVrhxxxxxA
\else
\ifnum#1=66 %
\hatcurRVrhxxxxxB
\else
??????\fi
\fi
}
\newcommand{\hatcurRVrk}[1]{\ifnum#1=65 %
\hatcurRVrkxxxxxA
\else
\ifnum#1=66 %
\hatcurRVrkxxxxxB
\else
??????\fi
\fi
}
\newcommand{\hatcurRVtrone}[1]{\ifnum#1=65 %
\hatcurRVtronexxxxxA
\else
\ifnum#1=66 %
\hatcurRVtronexxxxxB
\else
??????\fi
\fi
}
\newcommand{\hatcurRVtrtwo}[1]{\ifnum#1=65 %
\hatcurRVtrtwoxxxxxA
\else
\ifnum#1=66 %
\hatcurRVtrtwoxxxxxB
\else
??????\fi
\fi
}
\newcommand{\hatcurSMEiilogg}[1]{\ifnum#1=65 %
\hatcurSMEiiloggxxxxxA
\else
??????\fi
}
\newcommand{\hatcurSMEiiteff}[1]{\ifnum#1=65 %
\hatcurSMEiiteffxxxxxA
\else
??????\fi
}
\newcommand{\hatcurSMEiivsin}[1]{\ifnum#1=65 %
\hatcurSMEiivsinxxxxxA
\else
??????\fi
}
\newcommand{\hatcurSMEiizfeh}[1]{\ifnum#1=65 %
\hatcurSMEiizfehxxxxxA
\else
??????\fi
}
\newcommand{\hatcurSMEiizfehshort}[1]{\ifnum#1=65 %
\hatcurSMEiizfehshortxxxxxA
\else
??????\fi
}
\newcommand{\hatcurSMEilogg}[1]{\ifnum#1=65 %
\hatcurSMEiloggxxxxxA
\else
\ifnum#1=66 %
\hatcurSMEiloggxxxxxB
\else
??????\fi
\fi
}
\newcommand{\hatcurSMEiteff}[1]{\ifnum#1=65 %
\hatcurSMEiteffxxxxxA
\else
\ifnum#1=66 %
\hatcurSMEiteffxxxxxB
\else
??????\fi
\fi
}
\newcommand{\hatcurSMEivmac}[1]{\ifnum#1=65 %
\hatcurSMEivmacxxxxxA
\else
\ifnum#1=66 %
\hatcurSMEivmacxxxxxB
\else
??????\fi
\fi
}
\newcommand{\hatcurSMEivmic}[1]{\ifnum#1=65 %
\hatcurSMEivmicxxxxxA
\else
\ifnum#1=66 %
\hatcurSMEivmicxxxxxB
\else
??????\fi
\fi
}
\newcommand{\hatcurSMEivsin}[1]{\ifnum#1=65 %
\hatcurSMEivsinxxxxxA
\else
\ifnum#1=66 %
\hatcurSMEivsinxxxxxB
\else
??????\fi
\fi
}
\newcommand{\hatcurSMEizfeh}[1]{\ifnum#1=65 %
\hatcurSMEizfehxxxxxA
\else
\ifnum#1=66 %
\hatcurSMEizfehxxxxxB
\else
??????\fi
\fi
}
\newcommand{\hatcurSMEizfehshort}[1]{\ifnum#1=65 %
\hatcurSMEizfehshortxxxxxA
\else
\ifnum#1=66 %
\hatcurSMEizfehshortxxxxxB
\else
??????\fi
\fi
}
\newcommand{\hatcurXAv}[1]{\ifnum#1=65 %
\hatcurXAvxxxxxA
\else
\ifnum#1=66 %
\hatcurXAvxxxxxB
\else
??????\fi
\fi
}
\newcommand{\hatcurXdist}[1]{\ifnum#1=65 %
\hatcurXdistxxxxxA
\else
\ifnum#1=66 %
\hatcurXdistxxxxxB
\else
??????\fi
\fi
}
\newcommand{\hatcurXdistred}[1]{\ifnum#1=65 %
\hatcurXdistredxxxxxA
\else
\ifnum#1=66 %
\hatcurXdistredxxxxxB
\else
??????\fi
\fi
}
\newcommand{\hatcurXEBV}[1]{\ifnum#1=65 %
\hatcurXEBVxxxxxA
\else
\ifnum#1=66 %
\hatcurXEBVxxxxxB
\else
??????\fi
\fi
}
\newcommand{\hatcurXjhisored}[1]{\ifnum#1=65 %
\hatcurXjhisoredxxxxxA
\else
\ifnum#1=66 %
\hatcurXjhisoredxxxxxB
\else
??????\fi
\fi
}
\newcommand{\hatcurXjkisored}[1]{\ifnum#1=65 %
\hatcurXjkisoredxxxxxA
\else
\ifnum#1=66 %
\hatcurXjkisoredxxxxxB
\else
??????\fi
\fi
}
\newcommand{\hatcurXmhisored}[1]{\ifnum#1=65 %
\hatcurXmhisoredxxxxxA
\else
\ifnum#1=66 %
\hatcurXmhisoredxxxxxB
\else
??????\fi
\fi
}
\newcommand{\hatcurXmiisored}[1]{\ifnum#1=65 %
\hatcurXmiisoredxxxxxA
\else
\ifnum#1=66 %
\hatcurXmiisoredxxxxxB
\else
??????\fi
\fi
}
\newcommand{\hatcurXmjisored}[1]{\ifnum#1=65 %
\hatcurXmjisoredxxxxxA
\else
\ifnum#1=66 %
\hatcurXmjisoredxxxxxB
\else
??????\fi
\fi
}
\newcommand{\hatcurXmkisored}[1]{\ifnum#1=65 %
\hatcurXmkisoredxxxxxA
\else
\ifnum#1=66 %
\hatcurXmkisoredxxxxxB
\else
??????\fi
\fi
}
\newcommand{\hatcurXmvisored}[1]{\ifnum#1=65 %
\hatcurXmvisoredxxxxxA
\else
\ifnum#1=66 %
\hatcurXmvisoredxxxxxB
\else
??????\fi
\fi
}
\newcommand{\hatcurXsecdur}[1]{\ifnum#1=65 %
\hatcurXsecdurxxxxxA
\else
\ifnum#1=66 %
\hatcurXsecdurxxxxxB
\else
??????\fi
\fi
}
\newcommand{\hatcurXsecingdur}[1]{\ifnum#1=65 %
\hatcurXsecingdurxxxxxA
\else
\ifnum#1=66 %
\hatcurXsecingdurxxxxxB
\else
??????\fi
\fi
}
\newcommand{\hatcurXsecondary}[1]{\ifnum#1=65 %
\hatcurXsecondaryxxxxxA
\else
\ifnum#1=66 %
\hatcurXsecondaryxxxxxB
\else
??????\fi
\fi
}
\newcommand{\hatcurXsecphase}[1]{\ifnum#1=65 %
\hatcurXsecphasexxxxxA
\else
\ifnum#1=66 %
\hatcurXsecphasexxxxxB
\else
??????\fi
\fi
}
\newcommand{\hatcurXviisored}[1]{\ifnum#1=65 %
\hatcurXviisoredxxxxxA
\else
\ifnum#1=66 %
\hatcurXviisoredxxxxxB
\else
??????\fi
\fi
}
\newcommand{\hatcurXvkisored}[1]{\ifnum#1=65 %
\hatcurXvkisoredxxxxxA
\else
\ifnum#1=66 %
\hatcurXvkisoredxxxxxB
\else
??????\fi
\fi
}

\newcommand{\hatcurhtreccenxxxxxA}{HTR342-006}                       % Original HTR name of target
\newcommand{\hatcurfieldeccenxxxxxA}{\ensuremath{string}}            % HTR field
\newcommand{\hatcurCCraeccenxxxxxA}{\ensuremath{21^{\mathrm h}03^{\mathrm m}37.44{\mathrm s}}}                     % Right Ascension
\newcommand{\hatcurCCdececcenxxxxxA}{\ensuremath{+11{\arcdeg}59{\arcmin}21.9{\arcsec}}}                    % Declination
\newcommand{\hatcurCCmageccenxxxxxA}{13.145}                         % apparent V-band magnitude
\newcommand{\hatcurCCtwomasseccenxxxxxA}{2MASS~21033731+1159218}     % 2MASS identifier
\newcommand{\hatcurCCgsceccenxxxxxA}{GSC~1111-00383}                 % GSC(1.2) identifier
\newcommand{\hatcurCCtassmveccenxxxxxA}{\ensuremath{13.145\pm0.029}} % APASS V-band magnitude
\newcommand{\hatcurCCtassmvshorteccenxxxxxA}{\ensuremath{13.1}}      % APASS V-band magnitude
\newcommand{\hatcurCCtassmBeccenxxxxxA}{\ensuremath{13.818\pm0.021}} % APASS B-band magnitude
\newcommand{\hatcurCCtassmBshorteccenxxxxxA}{\ensuremath{13.8}}      % APASS B-band magnitude
\newcommand{\hatcurCCtassmIeccenxxxxxA}{\ensuremath{12.46\pm0.10}}   % TASS I-band magnitude
\newcommand{\hatcurCCtassmIshorteccenxxxxxA}{\ensuremath{12.5}}      % TASS I-band magnitude
\newcommand{\hatcurCCtassmgeccenxxxxxA}{\ensuremath{13.445\pm0.016}} % APASS g-band magnitude
\newcommand{\hatcurCCtassmgshorteccenxxxxxA}{\ensuremath{13.4}}      % APASS g-band magnitude
\newcommand{\hatcurCCtassmreccenxxxxxA}{\ensuremath{12.948\pm0.033}} % APASS r-band magnitude
\newcommand{\hatcurCCtassmrshorteccenxxxxxA}{\ensuremath{12.9}}      % APASS r-band magnitude
\newcommand{\hatcurCCtassmieccenxxxxxA}{\ensuremath{12.784\pm0.097}} % APASS i-band magnitude
\newcommand{\hatcurCCtassmishorteccenxxxxxA}{\ensuremath{12.8}}      % APASS i-band magnitude
\newcommand{\hatcurCCtwomassJmageccenxxxxxA}{\ensuremath{11.892\pm0.026}} % 2MASS ORIG MAG
\newcommand{\hatcurCCtwomassHmageccenxxxxxA}{\ensuremath{11.604\pm0.022}} % 2MASS ORIG MAG
\newcommand{\hatcurCCtwomassKmageccenxxxxxA}{\ensuremath{11.528\pm0.025}} % 2MASS ORIG MAG
\newcommand{\hatcurCCcitJmageccenxxxxxA}{\ensuremath{11.909\pm0.026}} % 2MASS CIT MAG
\newcommand{\hatcurCCcitHmageccenxxxxxA}{\ensuremath{11.599\pm0.022}} % 2MASS CIT MAG
\newcommand{\hatcurCCcitKmageccenxxxxxA}{\ensuremath{11.552\pm0.025}} % 2MASS CIT MAG
\newcommand{\hatcurCCbbJmageccenxxxxxA}{\ensuremath{11.958\pm0.028}} % 2MASS BB MAG
\newcommand{\hatcurCCbbHmageccenxxxxxA}{\ensuremath{11.620\pm0.023}} % 2MASS BB MAG
\newcommand{\hatcurCCbbKmageccenxxxxxA}{\ensuremath{11.572\pm0.025}} % 2MASS BB MAG
\newcommand{\hatcurCCesoJmageccenxxxxxA}{\ensuremath{11.960\pm0.029}} % 2MASS ESO MAG
\newcommand{\hatcurCCesoHmageccenxxxxxA}{\ensuremath{11.616\pm0.026}} % 2MASS ESO MAG
\newcommand{\hatcurCCesoKmageccenxxxxxA}{\ensuremath{11.571\pm0.026}} % 2MASS ESO MAG
\newcommand{\hatcurCCesoJHmageccenxxxxxA}{\ensuremath{0.344\pm0.037}} % 2MASS ESO JH COLOR
\newcommand{\hatcurCCesoJKmageccenxxxxxA}{\ensuremath{0.390\pm0.039}} % 2MASS ESO JK COLOR
\newcommand{\hatcurCCesoHKmageccenxxxxxA}{\ensuremath{0.045\pm0.037}} % 2MASS ESO HK COLOR
\newcommand{\hatcurLCdipeccenxxxxxA}{\ensuremath{13.5}}              % BLS detected dip (mmag)
\newcommand{\hatcurLCrprstareccenxxxxxA}{\ensuremath{0.1047\pm0.0025}} % Rp/R*
\newcommand{\hatcurLCbsqeccenxxxxxA}{\ensuremath{0.215_{-0.085}^{+0.073}}} % impact parameter square
\newcommand{\hatcurLCimpeccenxxxxxA}{\ensuremath{0.464_{-0.103}^{+0.072}}} % impact parameter
\newcommand{\hatcurLCzetaeccenxxxxxA}{\ensuremath{12.44\pm0.11}}     % zeta/R*
\newcommand{\hatcurLCdureccenxxxxxA}{\ensuremath{0.1818\pm0.0025}}   % transit duration (days)
\newcommand{\hatcurLCdurshorteccenxxxxxA}{\ensuremath{0.1818}}       % transit duration (days)
\newcommand{\hatcurLCdurhreccenxxxxxA}{\ensuremath{4.364\pm0.059}}   % transit duration (hours)
\newcommand{\hatcurLCdurhrshorteccenxxxxxA}{\ensuremath{4.364}}      % transit duration (hours)
\newcommand{\hatcurLCqeccenxxxxxA}{\ensuremath{0.06980\pm0.00094}}   % fractional transit duration (days)
\newcommand{\hatcurLCqshorteccenxxxxxA}{\ensuremath{0.070}}          % fractional transit duration (days)
\newcommand{\hatcurLCingdureccenxxxxxA}{\ensuremath{0.0215\pm0.0024}} % ingress/egress duration (days)
\newcommand{\hatcurLCPeccenxxxxxA}{\ensuremath{2.6054554\pm0.0000030}} % period (days)
\newcommand{\hatcurLCPprececcenxxxxxA}{\ensuremath{2.6054554}}       % period (days)
\newcommand{\hatcurLCPshorteccenxxxxxA}{\ensuremath{2.6055}}         % period (days)
\newcommand{\hatcurLCTeccenxxxxxA}{\ensuremath{2456320.74721\pm0.00065}} % epoch (BJD)
\newcommand{\hatcurLCTAeccenxxxxxA}{\ensuremath{2455090.9723\pm0.0015}} % TA (BJD)
\newcommand{\hatcurLCTBeccenxxxxxA}{\ensuremath{2456568.26548\pm0.00071}} % TB (BJD)
\newcommand{\hatcurLCrhoeccenxxxxxA}{\ensuremath{0.270_{-0.070}^{+0.095}}} % stellar density no isochrone constraint (cgs)
\newcommand{\hatcurSMEiteffeccenxxxxxA}{\ensuremath{5943\pm51}}      % Ini SME, stellar effective temperature
\newcommand{\hatcurSMEizfeheccenxxxxxA}{\ensuremath{0.170\pm0.080}}  % Ini SME, stellar metallicity
\newcommand{\hatcurSMEizfehshorteccenxxxxxA}{\ensuremath{0.17}}      % Ini SME, stellar metallicity
\newcommand{\hatcurSMEiloggeccenxxxxxA}{\ensuremath{4.18\pm0.10}}    % Ini SME, stellar surface gravity
\newcommand{\hatcurSMEivsineccenxxxxxA}{\ensuremath{6.90\pm0.50}}    % Ini SME, stellar rotational velocity
\newcommand{\hatcurSMEivmaceccenxxxxxA}{\ensuremath{0.0}}            % Ini SME, stellar macroturbulence
\newcommand{\hatcurSMEivmiceccenxxxxxA}{\ensuremath{0.0}}            % Ini SME, stellar microturbulence
\newcommand{\hatcurSMEiiteffeccenxxxxxA}{\ensuremath{5835\pm51}}     % Final SME, stellar effective temperature
\newcommand{\hatcurSMEiizfeheccenxxxxxA}{\ensuremath{0.100\pm0.080}} % Final SME, stellar metallicity
\newcommand{\hatcurSMEiizfehshorteccenxxxxxA}{\ensuremath{0.10}}     % Final SME, stellar metallicity
\newcommand{\hatcurSMEiiloggeccenxxxxxA}{\ensuremath{3.992\pm0.038}} % Final SME, stellar surface gravity
\newcommand{\hatcurSMEiivsineccenxxxxxA}{\ensuremath{7.10\pm0.50}}   % Final SME, stellar rotational velocity
\newcommand{\hatcurLBizeccenxxxxxA}{\ensuremath{0.1949}}             % Limb darkening parameters, Gamma1, z-band
\newcommand{\hatcurLBiizeccenxxxxxA}{\ensuremath{0.3379}}            % Limb darkening parameters, Gamma2, z-band
\newcommand{\hatcurLBiieccenxxxxxA}{\ensuremath{0.2544}}             % Limb darkening parameters, Gamma1, i-band
\newcommand{\hatcurLBiiieccenxxxxxA}{\ensuremath{0.3414}}            % Limb darkening parameters, Gamma2, i-band
\newcommand{\hatcurLBiIeccenxxxxxA}{\ensuremath{0.2336}}             % Limb darkening parameters, Gamma1, I-band
\newcommand{\hatcurLBiiIeccenxxxxxA}{\ensuremath{0.3416}}            % Limb darkening parameters, Gamma2, I-band
\newcommand{\hatcurLBigeccenxxxxxA}{\ensuremath{0.5412}}             % Limb darkening parameters, Gamma1, g-band
\newcommand{\hatcurLBiigeccenxxxxxA}{\ensuremath{0.2456}}            % Limb darkening parameters, Gamma2, g-band
\newcommand{\hatcurLBireccenxxxxxA}{\ensuremath{0.3439}}             % Limb darkening parameters, Gamma1, r-band
\newcommand{\hatcurLBiireccenxxxxxA}{\ensuremath{0.3359}}            % Limb darkening parameters, Gamma2, r-band
\newcommand{\hatcurLBiReccenxxxxxA}{\ensuremath{0.3190}}             % Limb darkening parameters, Gamma1, R-band
\newcommand{\hatcurLBiiReccenxxxxxA}{\ensuremath{0.3385}}            % Limb darkening parameters, Gamma2, R-band
\newcommand{\hatcurLBikepeccenxxxxxA}{\ensuremath{0.1000}}           % Limb darkening parameters, Gamma1, Kep-band
\newcommand{\hatcurLBiikepeccenxxxxxA}{\ensuremath{0.1000}}          % Limb darkening parameters, Gamma2, Kep-band
\newcommand{\hatcurISOmeccenxxxxxA}{\ensuremath{1.214\pm0.083}}      % stellar mass
\newcommand{\hatcurISOmshorteccenxxxxxA}{\ensuremath{1.21}}          % stellar mass
\newcommand{\hatcurISOmlongeccenxxxxxA}{\ensuremath{1.214\pm0.083}}  % stellar mass
\newcommand{\hatcurISOreccenxxxxxA}{\ensuremath{1.85\pm0.27}}        % stellar radius
\newcommand{\hatcurISOrshorteccenxxxxxA}{\ensuremath{1.85}}          % stellar radius
\newcommand{\hatcurISOrlongeccenxxxxxA}{\ensuremath{1.85\pm0.27}}    % stellar radius
\newcommand{\hatcurISOrhoeccenxxxxxA}{\ensuremath{0.268_{-0.070}^{+0.092}}} % stellar density (cgs)
\newcommand{\hatcurISOrholongeccenxxxxxA}{\ensuremath{0.268_{-0.070}^{+0.092}}} % stellar density (cgs)
\newcommand{\hatcurISOloggeccenxxxxxA}{\ensuremath{3.985\pm0.099}}   % stellar surface gravity from isochrones
\newcommand{\hatcurISOlumeccenxxxxxA}{\ensuremath{3.57_{-0.75}^{+0.99}}} % stellar luminosity
\newcommand{\hatcurISOlumshorteccenxxxxxA}{\ensuremath{3.57}}        % stellar luminosity
\newcommand{\hatcurISOmveccenxxxxxA}{\ensuremath{3.44\pm0.31}}       % stellar absolute magnitude
\newcommand{\hatcurISOvieccenxxxxxA}{\ensuremath{0.679\pm0.015}}     % stellar V-I index
\newcommand{\hatcurISOageeccenxxxxxA}{\ensuremath{5.39\pm0.96}}      % stellar age
\newcommand{\hatcurISOsigmaeccenxxxxxA}{\ensuremath{0.00020\pm0.00015}} % system mass-correction sigma parameter
\newcommand{\hatcurISOMJeccenxxxxxA}{\ensuremath{2.32\pm0.31}}       % stellar absolute J magnitude
\newcommand{\hatcurISOMHeccenxxxxxA}{\ensuremath{1.99\pm0.31}}       % stellar absolute H magnitude
\newcommand{\hatcurISOMKeccenxxxxxA}{\ensuremath{1.94\pm0.31}}       % stellar absolute K magnitude
\newcommand{\hatcurISOJKeccenxxxxxA}{\ensuremath{0.380\pm0.010}}     % J-K color index from isochrones.
\newcommand{\hatcurISOspececcenxxxxxA}{G}                            % stellar spectral type
\newcommand{\hatcurRVKeccenxxxxxA}{\ensuremath{69\pm12}}             % RV semi-amplitude [m/s]
\newcommand{\hatcurRVrkeccenxxxxxA}{\ensuremath{0.06\pm0.18}}        % sqrt(e)*cos(omega)
\newcommand{\hatcurRVrheccenxxxxxA}{\ensuremath{-0.02\pm0.27}}       % sqrt(e)*sin(omega)
\newcommand{\hatcurRVkeccenxxxxxA}{\ensuremath{0.013_{-0.046}^{+0.066}}} % e*cos(omega)
\newcommand{\hatcurRVheccenxxxxxA}{\ensuremath{-0.00\pm0.12}}        % e*sin(omega)
\newcommand{\hatcurRVtroneeccenxxxxxA}{\ensuremath{0\pm0}}           % RV linear trend tr1 factor
\newcommand{\hatcurRVtrtwoeccenxxxxxA}{\ensuremath{0\pm0}}           % RV linear trend tr2 factor
\newcommand{\hatcurRVgammaeccenxxxxxA}{\ensuremath{6.8\pm9.7}}       % RV gamma velocity, relative scale
\newcommand{\hatcurRVjittereccenxxxxxA}{\ensuremath{27.8\pm9.6}}     % RV jitter (m/s)
\newcommand{\hatcurRVjittertwosiglimeccenxxxxxA}{\ensuremath{<47.3}} % RV jitter (m/s) 95 percent confidence upper limit
\newcommand{\hatcurRVfitrmseccenxxxxxA}{\ensuremath{.1fym}}          % 
\newcommand{\hatcurRVecceneccenxxxxxA}{\ensuremath{0.081\pm0.096}}   % eccentricity
\newcommand{\hatcurRVeccentwosiglimeccenxxxxxA}{\ensuremath{<0.304}} % eccentricity
\newcommand{\hatcurRVomegaeccenxxxxxA}{\ensuremath{190\pm110}}       % argument of pericenter
\newcommand{\hatcurPPieccenxxxxxA}{\ensuremath{84.3\pm2.0}}          % orbital inclination
\newcommand{\hatcurPPgeccenxxxxxA}{\ensuremath{3.6\pm1.2}}           % planetary surface gravity (m/s^2)
\newcommand{\hatcurPPloggeccenxxxxxA}{\ensuremath{2.56\pm0.13}}      % planetary surface gravity (log cgs)
\newcommand{\hatcurPPareccenxxxxxA}{\ensuremath{4.58\pm0.59}}        % relative orbital radius (a/R*)
\newcommand{\hatcurPPareleccenxxxxxA}{\ensuremath{0.03953\pm0.00088}} % semimajor axis (AU)
\newcommand{\hatcurPPrhoeccenxxxxxA}{\ensuremath{0.096_{-0.032}^{+0.045}}} % planetary density (cgs)
\newcommand{\hatcurPPmeccenxxxxxA}{\ensuremath{0.530\pm0.093}}       % planetary mass (M_jup)
\newcommand{\hatcurPPmshorteccenxxxxxA}{\ensuremath{0.53}}           % planetary mass (M_jup)
\newcommand{\hatcurPPmlongeccenxxxxxA}{\ensuremath{0.530\pm0.093}}   % planetary mass (M_jup)
\newcommand{\hatcurPPmeeccenxxxxxA}{\ensuremath{169\pm30}}           % planetary mass (M_earth)
\newcommand{\hatcurPPmeshorteccenxxxxxA}{\ensuremath{168.5}}         % planetary mass (M_earth)
\newcommand{\hatcurPPmelongeccenxxxxxA}{\ensuremath{169\pm30}}       % planetary mass (M_earth)
\newcommand{\hatcurPPreccenxxxxxA}{\ensuremath{1.89\pm0.28}}         % planetary radius (R_jup)
\newcommand{\hatcurPPrshorteccenxxxxxA}{\ensuremath{1.89}}           % planetary radius (R_jup)
\newcommand{\hatcurPPrlongeccenxxxxxA}{\ensuremath{1.89\pm0.28}}     % planetary radius (R_jup)
\newcommand{\hatcurPPreeccenxxxxxA}{\ensuremath{21.2\pm3.2}}         % planetary radius (R_earth)
\newcommand{\hatcurPPreshorteccenxxxxxA}{\ensuremath{21.2}}          % planetary radius (R_earth)
\newcommand{\hatcurPPrelongeccenxxxxxA}{\ensuremath{21.2\pm3.2}}     % planetary radius (R_earth)
\newcommand{\hatcurPPmrcorreccenxxxxxA}{\ensuremath{0.28}}           % mass/radius correlation
\newcommand{\hatcurPPteffeccenxxxxxA}{\ensuremath{1930\pm120}}       % planetary temperature (K)
\newcommand{\hatcurPPthetaeccenxxxxxA}{\ensuremath{0.0180\pm0.0041}} % Safranov number
\newcommand{\hatcurPPfluxperieccenxxxxxA}{\ensuremath{3.58_{-0.50}^{+1.42}}} % flux @ periastron (CGS)
\newcommand{\hatcurPPfluxperidimeccenxxxxxA}{\ensuremath{9}}         % flux @ periastron (CGS) units.
\newcommand{\hatcurPPfluxapeccenxxxxxA}{\ensuremath{2.78_{-0.72}^{+0.45}}} % flux @ apastron (CGS)
\newcommand{\hatcurPPfluxapdimeccenxxxxxA}{\ensuremath{9}}           % flux @ apastron (CGS) units.
\newcommand{\hatcurPPfluxavgeccenxxxxxA}{\ensuremath{3.12_{-0.56}^{+0.74}}} % flux on average (CGS)
\newcommand{\hatcurPPfluxavgdimeccenxxxxxA}{\ensuremath{9}}          % flux average (CGS) units.
\newcommand{\hatcurPPfluxavglogeccenxxxxxA}{\ensuremath{9.49\pm0.11}} % log10 flux on average (CGS)
\newcommand{\hatcurXsecphaseeccenxxxxxA}{\ensuremath{0.508\pm0.046}} % Phase of secondary eclipse
\newcommand{\hatcurXsecondaryeccenxxxxxA}{\ensuremath{2456322.07\pm0.12}} % Secondary eclipse epoch
\newcommand{\hatcurXsecdureccenxxxxxA}{\ensuremath{0.181\pm0.030}}   % sec eclipse duration (days)
\newcommand{\hatcurXsecingdureccenxxxxxA}{\ensuremath{0.021\pm0.012}} % sec I/E duration (days)
\newcommand{\hatcurPPphiconjeccenxxxxxA}{\ensuremath{0.07\pm0.29}}   % phase diff between conjunction and periastron
\newcommand{\hatcurPPperieccenxxxxxA}{\ensuremath{2456320.56\pm0.74}} % time of periastron passage.
\newcommand{\hatcurPPaequiveccenxxxxxA}{\ensuremath{0.0209\pm0.0027}} % equivalent semi-major axis
\newcommand{\hatcurPPtcirceccenxxxxxA}{\ensuremath{4.5_{-2.2}^{+3.2}}} % circularization timescale
\newcommand{\hatcurPPtinfalleccenxxxxxA}{\ensuremath{116_{-48}^{+83}}} % infall timescale
\newcommand{\hatcurXdisteccenxxxxxA}{\ensuremath{850\pm120}}         % distance (pc), no reddenning correction
\newcommand{\hatcurXAveccenxxxxxA}{\ensuremath{0.091\pm0.052}}       % Av (mag)
\newcommand{\hatcurXdistredeccenxxxxxA}{\ensuremath{840\pm120}}      % distance with Av correction (pc)
\newcommand{\hatcurXEBVeccenxxxxxA}{\ensuremath{0.029\pm0.017}}      % E(B-V) (mag)
\newcommand{\hatcurXmvisoredeccenxxxxxA}{\ensuremath{13.145\pm0.028}} % Expected m_v with reddening (mag)
\newcommand{\hatcurXmiisoredeccenxxxxxA}{\ensuremath{12.418\pm0.015}} % Expected m_i with reddening (mag)
\newcommand{\hatcurXmjisoredeccenxxxxxA}{\ensuremath{11.961\pm0.014}} % Expected m_j with reddening (mag)
\newcommand{\hatcurXmhisoredeccenxxxxxA}{\ensuremath{11.626\pm0.015}} % Expected m_h with reddening (mag)
\newcommand{\hatcurXmkisoredeccenxxxxxA}{\ensuremath{11.563\pm0.017}} % Expected m_k with reddening (mag)
\newcommand{\hatcurXviisoredeccenxxxxxA}{\ensuremath{0.727\pm0.021}} % Expected V-I with reddening (mag)
\newcommand{\hatcurXvkisoredeccenxxxxxA}{\ensuremath{1.583\pm0.034}} % Expected V-K with reddening (mag)
\newcommand{\hatcurXjhisoredeccenxxxxxA}{\ensuremath{0.3350\pm0.0067}} % Expected J-H with reddening (mag)
\newcommand{\hatcurXjkisoredeccenxxxxxA}{\ensuremath{0.3990\pm0.0078}} % Expected J-K with reddening (mag)
\newcommand{\hatcurCCpmraeccenxxxxxA}{\ensuremath{5.5\pm1.9}}        % proper motion, in RA
\newcommand{\hatcurCCpmdececcenxxxxxA}{\ensuremath{-4.0\pm1.9}}      % proper motion, in DEC
\newcommand{\hatcurCCpmeccenxxxxxA}{\ensuremath{6.8\pm2.7}}          % proper motion

\newcommand{\hatcurhtreccenxxxxxB}{HTR101-005}                       % Original HTR name of target
\newcommand{\hatcurfieldeccenxxxxxB}{\ensuremath{string}}            % HTR field
\newcommand{\hatcurCCraeccenxxxxxB}{\ensuremath{10^{\mathrm h}02^{\mathrm m}17.52{\mathrm s}}}                     % Right Ascension
\newcommand{\hatcurCCdececcenxxxxxB}{\ensuremath{+53{\arcdeg}57{\arcmin}03.1{\arcsec}}}                    % Declination
\newcommand{\hatcurCCmageccenxxxxxB}{12.993}                         % apparent V-band magnitude
\newcommand{\hatcurCCtwomasseccenxxxxxB}{2MASS~10021743+5357031}     % 2MASS identifier
\newcommand{\hatcurCCgsceccenxxxxxB}{GSC~3814-00307}                 % GSC(1.2) identifier
\newcommand{\hatcurCCtassmveccenxxxxxB}{\ensuremath{12.993\pm0.052}} % APASS V-band magnitude
\newcommand{\hatcurCCtassmvshorteccenxxxxxB}{\ensuremath{13.0}}      % APASS V-band magnitude
\newcommand{\hatcurCCtassmBeccenxxxxxB}{\ensuremath{13.552\pm0.027}} % APASS B-band magnitude
\newcommand{\hatcurCCtassmBshorteccenxxxxxB}{\ensuremath{13.6}}      % APASS B-band magnitude
\newcommand{\hatcurCCtassmIeccenxxxxxB}{\ensuremath{12.339\pm0.084}} % TASS I-band magnitude
\newcommand{\hatcurCCtassmIshorteccenxxxxxB}{\ensuremath{12.3}}      % TASS I-band magnitude
\newcommand{\hatcurCCtassmgeccenxxxxxB}{\ensuremath{13.209\pm0.021}} % APASS g-band magnitude
\newcommand{\hatcurCCtassmgshorteccenxxxxxB}{\ensuremath{13.2}}      % APASS g-band magnitude
\newcommand{\hatcurCCtassmreccenxxxxxB}{\ensuremath{12.859\pm0.064}} % APASS r-band magnitude
\newcommand{\hatcurCCtassmrshorteccenxxxxxB}{\ensuremath{12.9}}      % APASS r-band magnitude
\newcommand{\hatcurCCtassmieccenxxxxxB}{\ensuremath{12.771\pm0.064}} % APASS i-band magnitude
\newcommand{\hatcurCCtassmishorteccenxxxxxB}{\ensuremath{12.8}}      % APASS i-band magnitude
\newcommand{\hatcurCCtwomassJmageccenxxxxxB}{\ensuremath{12.001\pm0.022}} % 2MASS ORIG MAG
\newcommand{\hatcurCCtwomassHmageccenxxxxxB}{\ensuremath{11.735\pm0.022}} % 2MASS ORIG MAG
\newcommand{\hatcurCCtwomassKmageccenxxxxxB}{\ensuremath{11.675\pm0.022}} % 2MASS ORIG MAG
\newcommand{\hatcurCCcitJmageccenxxxxxB}{\ensuremath{12.020\pm0.023}} % 2MASS CIT MAG
\newcommand{\hatcurCCcitHmageccenxxxxxB}{\ensuremath{11.730\pm0.023}} % 2MASS CIT MAG
\newcommand{\hatcurCCcitKmageccenxxxxxB}{\ensuremath{11.699\pm0.022}} % 2MASS CIT MAG
\newcommand{\hatcurCCbbJmageccenxxxxxB}{\ensuremath{12.065\pm0.023}} % 2MASS BB MAG
\newcommand{\hatcurCCbbHmageccenxxxxxB}{\ensuremath{11.751\pm0.023}} % 2MASS BB MAG
\newcommand{\hatcurCCbbKmageccenxxxxxB}{\ensuremath{11.719\pm0.022}} % 2MASS BB MAG
\newcommand{\hatcurCCesoJmageccenxxxxxB}{\ensuremath{12.068\pm0.024}} % 2MASS ESO MAG
\newcommand{\hatcurCCesoHmageccenxxxxxB}{\ensuremath{11.746\pm0.024}} % 2MASS ESO MAG
\newcommand{\hatcurCCesoKmageccenxxxxxB}{\ensuremath{11.718\pm0.023}} % 2MASS ESO MAG
\newcommand{\hatcurCCesoJHmageccenxxxxxB}{\ensuremath{0.3220\pm0.0070}} % 2MASS ESO JH COLOR
\newcommand{\hatcurCCesoJKmageccenxxxxxB}{\ensuremath{0.3490\pm0.0090}} % 2MASS ESO JK COLOR
\newcommand{\hatcurCCesoHKmageccenxxxxxB}{\ensuremath{0.0270\pm0.0080}} % 2MASS ESO HK COLOR
\newcommand{\hatcurLCdipeccenxxxxxB}{\ensuremath{0.0}}               % BLS detected dip (mmag)
\newcommand{\hatcurLCrprstareccenxxxxxB}{\ensuremath{0.0873\pm0.0022}} % Rp/R*
\newcommand{\hatcurLCbsqeccenxxxxxB}{\ensuremath{0.091_{-0.061}^{+0.112}}} % impact parameter square
\newcommand{\hatcurLCimpeccenxxxxxB}{\ensuremath{0.30_{-0.13}^{+0.15}}} % impact parameter
\newcommand{\hatcurLCzetaeccenxxxxxB}{\ensuremath{11.204\pm0.097}}   % zeta/R*
\newcommand{\hatcurLCdureccenxxxxxB}{\ensuremath{0.1956\pm0.0027}}   % transit duration (days)
\newcommand{\hatcurLCdurshorteccenxxxxxB}{\ensuremath{0.1956}}       % transit duration (days)
\newcommand{\hatcurLCdurhreccenxxxxxB}{\ensuremath{4.694\pm0.065}}   % transit duration (hours)
\newcommand{\hatcurLCdurhrshorteccenxxxxxB}{\ensuremath{4.694}}      % transit duration (hours)
\newcommand{\hatcurLCqeccenxxxxxB}{\ensuremath{0.06580\pm0.00091}}   % fractional transit duration (days)
\newcommand{\hatcurLCqshorteccenxxxxxB}{\ensuremath{0.066}}          % fractional transit duration (days)
\newcommand{\hatcurLCingdureccenxxxxxB}{\ensuremath{0.0171\pm0.0024}} % ingress/egress duration (days)
\newcommand{\hatcurLCPeccenxxxxxB}{\ensuremath{2.9720867\pm0.0000062}} % period (days)
\newcommand{\hatcurLCPprececcenxxxxxB}{\ensuremath{2.9720867}}       % period (days)
\newcommand{\hatcurLCPshorteccenxxxxxB}{\ensuremath{2.9721}}         % period (days)
\newcommand{\hatcurLCTeccenxxxxxB}{\ensuremath{2457261.77119\pm0.00077}} % epoch (BJD)
\newcommand{\hatcurLCTAeccenxxxxxB}{\ensuremath{2455609.2909\pm0.0033}} % TA (BJD)
\newcommand{\hatcurLCTBeccenxxxxxB}{\ensuremath{2457365.79425\pm0.00086}} % TB (BJD)
\newcommand{\hatcurLChatnetmeccenxxxxxB}{\ensuremath{13.29965\pm0.00012}} % HATNet OOT level
\newcommand{\hatcurLCiblendeccenxxxxxB}{\ensuremath{1\pm0}}          % HATNet iblend factor
\newcommand{\hatcurLCrhoeccenxxxxxB}{\ensuremath{0.290_{-0.056}^{+0.037}}} % stellar density no isochrone constraint (cgs)
\newcommand{\hatcurSMEiteffeccenxxxxxB}{\ensuremath{6002\pm50}}      % Ini SME, stellar effective temperature
\newcommand{\hatcurSMEizfeheccenxxxxxB}{\ensuremath{0.035\pm0.080}}  % Ini SME, stellar metallicity
\newcommand{\hatcurSMEizfehshorteccenxxxxxB}{\ensuremath{0.04}}      % Ini SME, stellar metallicity
\newcommand{\hatcurSMEiloggeccenxxxxxB}{\ensuremath{3.96\pm0.10}}    % Ini SME, stellar surface gravity
\newcommand{\hatcurSMEivsineccenxxxxxB}{\ensuremath{7.57\pm0.50}}    % Ini SME, stellar rotational velocity
\newcommand{\hatcurSMEivmaceccenxxxxxB}{\ensuremath{0.0}}            % Ini SME, stellar macroturbulence
\newcommand{\hatcurSMEivmiceccenxxxxxB}{\ensuremath{0.0}}            % Ini SME, stellar microturbulence
\newcommand{\hatcurLBizeccenxxxxxB}{\ensuremath{0.1703}}             % Limb darkening parameters, Gamma1, z-band
\newcommand{\hatcurLBiizeccenxxxxxB}{\ensuremath{0.3491}}            % Limb darkening parameters, Gamma2, z-band
\newcommand{\hatcurLBiieccenxxxxxB}{\ensuremath{0.2249}}             % Limb darkening parameters, Gamma1, i-band
\newcommand{\hatcurLBiiieccenxxxxxB}{\ensuremath{0.3551}}            % Limb darkening parameters, Gamma2, i-band
\newcommand{\hatcurLBiIeccenxxxxxB}{\ensuremath{0.2054}}             % Limb darkening parameters, Gamma1, I-band
\newcommand{\hatcurLBiiIeccenxxxxxB}{\ensuremath{0.3547}}            % Limb darkening parameters, Gamma2, I-band
\newcommand{\hatcurLBigeccenxxxxxB}{\ensuremath{0.4936}}             % Limb darkening parameters, Gamma1, g-band
\newcommand{\hatcurLBiigeccenxxxxxB}{\ensuremath{0.2793}}            % Limb darkening parameters, Gamma2, g-band
\newcommand{\hatcurLBireccenxxxxxB}{\ensuremath{0.3077}}             % Limb darkening parameters, Gamma1, r-band
\newcommand{\hatcurLBiireccenxxxxxB}{\ensuremath{0.3559}}            % Limb darkening parameters, Gamma2, r-band
\newcommand{\hatcurLBiReccenxxxxxB}{\ensuremath{0.2845}}             % Limb darkening parameters, Gamma1, R-band
\newcommand{\hatcurLBiiReccenxxxxxB}{\ensuremath{0.3570}}            % Limb darkening parameters, Gamma2, R-band
\newcommand{\hatcurLBikepeccenxxxxxB}{\ensuremath{0.1000}}           % Limb darkening parameters, Gamma1, Kep-band
\newcommand{\hatcurLBiikepeccenxxxxxB}{\ensuremath{0.1000}}          % Limb darkening parameters, Gamma2, Kep-band
\newcommand{\hatcurISOmeccenxxxxxB}{\ensuremath{1.245_{-0.057}^{+0.106}}} % stellar mass
\newcommand{\hatcurISOmshorteccenxxxxxB}{\ensuremath{1.25}}          % stellar mass
\newcommand{\hatcurISOmlongeccenxxxxxB}{\ensuremath{1.245_{-0.057}^{+0.106}}} % stellar mass
\newcommand{\hatcurISOreccenxxxxxB}{\ensuremath{1.829_{-0.091}^{+0.164}}} % stellar radius
\newcommand{\hatcurISOrshorteccenxxxxxB}{\ensuremath{1.83}}          % stellar radius
\newcommand{\hatcurISOrlongeccenxxxxxB}{\ensuremath{1.829_{-0.091}^{+0.164}}} % stellar radius
\newcommand{\hatcurISOrhoeccenxxxxxB}{\ensuremath{0.290_{-0.060}^{+0.037}}} % stellar density (cgs)
\newcommand{\hatcurISOrholongeccenxxxxxB}{\ensuremath{0.290_{-0.060}^{+0.037}}} % stellar density (cgs)
\newcommand{\hatcurISOloggeccenxxxxxB}{\ensuremath{4.013\pm0.051}}   % stellar surface gravity from isochrones
\newcommand{\hatcurISOlumeccenxxxxxB}{\ensuremath{3.90_{-0.43}^{+0.74}}} % stellar luminosity
\newcommand{\hatcurISOlumshorteccenxxxxxB}{\ensuremath{3.90}}        % stellar luminosity
\newcommand{\hatcurISOmveccenxxxxxB}{\ensuremath{3.32\pm0.17}}       % stellar absolute magnitude
\newcommand{\hatcurISOvieccenxxxxxB}{\ensuremath{0.629\pm0.014}}     % stellar V-I index
\newcommand{\hatcurISOageeccenxxxxxB}{\ensuremath{4.73_{-1.16}^{+0.58}}} % stellar age
\newcommand{\hatcurISOsigmaeccenxxxxxB}{\ensuremath{0.000300\pm0.000043}} % system mass-correction sigma parameter
\newcommand{\hatcurISOMJeccenxxxxxB}{\ensuremath{2.28\pm0.16}}       % stellar absolute J magnitude
\newcommand{\hatcurISOMHeccenxxxxxB}{\ensuremath{1.98\pm0.16}}       % stellar absolute H magnitude
\newcommand{\hatcurISOMKeccenxxxxxB}{\ensuremath{1.92\pm0.16}}       % stellar absolute K magnitude
\newcommand{\hatcurISOJKeccenxxxxxB}{\ensuremath{0.360\pm0.010}}     % J-K color index from isochrones.
\newcommand{\hatcurISOspececcenxxxxxB}{F}                            % stellar spectral type
\newcommand{\hatcurRVKeccenxxxxxB}{\ensuremath{91.5\pm3.3}}          % RV semi-amplitude [m/s]
\newcommand{\hatcurRVrkeccenxxxxxB}{\ensuremath{0.236\pm0.019}}      % sqrt(e)*cos(omega)
\newcommand{\hatcurRVrheccenxxxxxB}{\ensuremath{-0.081_{-0.072}^{+0.117}}} % sqrt(e)*sin(omega)
\newcommand{\hatcurRVkeccenxxxxxB}{\ensuremath{0.0614\pm0.0077}}     % e*cos(omega)
\newcommand{\hatcurRVheccenxxxxxB}{\ensuremath{-0.020\pm0.027}}      % e*sin(omega)
\newcommand{\hatcurRVtroneeccenxxxxxB}{\ensuremath{0\pm0}}           % RV linear trend tr1 factor
\newcommand{\hatcurRVtrtwoeccenxxxxxB}{\ensuremath{0\pm0}}           % RV linear trend tr2 factor
\newcommand{\hatcurRVgammaAeccenxxxxxB}{\ensuremath{-3.9\pm1.7}}     % RV gamma velocity, relative scale
\newcommand{\hatcurRVjitterAeccenxxxxxB}{\ensuremath{0.1\pm1.6}}     % RV jitter (m/s)
\newcommand{\hatcurRVjittertwosiglimAeccenxxxxxB}{\ensuremath{<3.8}} % RV jitter (m/s) 95 percent confidence upper limit
\newcommand{\hatcurRVfitrmsAeccenxxxxxB}{\ensuremath{0.0}}           % RVfitrms
\newcommand{\hatcurRVgammaBeccenxxxxxB}{\ensuremath{129\pm21}}       % RV gamma velocity, relative scale
\newcommand{\hatcurRVjitterBeccenxxxxxB}{\ensuremath{0.1\pm6.0}}     % RV jitter (m/s)
\newcommand{\hatcurRVjittertwosiglimBeccenxxxxxB}{\ensuremath{<13.1}} % RV jitter (m/s) 95 percent confidence upper limit
\newcommand{\hatcurRVfitrmsBeccenxxxxxB}{\ensuremath{0.0}}           % RVfitrms
\newcommand{\hatcurRVecceneccenxxxxxB}{\ensuremath{0.068\pm0.012}}   % eccentricity
\newcommand{\hatcurRVeccentwosiglimeccenxxxxxB}{\ensuremath{<0.090}} % eccentricity
\newcommand{\hatcurRVomegaeccenxxxxxB}{\ensuremath{330\pm150}}       % argument of pericenter
\newcommand{\hatcurPPieccenxxxxxB}{\ensuremath{86.8_{-2.2}^{+1.4}}}  % orbital inclination
\newcommand{\hatcurPPgeccenxxxxxB}{\ensuremath{7.7\pm1.2}}           % planetary surface gravity (m/s^2)
\newcommand{\hatcurPPloggeccenxxxxxB}{\ensuremath{2.885\pm0.071}}    % planetary surface gravity (log cgs)
\newcommand{\hatcurPPareccenxxxxxB}{\ensuremath{5.13_{-0.38}^{+0.21}}} % relative orbital radius (a/R*)
\newcommand{\hatcurPPareleccenxxxxxB}{\ensuremath{0.04352_{-0.00068}^{+0.00120}}} % semimajor axis (AU)
\newcommand{\hatcurPPrhoeccenxxxxxB}{\ensuremath{0.249\pm0.055}}     % planetary density (cgs)
\newcommand{\hatcurPPmeccenxxxxxB}{\ensuremath{0.754_{-0.035}^{+0.046}}} % planetary mass (M_jup)
\newcommand{\hatcurPPmshorteccenxxxxxB}{\ensuremath{0.75}}           % planetary mass (M_jup)
\newcommand{\hatcurPPmlongeccenxxxxxB}{\ensuremath{0.754_{-0.035}^{+0.046}}} % planetary mass (M_jup)
\newcommand{\hatcurPPmeeccenxxxxxB}{\ensuremath{240_{-11}^{+15}}}    % planetary mass (M_earth)
\newcommand{\hatcurPPmeshorteccenxxxxxB}{\ensuremath{239.6}}         % planetary mass (M_earth)
\newcommand{\hatcurPPmelongeccenxxxxxB}{\ensuremath{240_{-11}^{+15}}} % planetary mass (M_earth)
\newcommand{\hatcurPPreccenxxxxxB}{\ensuremath{1.552_{-0.092}^{+0.156}}} % planetary radius (R_jup)
\newcommand{\hatcurPPrshorteccenxxxxxB}{\ensuremath{1.55}}           % planetary radius (R_jup)
\newcommand{\hatcurPPrlongeccenxxxxxB}{\ensuremath{1.552_{-0.092}^{+0.156}}} % planetary radius (R_jup)
\newcommand{\hatcurPPreeccenxxxxxB}{\ensuremath{17.4_{-1.0}^{+1.8}}} % planetary radius (R_earth)
\newcommand{\hatcurPPreshorteccenxxxxxB}{\ensuremath{17.4}}          % planetary radius (R_earth)
\newcommand{\hatcurPPrelongeccenxxxxxB}{\ensuremath{17.4_{-1.0}^{+1.8}}} % planetary radius (R_earth)
\newcommand{\hatcurPPmrcorreccenxxxxxB}{\ensuremath{0.34}}           % mass/radius correlation
\newcommand{\hatcurPPteffeccenxxxxxB}{\ensuremath{1875_{-42}^{+73}}} % planetary temperature (K)
\newcommand{\hatcurPPthetaeccenxxxxxB}{\ensuremath{0.0334\pm0.0032}} % Safranov number
\newcommand{\hatcurPPfluxperieccenxxxxxB}{\ensuremath{3.20_{-0.25}^{+0.54}}} % flux @ periastron (CGS)
\newcommand{\hatcurPPfluxperidimeccenxxxxxB}{\ensuremath{9}}         % flux @ periastron (CGS) units.
\newcommand{\hatcurPPfluxapeccenxxxxxB}{\ensuremath{2.45_{-0.24}^{+0.40}}} % flux @ apastron (CGS)
\newcommand{\hatcurPPfluxapdimeccenxxxxxB}{\ensuremath{9}}           % flux @ apastron (CGS) units.
\newcommand{\hatcurPPfluxavgeccenxxxxxB}{\ensuremath{2.79_{-0.24}^{+0.46}}} % flux on average (CGS)
\newcommand{\hatcurPPfluxavgdimeccenxxxxxB}{\ensuremath{9}}          % flux average (CGS) units.
\newcommand{\hatcurPPfluxavglogeccenxxxxxB}{\ensuremath{9.445_{-0.039}^{+0.066}}} % log10 flux on average (CGS)
\newcommand{\hatcurXsecphaseeccenxxxxxB}{\ensuremath{0.5391\pm0.0049}} % Phase of secondary eclipse
\newcommand{\hatcurXsecondaryeccenxxxxxB}{\ensuremath{2457263.373\pm0.015}} % Secondary eclipse epoch
\newcommand{\hatcurXsecdureccenxxxxxB}{\ensuremath{0.1890\pm0.0097}} % sec eclipse duration (days)
\newcommand{\hatcurXsecingdureccenxxxxxB}{\ensuremath{0.0164\pm0.0026}} % sec I/E duration (days)
\newcommand{\hatcurPPphiconjeccenxxxxxB}{\ensuremath{0.282_{-0.075}^{+0.042}}} % phase diff between conjunction and periastron
\newcommand{\hatcurPPperieccenxxxxxB}{\ensuremath{2457260.93\pm0.17}} % time of periastron passage.
\newcommand{\hatcurPPaequiveccenxxxxxB}{\ensuremath{0.02220_{-0.00170}^{+0.00100}}} % equivalent semi-major axis
\newcommand{\hatcurPPtcirceccenxxxxxB}{\ensuremath{32\pm11}}         % circularization timescale
\newcommand{\hatcurPPtinfalleccenxxxxxB}{\ensuremath{167_{-53}^{+40}}} % infall timescale
\newcommand{\hatcurXdisteccenxxxxxB}{\ensuremath{910_{-48}^{+80}}}   % distance (pc), no reddenning correction
\newcommand{\hatcurXAveccenxxxxxB}{\ensuremath{0.0000\pm0.0058}}     % Av (mag)
\newcommand{\hatcurXdistredeccenxxxxxB}{\ensuremath{901_{-48}^{+80}}} % distance with Av correction (pc)
\newcommand{\hatcurXEBVeccenxxxxxB}{\ensuremath{0.0000\pm0.0019}}    % E(B-V) (mag)
\newcommand{\hatcurXmvisoredeccenxxxxxB}{\ensuremath{13.098\pm0.029}} % Expected m_v with reddening (mag)
\newcommand{\hatcurXmiisoredeccenxxxxxB}{\ensuremath{12.468\pm0.018}} % Expected m_i with reddening (mag)
\newcommand{\hatcurXmjisoredeccenxxxxxB}{\ensuremath{12.053\pm0.014}} % Expected m_j with reddening (mag)
\newcommand{\hatcurXmhisoredeccenxxxxxB}{\ensuremath{11.750\pm0.014}} % Expected m_h with reddening (mag)
\newcommand{\hatcurXmkisoredeccenxxxxxB}{\ensuremath{11.697\pm0.014}} % Expected m_k with reddening (mag)
\newcommand{\hatcurXviisoredeccenxxxxxB}{\ensuremath{0.629\pm0.014}} % Expected V-I with reddening (mag)
\newcommand{\hatcurXvkisoredeccenxxxxxB}{\ensuremath{1.401\pm0.031}} % Expected V-K with reddening (mag)
\newcommand{\hatcurXjhisoredeccenxxxxxB}{\ensuremath{0.3030\pm0.0081}} % Expected J-H with reddening (mag)
\newcommand{\hatcurXjkisoredeccenxxxxxB}{\ensuremath{0.3560\pm0.0090}} % Expected J-K with reddening (mag)
\newcommand{\hatcurCCpmraeccenxxxxxB}{\ensuremath{-9.2\pm1.8}}       % proper motion, in RA
\newcommand{\hatcurCCpmdececcenxxxxxB}{\ensuremath{-11.4\pm2.4}}     % proper motion, in DEC
\newcommand{\hatcurCCpmeccenxxxxxB}{\ensuremath{14.6\pm3.0}}         % proper motion

\newcommand{\hatcurCCbbHmageccen}[1]{\ifnum#1=65 %
\hatcurCCbbHmageccenxxxxxA
\else
\ifnum#1=66 %
\hatcurCCbbHmageccenxxxxxB
\else
??????\fi
\fi
}
\newcommand{\hatcurCCbbJmageccen}[1]{\ifnum#1=65 %
\hatcurCCbbJmageccenxxxxxA
\else
\ifnum#1=66 %
\hatcurCCbbJmageccenxxxxxB
\else
??????\fi
\fi
}
\newcommand{\hatcurCCbbKmageccen}[1]{\ifnum#1=65 %
\hatcurCCbbKmageccenxxxxxA
\else
\ifnum#1=66 %
\hatcurCCbbKmageccenxxxxxB
\else
??????\fi
\fi
}
\newcommand{\hatcurCCcitHmageccen}[1]{\ifnum#1=65 %
\hatcurCCcitHmageccenxxxxxA
\else
\ifnum#1=66 %
\hatcurCCcitHmageccenxxxxxB
\else
??????\fi
\fi
}
\newcommand{\hatcurCCcitJmageccen}[1]{\ifnum#1=65 %
\hatcurCCcitJmageccenxxxxxA
\else
\ifnum#1=66 %
\hatcurCCcitJmageccenxxxxxB
\else
??????\fi
\fi
}
\newcommand{\hatcurCCcitKmageccen}[1]{\ifnum#1=65 %
\hatcurCCcitKmageccenxxxxxA
\else
\ifnum#1=66 %
\hatcurCCcitKmageccenxxxxxB
\else
??????\fi
\fi
}
\newcommand{\hatcurCCdececcen}[1]{\ifnum#1=65 %
\hatcurCCdececcenxxxxxA
\else
\ifnum#1=66 %
\hatcurCCdececcenxxxxxB
\else
??????\fi
\fi
}
\newcommand{\hatcurCCesoHKmageccen}[1]{\ifnum#1=65 %
\hatcurCCesoHKmageccenxxxxxA
\else
\ifnum#1=66 %
\hatcurCCesoHKmageccenxxxxxB
\else
??????\fi
\fi
}
\newcommand{\hatcurCCesoHmageccen}[1]{\ifnum#1=65 %
\hatcurCCesoHmageccenxxxxxA
\else
\ifnum#1=66 %
\hatcurCCesoHmageccenxxxxxB
\else
??????\fi
\fi
}
\newcommand{\hatcurCCesoJHmageccen}[1]{\ifnum#1=65 %
\hatcurCCesoJHmageccenxxxxxA
\else
\ifnum#1=66 %
\hatcurCCesoJHmageccenxxxxxB
\else
??????\fi
\fi
}
\newcommand{\hatcurCCesoJKmageccen}[1]{\ifnum#1=65 %
\hatcurCCesoJKmageccenxxxxxA
\else
\ifnum#1=66 %
\hatcurCCesoJKmageccenxxxxxB
\else
??????\fi
\fi
}
\newcommand{\hatcurCCesoJmageccen}[1]{\ifnum#1=65 %
\hatcurCCesoJmageccenxxxxxA
\else
\ifnum#1=66 %
\hatcurCCesoJmageccenxxxxxB
\else
??????\fi
\fi
}
\newcommand{\hatcurCCesoKmageccen}[1]{\ifnum#1=65 %
\hatcurCCesoKmageccenxxxxxA
\else
\ifnum#1=66 %
\hatcurCCesoKmageccenxxxxxB
\else
??????\fi
\fi
}
\newcommand{\hatcurCCgsceccen}[1]{\ifnum#1=65 %
\hatcurCCgsceccenxxxxxA
\else
\ifnum#1=66 %
\hatcurCCgsceccenxxxxxB
\else
??????\fi
\fi
}
\newcommand{\hatcurCCmageccen}[1]{\ifnum#1=65 %
\hatcurCCmageccenxxxxxA
\else
\ifnum#1=66 %
\hatcurCCmageccenxxxxxB
\else
??????\fi
\fi
}
\newcommand{\hatcurCCpmdececcen}[1]{\ifnum#1=65 %
\hatcurCCpmdececcenxxxxxA
\else
\ifnum#1=66 %
\hatcurCCpmdececcenxxxxxB
\else
??????\fi
\fi
}
\newcommand{\hatcurCCpmeccen}[1]{\ifnum#1=65 %
\hatcurCCpmeccenxxxxxA
\else
\ifnum#1=66 %
\hatcurCCpmeccenxxxxxB
\else
??????\fi
\fi
}
\newcommand{\hatcurCCpmraeccen}[1]{\ifnum#1=65 %
\hatcurCCpmraeccenxxxxxA
\else
\ifnum#1=66 %
\hatcurCCpmraeccenxxxxxB
\else
??????\fi
\fi
}
\newcommand{\hatcurCCraeccen}[1]{\ifnum#1=65 %
\hatcurCCraeccenxxxxxA
\else
\ifnum#1=66 %
\hatcurCCraeccenxxxxxB
\else
??????\fi
\fi
}
\newcommand{\hatcurCCtassmBeccen}[1]{\ifnum#1=65 %
\hatcurCCtassmBeccenxxxxxA
\else
\ifnum#1=66 %
\hatcurCCtassmBeccenxxxxxB
\else
??????\fi
\fi
}
\newcommand{\hatcurCCtassmBshorteccen}[1]{\ifnum#1=65 %
\hatcurCCtassmBshorteccenxxxxxA
\else
\ifnum#1=66 %
\hatcurCCtassmBshorteccenxxxxxB
\else
??????\fi
\fi
}
\newcommand{\hatcurCCtassmgeccen}[1]{\ifnum#1=65 %
\hatcurCCtassmgeccenxxxxxA
\else
\ifnum#1=66 %
\hatcurCCtassmgeccenxxxxxB
\else
??????\fi
\fi
}
\newcommand{\hatcurCCtassmgshorteccen}[1]{\ifnum#1=65 %
\hatcurCCtassmgshorteccenxxxxxA
\else
\ifnum#1=66 %
\hatcurCCtassmgshorteccenxxxxxB
\else
??????\fi
\fi
}
\newcommand{\hatcurCCtassmieccen}[1]{\ifnum#1=65 %
\hatcurCCtassmieccenxxxxxA
\else
\ifnum#1=66 %
\hatcurCCtassmieccenxxxxxB
\else
??????\fi
\fi
}
\newcommand{\hatcurCCtassmIeccen}[1]{\ifnum#1=65 %
\hatcurCCtassmIeccenxxxxxA
\else
\ifnum#1=66 %
\hatcurCCtassmIeccenxxxxxB
\else
??????\fi
\fi
}
\newcommand{\hatcurCCtassmishorteccen}[1]{\ifnum#1=65 %
\hatcurCCtassmishorteccenxxxxxA
\else
\ifnum#1=66 %
\hatcurCCtassmishorteccenxxxxxB
\else
??????\fi
\fi
}
\newcommand{\hatcurCCtassmIshorteccen}[1]{\ifnum#1=65 %
\hatcurCCtassmIshorteccenxxxxxA
\else
\ifnum#1=66 %
\hatcurCCtassmIshorteccenxxxxxB
\else
??????\fi
\fi
}
\newcommand{\hatcurCCtassmreccen}[1]{\ifnum#1=65 %
\hatcurCCtassmreccenxxxxxA
\else
\ifnum#1=66 %
\hatcurCCtassmreccenxxxxxB
\else
??????\fi
\fi
}
\newcommand{\hatcurCCtassmrshorteccen}[1]{\ifnum#1=65 %
\hatcurCCtassmrshorteccenxxxxxA
\else
\ifnum#1=66 %
\hatcurCCtassmrshorteccenxxxxxB
\else
??????\fi
\fi
}
\newcommand{\hatcurCCtassmveccen}[1]{\ifnum#1=65 %
\hatcurCCtassmveccenxxxxxA
\else
\ifnum#1=66 %
\hatcurCCtassmveccenxxxxxB
\else
??????\fi
\fi
}
\newcommand{\hatcurCCtassmvshorteccen}[1]{\ifnum#1=65 %
\hatcurCCtassmvshorteccenxxxxxA
\else
\ifnum#1=66 %
\hatcurCCtassmvshorteccenxxxxxB
\else
??????\fi
\fi
}
\newcommand{\hatcurCCtwomasseccen}[1]{\ifnum#1=65 %
\hatcurCCtwomasseccenxxxxxA
\else
\ifnum#1=66 %
\hatcurCCtwomasseccenxxxxxB
\else
??????\fi
\fi
}
\newcommand{\hatcurCCtwomassHmageccen}[1]{\ifnum#1=65 %
\hatcurCCtwomassHmageccenxxxxxA
\else
\ifnum#1=66 %
\hatcurCCtwomassHmageccenxxxxxB
\else
??????\fi
\fi
}
\newcommand{\hatcurCCtwomassJmageccen}[1]{\ifnum#1=65 %
\hatcurCCtwomassJmageccenxxxxxA
\else
\ifnum#1=66 %
\hatcurCCtwomassJmageccenxxxxxB
\else
??????\fi
\fi
}
\newcommand{\hatcurCCtwomassKmageccen}[1]{\ifnum#1=65 %
\hatcurCCtwomassKmageccenxxxxxA
\else
\ifnum#1=66 %
\hatcurCCtwomassKmageccenxxxxxB
\else
??????\fi
\fi
}
\newcommand{\hatcurfieldeccen}[1]{\ifnum#1=65 %
\hatcurfieldeccenxxxxxA
\else
\ifnum#1=66 %
\hatcurfieldeccenxxxxxB
\else
??????\fi
\fi
}
\newcommand{\hatcurhtreccen}[1]{\ifnum#1=65 %
\hatcurhtreccenxxxxxA
\else
\ifnum#1=66 %
\hatcurhtreccenxxxxxB
\else
??????\fi
\fi
}
\newcommand{\hatcurISOageeccen}[1]{\ifnum#1=65 %
\hatcurISOageeccenxxxxxA
\else
\ifnum#1=66 %
\hatcurISOageeccenxxxxxB
\else
??????\fi
\fi
}
\newcommand{\hatcurISOJKeccen}[1]{\ifnum#1=65 %
\hatcurISOJKeccenxxxxxA
\else
\ifnum#1=66 %
\hatcurISOJKeccenxxxxxB
\else
??????\fi
\fi
}
\newcommand{\hatcurISOloggeccen}[1]{\ifnum#1=65 %
\hatcurISOloggeccenxxxxxA
\else
\ifnum#1=66 %
\hatcurISOloggeccenxxxxxB
\else
??????\fi
\fi
}
\newcommand{\hatcurISOlumeccen}[1]{\ifnum#1=65 %
\hatcurISOlumeccenxxxxxA
\else
\ifnum#1=66 %
\hatcurISOlumeccenxxxxxB
\else
??????\fi
\fi
}
\newcommand{\hatcurISOlumshorteccen}[1]{\ifnum#1=65 %
\hatcurISOlumshorteccenxxxxxA
\else
\ifnum#1=66 %
\hatcurISOlumshorteccenxxxxxB
\else
??????\fi
\fi
}
\newcommand{\hatcurISOmeccen}[1]{\ifnum#1=65 %
\hatcurISOmeccenxxxxxA
\else
\ifnum#1=66 %
\hatcurISOmeccenxxxxxB
\else
??????\fi
\fi
}
\newcommand{\hatcurISOMHeccen}[1]{\ifnum#1=65 %
\hatcurISOMHeccenxxxxxA
\else
\ifnum#1=66 %
\hatcurISOMHeccenxxxxxB
\else
??????\fi
\fi
}
\newcommand{\hatcurISOMJeccen}[1]{\ifnum#1=65 %
\hatcurISOMJeccenxxxxxA
\else
\ifnum#1=66 %
\hatcurISOMJeccenxxxxxB
\else
??????\fi
\fi
}
\newcommand{\hatcurISOMKeccen}[1]{\ifnum#1=65 %
\hatcurISOMKeccenxxxxxA
\else
\ifnum#1=66 %
\hatcurISOMKeccenxxxxxB
\else
??????\fi
\fi
}
\newcommand{\hatcurISOmlongeccen}[1]{\ifnum#1=65 %
\hatcurISOmlongeccenxxxxxA
\else
\ifnum#1=66 %
\hatcurISOmlongeccenxxxxxB
\else
??????\fi
\fi
}
\newcommand{\hatcurISOmshorteccen}[1]{\ifnum#1=65 %
\hatcurISOmshorteccenxxxxxA
\else
\ifnum#1=66 %
\hatcurISOmshorteccenxxxxxB
\else
??????\fi
\fi
}
\newcommand{\hatcurISOmveccen}[1]{\ifnum#1=65 %
\hatcurISOmveccenxxxxxA
\else
\ifnum#1=66 %
\hatcurISOmveccenxxxxxB
\else
??????\fi
\fi
}
\newcommand{\hatcurISOreccen}[1]{\ifnum#1=65 %
\hatcurISOreccenxxxxxA
\else
\ifnum#1=66 %
\hatcurISOreccenxxxxxB
\else
??????\fi
\fi
}
\newcommand{\hatcurISOrhoeccen}[1]{\ifnum#1=65 %
\hatcurISOrhoeccenxxxxxA
\else
\ifnum#1=66 %
\hatcurISOrhoeccenxxxxxB
\else
??????\fi
\fi
}
\newcommand{\hatcurISOrholongeccen}[1]{\ifnum#1=65 %
\hatcurISOrholongeccenxxxxxA
\else
\ifnum#1=66 %
\hatcurISOrholongeccenxxxxxB
\else
??????\fi
\fi
}
\newcommand{\hatcurISOrlongeccen}[1]{\ifnum#1=65 %
\hatcurISOrlongeccenxxxxxA
\else
\ifnum#1=66 %
\hatcurISOrlongeccenxxxxxB
\else
??????\fi
\fi
}
\newcommand{\hatcurISOrshorteccen}[1]{\ifnum#1=65 %
\hatcurISOrshorteccenxxxxxA
\else
\ifnum#1=66 %
\hatcurISOrshorteccenxxxxxB
\else
??????\fi
\fi
}
\newcommand{\hatcurISOsigmaeccen}[1]{\ifnum#1=65 %
\hatcurISOsigmaeccenxxxxxA
\else
\ifnum#1=66 %
\hatcurISOsigmaeccenxxxxxB
\else
??????\fi
\fi
}
\newcommand{\hatcurISOspececcen}[1]{\ifnum#1=65 %
\hatcurISOspececcenxxxxxA
\else
\ifnum#1=66 %
\hatcurISOspececcenxxxxxB
\else
??????\fi
\fi
}
\newcommand{\hatcurISOvieccen}[1]{\ifnum#1=65 %
\hatcurISOvieccenxxxxxA
\else
\ifnum#1=66 %
\hatcurISOvieccenxxxxxB
\else
??????\fi
\fi
}
\newcommand{\hatcurLBigeccen}[1]{\ifnum#1=65 %
\hatcurLBigeccenxxxxxA
\else
\ifnum#1=66 %
\hatcurLBigeccenxxxxxB
\else
??????\fi
\fi
}
\newcommand{\hatcurLBiieccen}[1]{\ifnum#1=65 %
\hatcurLBiieccenxxxxxA
\else
\ifnum#1=66 %
\hatcurLBiieccenxxxxxB
\else
??????\fi
\fi
}
\newcommand{\hatcurLBiIeccen}[1]{\ifnum#1=65 %
\hatcurLBiIeccenxxxxxA
\else
\ifnum#1=66 %
\hatcurLBiIeccenxxxxxB
\else
??????\fi
\fi
}
\newcommand{\hatcurLBiigeccen}[1]{\ifnum#1=65 %
\hatcurLBiigeccenxxxxxA
\else
\ifnum#1=66 %
\hatcurLBiigeccenxxxxxB
\else
??????\fi
\fi
}
\newcommand{\hatcurLBiiieccen}[1]{\ifnum#1=65 %
\hatcurLBiiieccenxxxxxA
\else
\ifnum#1=66 %
\hatcurLBiiieccenxxxxxB
\else
??????\fi
\fi
}
\newcommand{\hatcurLBiiIeccen}[1]{\ifnum#1=65 %
\hatcurLBiiIeccenxxxxxA
\else
\ifnum#1=66 %
\hatcurLBiiIeccenxxxxxB
\else
??????\fi
\fi
}
\newcommand{\hatcurLBiikepeccen}[1]{\ifnum#1=65 %
\hatcurLBiikepeccenxxxxxA
\else
\ifnum#1=66 %
\hatcurLBiikepeccenxxxxxB
\else
??????\fi
\fi
}
\newcommand{\hatcurLBiireccen}[1]{\ifnum#1=65 %
\hatcurLBiireccenxxxxxA
\else
\ifnum#1=66 %
\hatcurLBiireccenxxxxxB
\else
??????\fi
\fi
}
\newcommand{\hatcurLBiiReccen}[1]{\ifnum#1=65 %
\hatcurLBiiReccenxxxxxA
\else
\ifnum#1=66 %
\hatcurLBiiReccenxxxxxB
\else
??????\fi
\fi
}
\newcommand{\hatcurLBiizeccen}[1]{\ifnum#1=65 %
\hatcurLBiizeccenxxxxxA
\else
\ifnum#1=66 %
\hatcurLBiizeccenxxxxxB
\else
??????\fi
\fi
}
\newcommand{\hatcurLBikepeccen}[1]{\ifnum#1=65 %
\hatcurLBikepeccenxxxxxA
\else
\ifnum#1=66 %
\hatcurLBikepeccenxxxxxB
\else
??????\fi
\fi
}
\newcommand{\hatcurLBireccen}[1]{\ifnum#1=65 %
\hatcurLBireccenxxxxxA
\else
\ifnum#1=66 %
\hatcurLBireccenxxxxxB
\else
??????\fi
\fi
}
\newcommand{\hatcurLBiReccen}[1]{\ifnum#1=65 %
\hatcurLBiReccenxxxxxA
\else
\ifnum#1=66 %
\hatcurLBiReccenxxxxxB
\else
??????\fi
\fi
}
\newcommand{\hatcurLBizeccen}[1]{\ifnum#1=65 %
\hatcurLBizeccenxxxxxA
\else
\ifnum#1=66 %
\hatcurLBizeccenxxxxxB
\else
??????\fi
\fi
}
\newcommand{\hatcurLCbsqeccen}[1]{\ifnum#1=65 %
\hatcurLCbsqeccenxxxxxA
\else
\ifnum#1=66 %
\hatcurLCbsqeccenxxxxxB
\else
??????\fi
\fi
}
\newcommand{\hatcurLCdipeccen}[1]{\ifnum#1=65 %
\hatcurLCdipeccenxxxxxA
\else
\ifnum#1=66 %
\hatcurLCdipeccenxxxxxB
\else
??????\fi
\fi
}
\newcommand{\hatcurLCdureccen}[1]{\ifnum#1=65 %
\hatcurLCdureccenxxxxxA
\else
\ifnum#1=66 %
\hatcurLCdureccenxxxxxB
\else
??????\fi
\fi
}
\newcommand{\hatcurLCdurhreccen}[1]{\ifnum#1=65 %
\hatcurLCdurhreccenxxxxxA
\else
\ifnum#1=66 %
\hatcurLCdurhreccenxxxxxB
\else
??????\fi
\fi
}
\newcommand{\hatcurLCdurhrshorteccen}[1]{\ifnum#1=65 %
\hatcurLCdurhrshorteccenxxxxxA
\else
\ifnum#1=66 %
\hatcurLCdurhrshorteccenxxxxxB
\else
??????\fi
\fi
}
\newcommand{\hatcurLCdurshorteccen}[1]{\ifnum#1=65 %
\hatcurLCdurshorteccenxxxxxA
\else
\ifnum#1=66 %
\hatcurLCdurshorteccenxxxxxB
\else
??????\fi
\fi
}
\newcommand{\hatcurLChatnetmeccen}[1]{\ifnum#1=65 %
\hatcurLChatnetmeccenxxxxxA
\else
\ifnum#1=66 %
\hatcurLChatnetmeccenxxxxxB
\else
??????\fi
\fi
}
\newcommand{\hatcurLCiblendeccen}[1]{\ifnum#1=65 %
\hatcurLCiblendeccenxxxxxA
\else
\ifnum#1=66 %
\hatcurLCiblendeccenxxxxxB
\else
??????\fi
\fi
}
\newcommand{\hatcurLCimpeccen}[1]{\ifnum#1=65 %
\hatcurLCimpeccenxxxxxA
\else
\ifnum#1=66 %
\hatcurLCimpeccenxxxxxB
\else
??????\fi
\fi
}
\newcommand{\hatcurLCingdureccen}[1]{\ifnum#1=65 %
\hatcurLCingdureccenxxxxxA
\else
\ifnum#1=66 %
\hatcurLCingdureccenxxxxxB
\else
??????\fi
\fi
}
\newcommand{\hatcurLCPeccen}[1]{\ifnum#1=65 %
\hatcurLCPeccenxxxxxA
\else
\ifnum#1=66 %
\hatcurLCPeccenxxxxxB
\else
??????\fi
\fi
}
\newcommand{\hatcurLCPprececcen}[1]{\ifnum#1=65 %
\hatcurLCPprececcenxxxxxA
\else
\ifnum#1=66 %
\hatcurLCPprececcenxxxxxB
\else
??????\fi
\fi
}
\newcommand{\hatcurLCPshorteccen}[1]{\ifnum#1=65 %
\hatcurLCPshorteccenxxxxxA
\else
\ifnum#1=66 %
\hatcurLCPshorteccenxxxxxB
\else
??????\fi
\fi
}
\newcommand{\hatcurLCqeccen}[1]{\ifnum#1=65 %
\hatcurLCqeccenxxxxxA
\else
\ifnum#1=66 %
\hatcurLCqeccenxxxxxB
\else
??????\fi
\fi
}
\newcommand{\hatcurLCqshorteccen}[1]{\ifnum#1=65 %
\hatcurLCqshorteccenxxxxxA
\else
\ifnum#1=66 %
\hatcurLCqshorteccenxxxxxB
\else
??????\fi
\fi
}
\newcommand{\hatcurLCrhoeccen}[1]{\ifnum#1=65 %
\hatcurLCrhoeccenxxxxxA
\else
\ifnum#1=66 %
\hatcurLCrhoeccenxxxxxB
\else
??????\fi
\fi
}
\newcommand{\hatcurLCrprstareccen}[1]{\ifnum#1=65 %
\hatcurLCrprstareccenxxxxxA
\else
\ifnum#1=66 %
\hatcurLCrprstareccenxxxxxB
\else
??????\fi
\fi
}
\newcommand{\hatcurLCTAeccen}[1]{\ifnum#1=65 %
\hatcurLCTAeccenxxxxxA
\else
\ifnum#1=66 %
\hatcurLCTAeccenxxxxxB
\else
??????\fi
\fi
}
\newcommand{\hatcurLCTBeccen}[1]{\ifnum#1=65 %
\hatcurLCTBeccenxxxxxA
\else
\ifnum#1=66 %
\hatcurLCTBeccenxxxxxB
\else
??????\fi
\fi
}
\newcommand{\hatcurLCTeccen}[1]{\ifnum#1=65 %
\hatcurLCTeccenxxxxxA
\else
\ifnum#1=66 %
\hatcurLCTeccenxxxxxB
\else
??????\fi
\fi
}
\newcommand{\hatcurLCzetaeccen}[1]{\ifnum#1=65 %
\hatcurLCzetaeccenxxxxxA
\else
\ifnum#1=66 %
\hatcurLCzetaeccenxxxxxB
\else
??????\fi
\fi
}
\newcommand{\hatcurPPaequiveccen}[1]{\ifnum#1=65 %
\hatcurPPaequiveccenxxxxxA
\else
\ifnum#1=66 %
\hatcurPPaequiveccenxxxxxB
\else
??????\fi
\fi
}
\newcommand{\hatcurPPareccen}[1]{\ifnum#1=65 %
\hatcurPPareccenxxxxxA
\else
\ifnum#1=66 %
\hatcurPPareccenxxxxxB
\else
??????\fi
\fi
}
\newcommand{\hatcurPPareleccen}[1]{\ifnum#1=65 %
\hatcurPPareleccenxxxxxA
\else
\ifnum#1=66 %
\hatcurPPareleccenxxxxxB
\else
??????\fi
\fi
}
\newcommand{\hatcurPPfluxapdimeccen}[1]{\ifnum#1=65 %
\hatcurPPfluxapdimeccenxxxxxA
\else
\ifnum#1=66 %
\hatcurPPfluxapdimeccenxxxxxB
\else
??????\fi
\fi
}
\newcommand{\hatcurPPfluxapeccen}[1]{\ifnum#1=65 %
\hatcurPPfluxapeccenxxxxxA
\else
\ifnum#1=66 %
\hatcurPPfluxapeccenxxxxxB
\else
??????\fi
\fi
}
\newcommand{\hatcurPPfluxavgdimeccen}[1]{\ifnum#1=65 %
\hatcurPPfluxavgdimeccenxxxxxA
\else
\ifnum#1=66 %
\hatcurPPfluxavgdimeccenxxxxxB
\else
??????\fi
\fi
}
\newcommand{\hatcurPPfluxavgeccen}[1]{\ifnum#1=65 %
\hatcurPPfluxavgeccenxxxxxA
\else
\ifnum#1=66 %
\hatcurPPfluxavgeccenxxxxxB
\else
??????\fi
\fi
}
\newcommand{\hatcurPPfluxavglogeccen}[1]{\ifnum#1=65 %
\hatcurPPfluxavglogeccenxxxxxA
\else
\ifnum#1=66 %
\hatcurPPfluxavglogeccenxxxxxB
\else
??????\fi
\fi
}
\newcommand{\hatcurPPfluxperidimeccen}[1]{\ifnum#1=65 %
\hatcurPPfluxperidimeccenxxxxxA
\else
\ifnum#1=66 %
\hatcurPPfluxperidimeccenxxxxxB
\else
??????\fi
\fi
}
\newcommand{\hatcurPPfluxperieccen}[1]{\ifnum#1=65 %
\hatcurPPfluxperieccenxxxxxA
\else
\ifnum#1=66 %
\hatcurPPfluxperieccenxxxxxB
\else
??????\fi
\fi
}
\newcommand{\hatcurPPgeccen}[1]{\ifnum#1=65 %
\hatcurPPgeccenxxxxxA
\else
\ifnum#1=66 %
\hatcurPPgeccenxxxxxB
\else
??????\fi
\fi
}
\newcommand{\hatcurPPieccen}[1]{\ifnum#1=65 %
\hatcurPPieccenxxxxxA
\else
\ifnum#1=66 %
\hatcurPPieccenxxxxxB
\else
??????\fi
\fi
}
\newcommand{\hatcurPPloggeccen}[1]{\ifnum#1=65 %
\hatcurPPloggeccenxxxxxA
\else
\ifnum#1=66 %
\hatcurPPloggeccenxxxxxB
\else
??????\fi
\fi
}
\newcommand{\hatcurPPmeccen}[1]{\ifnum#1=65 %
\hatcurPPmeccenxxxxxA
\else
\ifnum#1=66 %
\hatcurPPmeccenxxxxxB
\else
??????\fi
\fi
}
\newcommand{\hatcurPPmeeccen}[1]{\ifnum#1=65 %
\hatcurPPmeeccenxxxxxA
\else
\ifnum#1=66 %
\hatcurPPmeeccenxxxxxB
\else
??????\fi
\fi
}
\newcommand{\hatcurPPmelongeccen}[1]{\ifnum#1=65 %
\hatcurPPmelongeccenxxxxxA
\else
\ifnum#1=66 %
\hatcurPPmelongeccenxxxxxB
\else
??????\fi
\fi
}
\newcommand{\hatcurPPmeshorteccen}[1]{\ifnum#1=65 %
\hatcurPPmeshorteccenxxxxxA
\else
\ifnum#1=66 %
\hatcurPPmeshorteccenxxxxxB
\else
??????\fi
\fi
}
\newcommand{\hatcurPPmlongeccen}[1]{\ifnum#1=65 %
\hatcurPPmlongeccenxxxxxA
\else
\ifnum#1=66 %
\hatcurPPmlongeccenxxxxxB
\else
??????\fi
\fi
}
\newcommand{\hatcurPPmrcorreccen}[1]{\ifnum#1=65 %
\hatcurPPmrcorreccenxxxxxA
\else
\ifnum#1=66 %
\hatcurPPmrcorreccenxxxxxB
\else
??????\fi
\fi
}
\newcommand{\hatcurPPmshorteccen}[1]{\ifnum#1=65 %
\hatcurPPmshorteccenxxxxxA
\else
\ifnum#1=66 %
\hatcurPPmshorteccenxxxxxB
\else
??????\fi
\fi
}
\newcommand{\hatcurPPperieccen}[1]{\ifnum#1=65 %
\hatcurPPperieccenxxxxxA
\else
\ifnum#1=66 %
\hatcurPPperieccenxxxxxB
\else
??????\fi
\fi
}
\newcommand{\hatcurPPphiconjeccen}[1]{\ifnum#1=65 %
\hatcurPPphiconjeccenxxxxxA
\else
\ifnum#1=66 %
\hatcurPPphiconjeccenxxxxxB
\else
??????\fi
\fi
}
\newcommand{\hatcurPPreccen}[1]{\ifnum#1=65 %
\hatcurPPreccenxxxxxA
\else
\ifnum#1=66 %
\hatcurPPreccenxxxxxB
\else
??????\fi
\fi
}
\newcommand{\hatcurPPreeccen}[1]{\ifnum#1=65 %
\hatcurPPreeccenxxxxxA
\else
\ifnum#1=66 %
\hatcurPPreeccenxxxxxB
\else
??????\fi
\fi
}
\newcommand{\hatcurPPrelongeccen}[1]{\ifnum#1=65 %
\hatcurPPrelongeccenxxxxxA
\else
\ifnum#1=66 %
\hatcurPPrelongeccenxxxxxB
\else
??????\fi
\fi
}
\newcommand{\hatcurPPreshorteccen}[1]{\ifnum#1=65 %
\hatcurPPreshorteccenxxxxxA
\else
\ifnum#1=66 %
\hatcurPPreshorteccenxxxxxB
\else
??????\fi
\fi
}
\newcommand{\hatcurPPrhoeccen}[1]{\ifnum#1=65 %
\hatcurPPrhoeccenxxxxxA
\else
\ifnum#1=66 %
\hatcurPPrhoeccenxxxxxB
\else
??????\fi
\fi
}
\newcommand{\hatcurPPrlongeccen}[1]{\ifnum#1=65 %
\hatcurPPrlongeccenxxxxxA
\else
\ifnum#1=66 %
\hatcurPPrlongeccenxxxxxB
\else
??????\fi
\fi
}
\newcommand{\hatcurPPrshorteccen}[1]{\ifnum#1=65 %
\hatcurPPrshorteccenxxxxxA
\else
\ifnum#1=66 %
\hatcurPPrshorteccenxxxxxB
\else
??????\fi
\fi
}
\newcommand{\hatcurPPtcirceccen}[1]{\ifnum#1=65 %
\hatcurPPtcirceccenxxxxxA
\else
\ifnum#1=66 %
\hatcurPPtcirceccenxxxxxB
\else
??????\fi
\fi
}
\newcommand{\hatcurPPteffeccen}[1]{\ifnum#1=65 %
\hatcurPPteffeccenxxxxxA
\else
\ifnum#1=66 %
\hatcurPPteffeccenxxxxxB
\else
??????\fi
\fi
}
\newcommand{\hatcurPPthetaeccen}[1]{\ifnum#1=65 %
\hatcurPPthetaeccenxxxxxA
\else
\ifnum#1=66 %
\hatcurPPthetaeccenxxxxxB
\else
??????\fi
\fi
}
\newcommand{\hatcurPPtinfalleccen}[1]{\ifnum#1=65 %
\hatcurPPtinfalleccenxxxxxA
\else
\ifnum#1=66 %
\hatcurPPtinfalleccenxxxxxB
\else
??????\fi
\fi
}
\newcommand{\hatcurRVecceneccen}[1]{\ifnum#1=65 %
\hatcurRVecceneccenxxxxxA
\else
\ifnum#1=66 %
\hatcurRVecceneccenxxxxxB
\else
??????\fi
\fi
}
\newcommand{\hatcurRVeccentwosiglimeccen}[1]{\ifnum#1=65 %
\hatcurRVeccentwosiglimeccenxxxxxA
\else
\ifnum#1=66 %
\hatcurRVeccentwosiglimeccenxxxxxB
\else
??????\fi
\fi
}
\newcommand{\hatcurRVfitrmsAeccen}[1]{\ifnum#1=66 %
\hatcurRVfitrmsAeccenxxxxxB
\else
??????\fi
}
\newcommand{\hatcurRVfitrmsBeccen}[1]{\ifnum#1=66 %
\hatcurRVfitrmsBeccenxxxxxB
\else
??????\fi
}
\newcommand{\hatcurRVfitrmseccen}[1]{\ifnum#1=65 %
\hatcurRVfitrmseccenxxxxxA
\else
??????\fi
}
\newcommand{\hatcurRVgammaAeccen}[1]{\ifnum#1=66 %
\hatcurRVgammaAeccenxxxxxB
\else
??????\fi
}
\newcommand{\hatcurRVgammaBeccen}[1]{\ifnum#1=66 %
\hatcurRVgammaBeccenxxxxxB
\else
??????\fi
}
\newcommand{\hatcurRVgammaeccen}[1]{\ifnum#1=65 %
\hatcurRVgammaeccenxxxxxA
\else
??????\fi
}
\newcommand{\hatcurRVheccen}[1]{\ifnum#1=65 %
\hatcurRVheccenxxxxxA
\else
\ifnum#1=66 %
\hatcurRVheccenxxxxxB
\else
??????\fi
\fi
}
\newcommand{\hatcurRVjitterAeccen}[1]{\ifnum#1=66 %
\hatcurRVjitterAeccenxxxxxB
\else
??????\fi
}
\newcommand{\hatcurRVjitterBeccen}[1]{\ifnum#1=66 %
\hatcurRVjitterBeccenxxxxxB
\else
??????\fi
}
\newcommand{\hatcurRVjittereccen}[1]{\ifnum#1=65 %
\hatcurRVjittereccenxxxxxA
\else
??????\fi
}
\newcommand{\hatcurRVjittertwosiglimAeccen}[1]{\ifnum#1=66 %
\hatcurRVjittertwosiglimAeccenxxxxxB
\else
??????\fi
}
\newcommand{\hatcurRVjittertwosiglimBeccen}[1]{\ifnum#1=66 %
\hatcurRVjittertwosiglimBeccenxxxxxB
\else
??????\fi
}
\newcommand{\hatcurRVjittertwosiglimeccen}[1]{\ifnum#1=65 %
\hatcurRVjittertwosiglimeccenxxxxxA
\else
??????\fi
}
\newcommand{\hatcurRVkeccen}[1]{\ifnum#1=65 %
\hatcurRVkeccenxxxxxA
\else
\ifnum#1=66 %
\hatcurRVkeccenxxxxxB
\else
??????\fi
\fi
}
\newcommand{\hatcurRVKeccen}[1]{\ifnum#1=65 %
\hatcurRVKeccenxxxxxA
\else
\ifnum#1=66 %
\hatcurRVKeccenxxxxxB
\else
??????\fi
\fi
}
\newcommand{\hatcurRVomegaeccen}[1]{\ifnum#1=65 %
\hatcurRVomegaeccenxxxxxA
\else
\ifnum#1=66 %
\hatcurRVomegaeccenxxxxxB
\else
??????\fi
\fi
}
\newcommand{\hatcurRVrheccen}[1]{\ifnum#1=65 %
\hatcurRVrheccenxxxxxA
\else
\ifnum#1=66 %
\hatcurRVrheccenxxxxxB
\else
??????\fi
\fi
}
\newcommand{\hatcurRVrkeccen}[1]{\ifnum#1=65 %
\hatcurRVrkeccenxxxxxA
\else
\ifnum#1=66 %
\hatcurRVrkeccenxxxxxB
\else
??????\fi
\fi
}
\newcommand{\hatcurRVtroneeccen}[1]{\ifnum#1=65 %
\hatcurRVtroneeccenxxxxxA
\else
\ifnum#1=66 %
\hatcurRVtroneeccenxxxxxB
\else
??????\fi
\fi
}
\newcommand{\hatcurRVtrtwoeccen}[1]{\ifnum#1=65 %
\hatcurRVtrtwoeccenxxxxxA
\else
\ifnum#1=66 %
\hatcurRVtrtwoeccenxxxxxB
\else
??????\fi
\fi
}
\newcommand{\hatcurSMEiiloggeccen}[1]{\ifnum#1=65 %
\hatcurSMEiiloggeccenxxxxxA
\else
??????\fi
}
\newcommand{\hatcurSMEiiteffeccen}[1]{\ifnum#1=65 %
\hatcurSMEiiteffeccenxxxxxA
\else
??????\fi
}
\newcommand{\hatcurSMEiivsineccen}[1]{\ifnum#1=65 %
\hatcurSMEiivsineccenxxxxxA
\else
??????\fi
}
\newcommand{\hatcurSMEiizfeheccen}[1]{\ifnum#1=65 %
\hatcurSMEiizfeheccenxxxxxA
\else
??????\fi
}
\newcommand{\hatcurSMEiizfehshorteccen}[1]{\ifnum#1=65 %
\hatcurSMEiizfehshorteccenxxxxxA
\else
??????\fi
}
\newcommand{\hatcurSMEiloggeccen}[1]{\ifnum#1=65 %
\hatcurSMEiloggeccenxxxxxA
\else
\ifnum#1=66 %
\hatcurSMEiloggeccenxxxxxB
\else
??????\fi
\fi
}
\newcommand{\hatcurSMEiteffeccen}[1]{\ifnum#1=65 %
\hatcurSMEiteffeccenxxxxxA
\else
\ifnum#1=66 %
\hatcurSMEiteffeccenxxxxxB
\else
??????\fi
\fi
}
\newcommand{\hatcurSMEivmaceccen}[1]{\ifnum#1=65 %
\hatcurSMEivmaceccenxxxxxA
\else
\ifnum#1=66 %
\hatcurSMEivmaceccenxxxxxB
\else
??????\fi
\fi
}
\newcommand{\hatcurSMEivmiceccen}[1]{\ifnum#1=65 %
\hatcurSMEivmiceccenxxxxxA
\else
\ifnum#1=66 %
\hatcurSMEivmiceccenxxxxxB
\else
??????\fi
\fi
}
\newcommand{\hatcurSMEivsineccen}[1]{\ifnum#1=65 %
\hatcurSMEivsineccenxxxxxA
\else
\ifnum#1=66 %
\hatcurSMEivsineccenxxxxxB
\else
??????\fi
\fi
}
\newcommand{\hatcurSMEizfeheccen}[1]{\ifnum#1=65 %
\hatcurSMEizfeheccenxxxxxA
\else
\ifnum#1=66 %
\hatcurSMEizfeheccenxxxxxB
\else
??????\fi
\fi
}
\newcommand{\hatcurSMEizfehshorteccen}[1]{\ifnum#1=65 %
\hatcurSMEizfehshorteccenxxxxxA
\else
\ifnum#1=66 %
\hatcurSMEizfehshorteccenxxxxxB
\else
??????\fi
\fi
}
\newcommand{\hatcurXAveccen}[1]{\ifnum#1=65 %
\hatcurXAveccenxxxxxA
\else
\ifnum#1=66 %
\hatcurXAveccenxxxxxB
\else
??????\fi
\fi
}
\newcommand{\hatcurXdisteccen}[1]{\ifnum#1=65 %
\hatcurXdisteccenxxxxxA
\else
\ifnum#1=66 %
\hatcurXdisteccenxxxxxB
\else
??????\fi
\fi
}
\newcommand{\hatcurXdistredeccen}[1]{\ifnum#1=65 %
\hatcurXdistredeccenxxxxxA
\else
\ifnum#1=66 %
\hatcurXdistredeccenxxxxxB
\else
??????\fi
\fi
}
\newcommand{\hatcurXEBVeccen}[1]{\ifnum#1=65 %
\hatcurXEBVeccenxxxxxA
\else
\ifnum#1=66 %
\hatcurXEBVeccenxxxxxB
\else
??????\fi
\fi
}
\newcommand{\hatcurXjhisoredeccen}[1]{\ifnum#1=65 %
\hatcurXjhisoredeccenxxxxxA
\else
\ifnum#1=66 %
\hatcurXjhisoredeccenxxxxxB
\else
??????\fi
\fi
}
\newcommand{\hatcurXjkisoredeccen}[1]{\ifnum#1=65 %
\hatcurXjkisoredeccenxxxxxA
\else
\ifnum#1=66 %
\hatcurXjkisoredeccenxxxxxB
\else
??????\fi
\fi
}
\newcommand{\hatcurXmhisoredeccen}[1]{\ifnum#1=65 %
\hatcurXmhisoredeccenxxxxxA
\else
\ifnum#1=66 %
\hatcurXmhisoredeccenxxxxxB
\else
??????\fi
\fi
}
\newcommand{\hatcurXmiisoredeccen}[1]{\ifnum#1=65 %
\hatcurXmiisoredeccenxxxxxA
\else
\ifnum#1=66 %
\hatcurXmiisoredeccenxxxxxB
\else
??????\fi
\fi
}
\newcommand{\hatcurXmjisoredeccen}[1]{\ifnum#1=65 %
\hatcurXmjisoredeccenxxxxxA
\else
\ifnum#1=66 %
\hatcurXmjisoredeccenxxxxxB
\else
??????\fi
\fi
}
\newcommand{\hatcurXmkisoredeccen}[1]{\ifnum#1=65 %
\hatcurXmkisoredeccenxxxxxA
\else
\ifnum#1=66 %
\hatcurXmkisoredeccenxxxxxB
\else
??????\fi
\fi
}
\newcommand{\hatcurXmvisoredeccen}[1]{\ifnum#1=65 %
\hatcurXmvisoredeccenxxxxxA
\else
\ifnum#1=66 %
\hatcurXmvisoredeccenxxxxxB
\else
??????\fi
\fi
}
\newcommand{\hatcurXsecdureccen}[1]{\ifnum#1=65 %
\hatcurXsecdureccenxxxxxA
\else
\ifnum#1=66 %
\hatcurXsecdureccenxxxxxB
\else
??????\fi
\fi
}
\newcommand{\hatcurXsecingdureccen}[1]{\ifnum#1=65 %
\hatcurXsecingdureccenxxxxxA
\else
\ifnum#1=66 %
\hatcurXsecingdureccenxxxxxB
\else
??????\fi
\fi
}
\newcommand{\hatcurXsecondaryeccen}[1]{\ifnum#1=65 %
\hatcurXsecondaryeccenxxxxxA
\else
\ifnum#1=66 %
\hatcurXsecondaryeccenxxxxxB
\else
??????\fi
\fi
}
\newcommand{\hatcurXsecphaseeccen}[1]{\ifnum#1=65 %
\hatcurXsecphaseeccenxxxxxA
\else
\ifnum#1=66 %
\hatcurXsecphaseeccenxxxxxB
\else
??????\fi
\fi
}
\newcommand{\hatcurXviisoredeccen}[1]{\ifnum#1=65 %
\hatcurXviisoredeccenxxxxxA
\else
\ifnum#1=66 %
\hatcurXviisoredeccenxxxxxB
\else
??????\fi
\fi
}
\newcommand{\hatcurXvkisoredeccen}[1]{\ifnum#1=65 %
\hatcurXvkisoredeccenxxxxxA
\else
\ifnum#1=66 %
\hatcurXvkisoredeccenxxxxxB
\else
??????\fi
\fi
}

\newcommand{\hatcurxxxxxA}{HAT-P-65}
\newcommand{\hatcurbxxxxxA}{HAT-P-65b}
\newcommand{\hatcurcxxxxxA}{HAT-P-65c}

\newcommand{\hatcurplanetnumxxxxxA}{65}

\newcommand{\hatcurtwomassshortxxxxxA}{21033731+1159218}

\newcommand{\hatcurSindexxxxxxA}{\ensuremath{\cdots}}
\newcommand{\hatcurRHKindexxxxxxA}{\ensuremath{\cdots}}

\newcommand{\hatcurRVgammaabsxxxxxA}{\ensuremath{-47.77\pm0.10}}                           % Absolute Gamma velocity
\newcommand{\hatcurRVgammainstxxxxxA}{TRES}                           % Absolute Gamma velocity

\newcommand{\hatcurCCtassvixxxxxA}{NULL}                  % TASS V-I

\newcommand{\hatcurSMEversionxxxxxA}{ii}                                       % Final SME version:i or ii?

\newcommand{\hatcurisoshortxxxxxA}{YY}
\newcommand{\hatcurisofullxxxxxA}{Yonsei-Yale (YY)}
\newcommand{\hatcurisocitexxxxxA}{yi:2001}
\newcommand{\hatcurlumindxxxxxA}{\arstar}
\newcommand{\hatcurjhkfilsetxxxxxA}{ESO}
\newcommand{\hatcurSMEteffxxxxxA}{\ifthenelse{\equal{\hatcurSMEversionxxxxxA}{i}}{\hatcurSMEiteff{\hatcurplanetnumxxxxxA}}{\hatcurSMEiiteff{\hatcurplanetnumxxxxxA}}}
\newcommand{\hatcurSMEzfehxxxxxA}{\ifthenelse{\equal{\hatcurSMEversionxxxxxA}{i}}{\hatcurSMEizfeh{\hatcurplanetnumxxxxxA}}{\hatcurSMEiizfeh{\hatcurplanetnumxxxxxA}}}
\newcommand{\hatcurSMEzfehshortxxxxxA}{\ifthenelse{\equal{\hatcurSMEversionxxxxxA}{i}}{\hatcurSMEizfehshort{\hatcurplanetnumxxxxxA}}{\hatcurSMEiizfehshort{\hatcurplanetnumxxxxxA}}}
\newcommand{\hatcurSMEloggxxxxxA}{\ifthenelse{\equal{\hatcurSMEversionxxxxxA}{i}}{\hatcurSMEilogg{\hatcurplanetnumxxxxxA}}{\hatcurSMEiilogg{\hatcurplanetnumxxxxxA}}}
\newcommand{\hatcurSMEvsinxxxxxA}{\ifthenelse{\equal{\hatcurSMEversionxxxxxA}{i}}{\hatcurSMEivsin{\hatcurplanetnumxxxxxA}}{\hatcurSMEiivsin{\hatcurplanetnumxxxxxA}}}
\newcommand{\hatcurSMEvmacxxxxxA}{\ifthenelse{\equal{\hatcurSMEversionxxxxxA}{i}}{\hatcurSMEivmac{\hatcurplanetnumxxxxxA}}{\hatcurSMEiivmac{\hatcurplanetnumxxxxxA}}}
\newcommand{\hatcurSMEvmicxxxxxA}{\ifthenelse{\equal{\hatcurSMEversionxxxxxA}{i}}{\hatcurSMEivmic{\hatcurplanetnumxxxxxA}}{\hatcurSMEiivmic{\hatcurplanetnumxxxxxA}}}

\newcommand{\hatcurxxxxxB}{HAT-P-66}
\newcommand{\hatcurbxxxxxB}{HAT-P-66b}
\newcommand{\hatcurcxxxxxB}{HAT-P-66c}

\newcommand{\hatcurplanetnumxxxxxB}{66}

\newcommand{\hatcurtwomassshortxxxxxB}{10021743+5357031}

\newcommand{\hatcurSindexxxxxxB}{\ensuremath{\cdots}}
\newcommand{\hatcurRHKindexxxxxxB}{\ensuremath{\cdots}}

\newcommand{\hatcurRVgammaabsxxxxxB}{\ensuremath{7.97\pm0.10}}                           % Absolute Gamma velocity
\newcommand{\hatcurRVgammainstxxxxxB}{TRES}                           % Absolute Gamma velocity

\newcommand{\hatcurCCtassvixxxxxB}{NULL}                  % TASS V-I

\newcommand{\hatcurSMEversionxxxxxB}{i}                                       % Final SME version:i or ii?

\newcommand{\hatcurisoshortxxxxxB}{YY}
\newcommand{\hatcurisofullxxxxxB}{Yonsei-Yale (YY)}
\newcommand{\hatcurisocitexxxxxB}{yi:2001}
\newcommand{\hatcurlumindxxxxxB}{\arstar}
\newcommand{\hatcurjhkfilsetxxxxxB}{ESO}
\newcommand{\hatcurSMEteffxxxxxB}{\ifthenelse{\equal{\hatcurSMEversionxxxxxB}{i}}{\hatcurSMEiteff{\hatcurplanetnumxxxxxB}}{\hatcurSMEiiteff{\hatcurplanetnumxxxxxB}}}
\newcommand{\hatcurSMEzfehxxxxxB}{\ifthenelse{\equal{\hatcurSMEversionxxxxxB}{i}}{\hatcurSMEizfeh{\hatcurplanetnumxxxxxB}}{\hatcurSMEiizfeh{\hatcurplanetnumxxxxxB}}}
\newcommand{\hatcurSMEzfehshortxxxxxB}{\ifthenelse{\equal{\hatcurSMEversionxxxxxB}{i}}{\hatcurSMEizfehshort{\hatcurplanetnumxxxxxB}}{\hatcurSMEiizfehshort{\hatcurplanetnumxxxxxB}}}
\newcommand{\hatcurSMEloggxxxxxB}{\ifthenelse{\equal{\hatcurSMEversionxxxxxB}{i}}{\hatcurSMEilogg{\hatcurplanetnumxxxxxB}}{\hatcurSMEiilogg{\hatcurplanetnumxxxxxB}}}
\newcommand{\hatcurSMEvsinxxxxxB}{\ifthenelse{\equal{\hatcurSMEversionxxxxxB}{i}}{\hatcurSMEivsin{\hatcurplanetnumxxxxxB}}{\hatcurSMEiivsin{\hatcurplanetnumxxxxxB}}}
\newcommand{\hatcurSMEvmacxxxxxB}{\ifthenelse{\equal{\hatcurSMEversionxxxxxB}{i}}{\hatcurSMEivmac{\hatcurplanetnumxxxxxB}}{\hatcurSMEiivmac{\hatcurplanetnumxxxxxB}}}
\newcommand{\hatcurSMEvmicxxxxxB}{\ifthenelse{\equal{\hatcurSMEversionxxxxxB}{i}}{\hatcurSMEivmic{\hatcurplanetnumxxxxxB}}{\hatcurSMEiivmic{\hatcurplanetnumxxxxxB}}}

\newcommand{\hatcur}[1]{\ifnum#1=65 %
\hatcurxxxxxA
\else
\ifnum#1=66 %
\hatcurxxxxxB
\else
??????\fi
\fi
}
\newcommand{\hatcurb}[1]{\ifnum#1=65 %
\hatcurbxxxxxA
\else
\ifnum#1=66 %
\hatcurbxxxxxB
\else
??????\fi
\fi
}
\newcommand{\hatcurc}[1]{\ifnum#1=65 %
\hatcurcxxxxxA
\else
\ifnum#1=66 %
\hatcurcxxxxxB
\else
??????\fi
\fi
}
\newcommand{\hatcurCCtassvi}[1]{\ifnum#1=65 %
\hatcurCCtassvixxxxxA
\else
\ifnum#1=66 %
\hatcurCCtassvixxxxxB
\else
??????\fi
\fi
}
\newcommand{\hatcurisocite}[1]{\ifnum#1=65 %
\hatcurisocitexxxxxA
\else
\ifnum#1=66 %
\hatcurisocitexxxxxB
\else
??????\fi
\fi
}
\newcommand{\hatcurisofull}[1]{\ifnum#1=65 %
\hatcurisofullxxxxxA
\else
\ifnum#1=66 %
\hatcurisofullxxxxxB
\else
??????\fi
\fi
}
\newcommand{\hatcurisoshort}[1]{\ifnum#1=65 %
\hatcurisoshortxxxxxA
\else
\ifnum#1=66 %
\hatcurisoshortxxxxxB
\else
??????\fi
\fi
}
\newcommand{\hatcurjhkfilset}[1]{\ifnum#1=65 %
\hatcurjhkfilsetxxxxxA
\else
\ifnum#1=66 %
\hatcurjhkfilsetxxxxxB
\else
??????\fi
\fi
}
\newcommand{\hatcurlumind}[1]{\ifnum#1=65 %
\hatcurlumindxxxxxA
\else
\ifnum#1=66 %
\hatcurlumindxxxxxB
\else
??????\fi
\fi
}
\newcommand{\hatcurplanetnum}[1]{\ifnum#1=65 %
\hatcurplanetnumxxxxxA
\else
\ifnum#1=66 %
\hatcurplanetnumxxxxxB
\else
??????\fi
\fi
}
\newcommand{\hatcurRHKindex}[1]{\ifnum#1=65 %
\hatcurRHKindexxxxxxA
\else
\ifnum#1=66 %
\hatcurRHKindexxxxxxB
\else
??????\fi
\fi
}
\newcommand{\hatcurRVgammaabs}[1]{\ifnum#1=65 %
\hatcurRVgammaabsxxxxxA
\else
\ifnum#1=66 %
\hatcurRVgammaabsxxxxxB
\else
??????\fi
\fi
}
\newcommand{\hatcurRVgammainst}[1]{\ifnum#1=65 %
\hatcurRVgammainstxxxxxA
\else
\ifnum#1=66 %
\hatcurRVgammainstxxxxxB
\else
??????\fi
\fi
}
\newcommand{\hatcurSindex}[1]{\ifnum#1=65 %
\hatcurSindexxxxxxA
\else
\ifnum#1=66 %
\hatcurSindexxxxxxB
\else
??????\fi
\fi
}
\newcommand{\hatcurSMElogg}[1]{\ifnum#1=65 %
\hatcurSMEloggxxxxxA
\else
\ifnum#1=66 %
\hatcurSMEloggxxxxxB
\else
??????\fi
\fi
}
\newcommand{\hatcurSMEteff}[1]{\ifnum#1=65 %
\hatcurSMEteffxxxxxA
\else
\ifnum#1=66 %
\hatcurSMEteffxxxxxB
\else
??????\fi
\fi
}
\newcommand{\hatcurSMEversion}[1]{\ifnum#1=65 %
\hatcurSMEversionxxxxxA
\else
\ifnum#1=66 %
\hatcurSMEversionxxxxxB
\else
??????\fi
\fi
}
\newcommand{\hatcurSMEvmac}[1]{\ifnum#1=65 %
\hatcurSMEvmacxxxxxA
\else
\ifnum#1=66 %
\hatcurSMEvmacxxxxxB
\else
??????\fi
\fi
}
\newcommand{\hatcurSMEvmic}[1]{\ifnum#1=65 %
\hatcurSMEvmicxxxxxA
\else
\ifnum#1=66 %
\hatcurSMEvmicxxxxxB
\else
??????\fi
\fi
}
\newcommand{\hatcurSMEvsin}[1]{\ifnum#1=65 %
\hatcurSMEvsinxxxxxA
\else
\ifnum#1=66 %
\hatcurSMEvsinxxxxxB
\else
??????\fi
\fi
}
\newcommand{\hatcurSMEzfeh}[1]{\ifnum#1=65 %
\hatcurSMEzfehxxxxxA
\else
\ifnum#1=66 %
\hatcurSMEzfehxxxxxB
\else
??????\fi
\fi
}
\newcommand{\hatcurSMEzfehshort}[1]{\ifnum#1=65 %
\hatcurSMEzfehshortxxxxxA
\else
\ifnum#1=66 %
\hatcurSMEzfehshortxxxxxB
\else
??????\fi
\fi
}
\newcommand{\hatcurtwomassshort}[1]{\ifnum#1=65 %
\hatcurtwomassshortxxxxxA
\else
\ifnum#1=66 %
\hatcurtwomassshortxxxxxB
\else
??????\fi
\fi
}

\newcommand{\SpearmanrhoHAT}{0.344}
\newcommand{\SpearmanprobHAT}{1.4}
\newcommand{\SpearmanMprhoHAT}{0.428}
\newcommand{\SpearmanMpprobHAT}{0.84}
\newcommand{\SpearmanrhoWASP}{0.277}
\newcommand{\SpearmanprobWASP}{3.5}
\newcommand{\SpearmanMprhoWASP}{0.273}
\newcommand{\SpearmanMpprobWASP}{7.7}
\newcommand{\Spearmanrhocomb}{0.347}
\newcommand{\Spearmanprobcomb}{0.0041}
\newcommand{\SpearmanMprhocomb}{0.398}
\newcommand{\SpearmanMpprobcomb}{0.0068}

\newcommand{\Kendallrtrunc}{\ensuremath{0.228}}
\newcommand{\Kendallprobtrunc}{\ensuremath{2.88}}
\newcommand{\KendallrtruncMp}{\ensuremath{0.284}}
\newcommand{\KendallprobtruncMp}{\ensuremath{2.92}}
\newcommand{\Kendallrnotrunc}{\ensuremath{0.235}}
\newcommand{\Kendallprobnotrunc}{\ensuremath{1.52}}
\newcommand{\KendallrnotruncMp}{\ensuremath{0.297}}
\newcommand{\KendallprobnotruncMp}{\ensuremath{0.87}}

\newcounter{planetcounter}

\newboolean{emulateapj}
\setboolean{emulateapj}{true}

\newboolean{rvtablelong}
\setboolean{rvtablelong}{true}

\newboolean{astroph}
\setboolean{astroph}{true}
\shortauthors{Hartman et al.}
\shorttitle{\hatcur{65}\lowercase{b} and \hatcur{66}\lowercase{b}}
\ifthenelse{\boolean{emulateapj}}{
    \newcommand{\titledag}{$\dagger$}
}{
    \newcommand{\titledag}{\dagger}
}

\begin{document}

%% Titlepage
\title{%%
\hatcur{65}\lowercase{b} and \hatcur{66}\lowercase{b}: Two Transiting Inflated Hot Jupiters and Observational Evidence for the Re-Inflation of Close-In Giant Planets\altaffilmark{\titledag}
}

%% Authors
\author{
    J.~D.~Hartman\altaffilmark{1}, 
    G.~\'A.~Bakos\altaffilmark{1,*,**},
    W.~Bhatti\altaffilmark{1},
    K.~Penev\altaffilmark{1},
    A.~Bieryla\altaffilmark{2},
    D.~W.~Latham\altaffilmark{2},
    G.~Kov\'acs\altaffilmark{3},
    G.~Torres\altaffilmark{2},
    Z.~Csubry\altaffilmark{1},
    M.~de~Val-Borro\altaffilmark{1},
    L.~Buchhave\altaffilmark{4},
    T.~Kov\'acs\altaffilmark{3},
    S.~Quinn\altaffilmark{5},
    A.~W.~Howard\altaffilmark{6},
    H.~Isaacson\altaffilmark{7},
    B.~J.~Fulton\altaffilmark{6},
    M.~E.~Everett\altaffilmark{8},
    G.~Esquerdo\altaffilmark{2},
    B.~B\'eky\altaffilmark{9},
    T.~Szklenar\altaffilmark{10},
    E.~Falco\altaffilmark{2},
    A.~Santerne\altaffilmark{12},
    I.~Boisse\altaffilmark{11},
    G.~H\'ebrard\altaffilmark{13},
    A.~Burrows\altaffilmark{1},
    J.~L\'az\'ar\altaffilmark{10},
    I.~Papp\altaffilmark{10},
    P.~S\'ari\altaffilmark{10}
}

\altaffiltext{1}{Department of Astrophysical Sciences, Princeton
  University, Princeton, NJ 08544, USA; email: jhartman@astro.princeton.edu}
\altaffiltext{$*$}{%%
Alfred P. Sloan Research Fellow
}
\altaffiltext{$**$}{%%
Packard Fellow
}
\altaffiltext{2}{Harvard-Smithsonian Center for Astrophysics, Cambridge, MA 02138, USA}
\altaffiltext{3}{Konkoly Observatory of the Hungarian Academy of Sciences, Budapest, Hungary}
\altaffiltext{4}{Centre for Star and Planet Formation, Natural History Museum of Denmark, University of Copenhagen, DK-1350 Copenhagen, Denmark}
\altaffiltext{5}{Department of Physics and Astronomy, Georgia State University, Atlanta, GA 30303, USA}
\altaffiltext{6}{Institute for Astronomy, University of Hawaii, Honolulu, HI 96822, USA}
\altaffiltext{7}{Department of Astronomy, University of California, Berkeley, CA, USA}
\altaffiltext{8}{National Optical Astronomy Observatory, Tucson, AZ, USA}
\altaffiltext{9}{Google}
\altaffiltext{10}{Hungarian Astronomical Association, Budapest, Hungary}
\altaffiltext{11}{Aix Marseille Universit\'e, CNRS, LAM (Laboratoire d'Astrophysique de Marseille) UMR 7326, F-13388, Marseille, France}
\altaffiltext{12}{Instituto de Astrofisica e Ci\^encias do Espa\c{c}o, Universidade do Porto, CAUP, Rua das Estrelas, PT4150-762 Porto, Portugal}
\altaffiltext{13}{Institut d'Astrophysique de Paris, UMR7095 CNRS, Universit\'e Pierre \& Marie Curie, 98bis boulevard Arago, 75014 Paris, France}
\altaffiltext{$\dagger$}{%%
 Based on observations obtained with the Hungarian-made Automated
 Telescope Network. Based on observations obtained at the W.~M.~Keck
 Observatory, which is operated by the University of California and
 the California Institute of Technology. Keck time has been granted by
 NOAO (A289Hr, A245Hr) and NASA (N029Hr, N154Hr, N130Hr, N133Hr,
 N169Hr, N186Hr). Based on observations obtained with the Tillinghast
 Reflector 1.5\,m telescope and the 1.2\,m telescope, both operated by
 the Smithsonian Astrophysical Observatory at the Fred Lawrence
 Whipple Observatory in AZ. Based on observations made with
 the Nordic Optical Telescope, operated on the island of La Palma
 jointly by Denmark, Finland, Norway, Sweden, in the Spanish
 Observatorio del Roque de los Muchachos of the Intituto de
 Astrof\'isica de Canarias. Based on observations made with the SOPHIE spectrograph on the 1.93\,m telescope at Observatoire de Haute-Provence (OHP, CNRS/AMU), France (programs 15A.PNP.HEBR and 15B.PNP.HEBR). Data presented herein were obtained at the WIYN Observatory from
telescope time allocated to NN-EXPLORE through the scientific
partnership of the National Aeronautics and Space Administration,
the National Science Foundation, and the National Optical Astronomy
Observatory. This work was supported by a NASA WIYN PI Data Award,
administered by the NASA Exoplanet Science Institute.
}

%% EOF authors

% #####################################################################
%% abstract
\begin{abstract}
\setcounter{footnote}{1}
We present the discovery of the transiting exoplanets \hatcurb{65} and
\hatcurb{66}, with orbital periods of
\hatcurLCPshort{65}\,d and \hatcurLCPshort{66}\,d, masses of
\hatcurPPm{65}\,\mjup\ and \hatcurPPm{66}\,\mjup, and inflated radii
of \hatcurPPr{65}\,\rjup\ and \hatcurPPr{66}\,\rjup,
respectively. They orbit moderately bright ($V=\hatcurCCtassmv{65}$,
and $V=\hatcurCCtassmv{66}$) stars of mass \hatcurISOm{65}\,\msun\ and
\hatcurISOm{66}\,\msun. The stars 
are at the
main sequence turnoff. While it is
well known that the radii of close-in giant planets are correlated
with their equilibrium temperatures, whether or not the radii of
planets increase in time as their hosts evolve and become more
luminous is an open question. Looking at the broader sample of
well-characterized close-in transiting giant planets, we find that
there is a statistically significant correlation between planetary
radii and the fractional ages of their host stars, with a false alarm
probability of only \Spearmanprobcomb\%. 
We find that the correlation between the radii of planets
and the fractional ages of their hosts is fully explained by the
known correlation between planetary radii and their present day
equilibrium temperatures, however if the zero-age main sequence
equilibrium temperature is used in place of the present day
equilibrium temperature then a correlation with age must also be
included to explain the planetary radii. This suggests that, after
contracting during the pre-main-sequence, close-in giant planets are
re-inflated over time due to the increasing level of irradiation
received from their 
host stars. Prior
theoretical work indicates that such a dynamic response to irradiation
requires a significant fraction of the incident energy to be
deposited deep within the planetary interiors. 
\setcounter{footnote}{0}
\end{abstract}

% #####################################################################
%% keywords
\keywords{
    planetary systems ---
    stars: individual (
\setcounter{planetcounter}{1}
\hatcur{65},
\hatcurCCgsc{65}\loopcommanoperiod
\setcounter{planetcounter}{2}
\hatcur{66},
\hatcurCCgsc{66}\loopcommanoperiod
\setcounter{planetcounter}{3}
) 
    techniques: spectroscopic, photometric
}

%% EOF keywords
%% EOF titlepage

% #####################################################################
%% Introduction
\section{Introduction}
\label{sec:introduction}
%++++++++++++++++++++++++++++++++++++++++++++++++++++++++++++++++++++++
%++++++++++++++++++++++++++++++++++++++++++++++++++++++++++++++++++++++
%% EOF introduction

The first transiting exoplanet (TEP) discovered, HD~209458b
\citep{henry:2000,charbonneau:2000}, surprised the community in having
a radius much larger than expected based on theoretical planetary
structure models \citep[e.g.,][]{burrows:2000,bodenheimer:2001}. Since then many
more inflated transiting planets have been discovered, the largest
being WASP-79b with $R_{P} = 2.09 \pm
0.14$\,\rjup\ \citep{2012AAP...547A..61S}. It has also become apparent
that the degree of planet inflation is closely tied to a planet's
proximity to its host star
\citep[e.g.,][]{fortney:2007,2011AJ....142...86E,2010ApJ...724..866K,2011ApJ...734..109B,enoch:2012}. This
is expected on theoretical grounds, as some additional energy, beyond
the initial heat from formation, must be responsible for making the
planet so large, and in principle there is more than enough energy
available from stellar irradiation or tidal forces to inflate close-in
planets at $a < 0.1$\,AU \citep{bodenheimer:2001}. Whether and how the energy is
transfered into planetary interiors remains a mystery, however,
despite a large amount of theoretical work devoted to the subject
(see, e.g., \citealp{spiegel:2013} for a review). The problem is
intrinsically challenging, requiring the simultaneous treatment of
molecular chemistry, radiative transport, and turbulent
(magneto-)hydrodynamics, carried out over pressures, densities,
temperatures, and length-scales that span many orders of
magnitude. Theoretical models of planet inflation have thus, by
necessity, made numerous simplifying assumptions, often introducing
free parameters whose values are unknown, or poorly known. One way to
make further progress on this problem is to build up a larger sample
of inflated planets to identify patterns in their properties that may
be used to discriminate between different theories.

Recently \citet{lopez:2016} proposed an observational test to
distinguish between two broad classes of models. Noting that once a
star leaves the main sequence, the irradiation of its planets with
periods of tens of days becomes comparable to the irradiation of very
short period planets around main sequence stars, they suggested
searching for inflated planets with periods of tens of days around
giant stars. Planets at these orbital periods are not inflated when
found around main sequence stars \citep{demory:2011}, so finding them to be inflated
around giants would indicate that the enhanced irradiation is able to
directly inflate the planets. As shown, for example, by
\citet{liu:2008} and also by \citet{spiegel:2013}, this in turn would
imply that energy must be transferred deep into the planetary
interior, and would rule out models where the energy is deposited only
in the outer layers of the planet, and serves simply to slow the
planet's contraction from its initial highly inflated state. The
recently discovered planet EPIC~211351816.01 \citep[found using {\em
    K2}]{grunblatt:2016} is a possible example of a re-inflated planet
around a giant star, with the planet having a larger than usual radius
of $1.27 \pm 0.09$\,\rjup\ given its orbital period of
$8.4$\,days. The planet K2-39b \citep{vaneylen:2016}, on the other
hand, does not appear to be exceptionally inflated ($\rpl = 0.732 \pm
0.098$\,\rjup) despite being found on a very short period orbit around
a sub-giant star. This planet, however, is in the Super-Neptune mass
range ($\mpl = 0.158 \pm 0.031$\,\mjup) and may not have a
gas-dominated composition.

Here we present the discovery of two transiting inflated planets by
the Hungarian-made Automated Telescope Network
\citep[HATNet;][]{bakos:2004:hatnet}. As we will show, the planets
have radii of \hatcurPPr{65}\,\rjup\ and \hatcurPPr{66}\,\rjup, and
are around a pair of stars that are leaving the main sequence. HATNet,
together with its southern counterpart HATSouth
\citep{bakos:2013:hatsouth}, has now discovered 17 highly inflated
planets with $R \geq 1.5$\,\rjup\footnote{This radius is chosen simply for illustrative purposes, and is not meant to imply that planets with radii above this value are physically distinct from those with radii below this value.}. Adding those found by WASP
\citep{pollacco:2006}, {\em Kepler} \citep{borucki:2010}, TrES
\citep[e.g.,][]{2007ApJ...667L.195M} and KELT
\citep[e.g.,][]{2012ApJ...761..123S}, a total of 45 well-characterized
highly inflated planets are now known, allowing us to explore some of
their statistical properties. In this paper we find that inflated
planets are more commonly found around moderately evolved stars that
are more than 50\% of the way through their main sequence
lifetimes. Smaller radius close-in giant planets, by contrast, are
generally found around less evolved stars. Taken at face value, this
suggests that planets are re-inflated as they age, and indicates that
energy must be transferred deep into the planetary interiors
\citep[e.g.,][]{liu:2008}.

Of course, observational selection effects or systematic errors in the
determination of stellar and planetary properties could potentially be
responsible for the correlation as well. We therefore consider a
variety of potentially important effects, such as the effect of
stellar evolution on the detectability of transits and our ability to
confirm planets through follow-up observations, and systematic errors
in the orbital eccentricity, transit parameters, stellar atmospheric
parameters, or in the comparison to stellar evolution models. We
conclude that the net selection effect would, if anything, tend to
favor the discovery of large planets around less evolved stars, while
potential systematic errors are too small to explain the
correlation. We also show that the correlation remains significant
even after accounting for non-trivial truncations placed on the data
as a result of the observational selection biases. We are therefore
confident in the robustness of this result.

The organization of the paper is as follows. In Section~\ref{sec:obs}
we describe the photometric and spectroscopic observations made to
discover and characterize \hatcurb{65} and \hatcurb{66}. In
Section~\ref{sec:analysis} we present the analysis carried out to
determine the stellar and planetary parameters and to rule out blended
stellar eclipsing binary false positive scenarios. In
Section~\ref{sec:discussion} we place these planets into context, and
find that large radius planets are more commonly found around
moderately evolved, brighter stars. We provide a brief summary of
the results in Section~\ref{sec:summary}.

% #####################################################################
\section{Observations}
\label{sec:obs}
%++++++++++++++++++++++++++++++++++++++++++++++++++++++++++++++++++++++
%++++++++++++++++++++++++++++++++++++++++++++++++++++++++++++++++++++++

% =====================================================================
%% Photometric detection
\subsection{Photometric detection}
\label{sec:detection}
%++++++++++++++++++++++++++++++++++++++++++++++++++++++++++++++++++++++
%++++++++++++++++++++++++++++++++++++++++++++++++++++++++++++++++++++++

Both \hatcur{65} (${\rm R.A.}=$\hatcurCCra{65}, ${\rm
  Dec.}=$\hatcurCCdec{65} (J2000), $V=\hatcurCCtassmv{65}$\,mag,
spectral type G2) and \hatcur{66} (${\rm R.A.}=$\hatcurCCra{66}, ${\rm
  Dec.}=$\hatcurCCdec{66} (J2000), $V=\hatcurCCtassmv{66}$\,mag,
spectral type G0) were selected as candidate transiting planet systems
based on Sloan $r$-band photometric time series observations carried
out with the HATNet telescope network \citep{bakos:2004:hatnet}. 

HATNet consists of six 11\,cm aperture telephoto lenses, each coupled
to an APOGEE front-side-illuminated CCD camera, and each placed on a
fully-automated telescope mount. Four of the instruments are located
at Fred Lawrence Whipple Observatory (FLWO) in Arizona, USA, while two
are located on the roof of the Submillimeter Array hangar building at
Mauna Kea Observatory (MKO) on the island of Hawaii, USA. Each
instrument observes a $10.6^{\circ}\times 10.6^{\circ}$ field of view,
and continuously monitors one or two fields each night, where a field
corresponds to one of 838 fixed pointings used to cover the full 4$\pi$
celestial sphere.  A typical field is observed for approximately three
months using one or two instruments (e.g., field G342 containing
\hatcur{65}), while a handful of fields have been observed extensively
using all six instruments in the network and with observations
repeated in multiple seasons (e.g., field G101 containing
\hatcur{66}). The former observing strategy maximizes the sky coverage
of the survey, while maintaining nearly complete sensitivity to
transiting giant planets with orbital periods of a few days. The
latter strategy substantially increases the sensitivity to Neptune and
Super-Earth-size planets, as well as planets with periods greater than
10\,days, but with the trade-off of covering a smaller area of the
sky.

\reftabl{photobs} summarizes the properties of the HATNet
observations collected for each system, including which HAT
instruments were used, the date ranges over which each target was
observed, the median cadence of the observations, and the per-point
photometric precision after trend filtering.

We reduce the HATNet observations to light curves, for all stars in a
field with $r < 14.5$, following \citet{bakos:2004:hatnet}. We used
aperture photometry routines based on the FITSH software package
\citep{pal:2012}, and filtered systematic trends from the light curves
following \citet{kovacs:2005:TFA} (i.e., TFA) and
\citet{2010ApJ...710.1724B} (i.e., EPD).  Transits were identified in
the filtered light curves using the Box-Least Squares method
\citep[BLS;][]{kovacs:2002:BLS}. After identifying the transits we
then re-applied TFA while preserving the shape of the transit signal
as described in \citet{kovacs:2005:TFA}. This procedure is referred to
as signal-reconstruction TFA. The final trend-filtered, and
signal-reconstructed light curves are shown phase-folded in
\reffigl{hatnet}, while the measurements are available in
\reftabl{phfu}.

We searched the residual HATNet light curves of both objects for
additional periodic signals using BLS. Neither target shows evidence
for additional transits with BLS, however this conclusion depends on
the set of template light curves used in applying
signal-reconstruction TFA to remove systematics. For \hatcur{65} we
find that with an alternative set of templates the residuals display a
marginally significant transit signal with a period of $2.573$\,days,
which is only slightly different from the main transit period of
$\hatcurLCP{65}$\,days. The transits detected at this period come from
data points near orbital phase $0.25$ when phased at the primary
transit period. Since the detection of this additional signal depends
on the template set used, and since any planet orbiting with a period
so close to (but not equal to) that of the hot Jupiter \hatcurb{65}
would almost certainly be unstable, we suspect that the $P =
2.573$\,day transit signal is not of physical origin.

We also searched the residual light curves for periodic signals using
the Generalized Lomb-Scargle method \citep[GLS;][]{zechmeister:2009}.
For \hatcur{65} no statistically significant signal is detected in the
GLS periodogram either. The highest peak in the periodogram is at a
period of $0.035$\,d and has a semi-amplitude of 1.2\,mmag (using a
Markov-Chain Monte Carlo procedure to fit a sinusoid with a variable
period yields a 95\% confidence upper limit of 1.7\,mmag on the
semi-amplitude). For \hatcur{66}, for our default light curve (i.e.,
the one included in \reftabl{phfu}), we do see significant peaks in
the periodogram at periods of $P = 83.3029$\,d and $0.98664$\,d (and
its harmonics) and with formal false alarm probabilities of
$10^{-11}$, and semi-amplitudes of $\sim 0.02$\,mag. Given the
effective sampling rate of the observations, the two signals are
aliases of each other. Based on an inspection of the light curve, we
conclude that this detected variability is likely due to additional
systematic errors in the photometry which were not effectively removed
by our filtering procedures, and that the signal is not astrophysical
in nature. Indeed if we use an alternative TFA template set in
filtering the \hatcur{66} light curve, we detect no significant signal
in the GLS spectrum, and place an upper limit on the amplitude of any
periodic signal of 1\,mmag.

%
%
%% ----------------
\ifthenelse{\boolean{emulateapj}}{
    \begin{figure*}[!ht]
}{
    \begin{figure}[!ht]
}
\plottwo{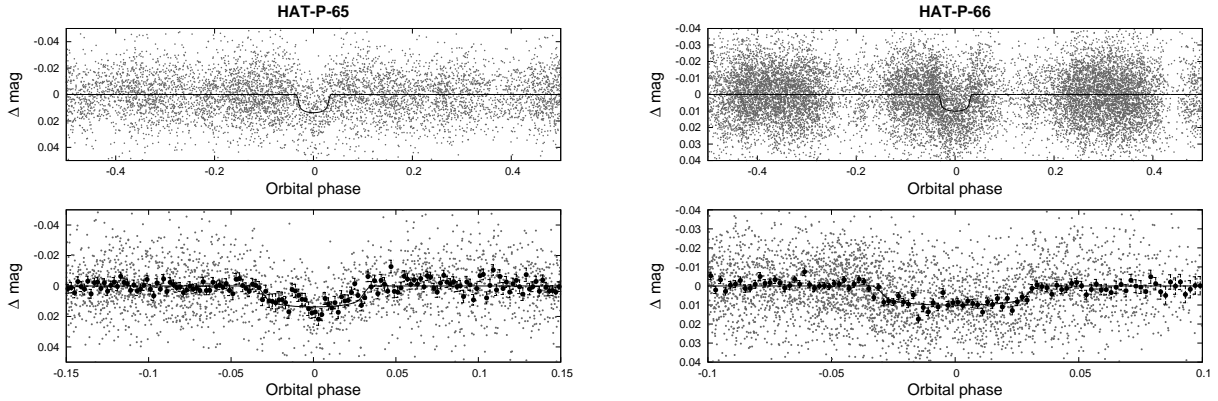}{\hatcurhtr{66}-hatnet.eps}
\caption[]{
    Phase-folded unbinned HATNet light curves for \hatcur{65} (left)
    and \hatcur{66} (right). In each case we show two panels. The top
    panel shows the full light curve, while the bottom panel shows the
    light curve zoomed-in on the transit. The solid lines show the
    model fits to the light curves. The dark filled circles in the
    bottom panels show the light curves binned in phase with a bin
    size of 0.002.
\label{fig:hatnet}}
\ifthenelse{\boolean{emulateapj}}{
    \end{figure*}
}{
    \end{figure}
}
%% ----------------

%% --------------------------------------------------------------------
%% Table summarizing photometric observations
%%
\ifthenelse{\boolean{emulateapj}}{
    \begin{deluxetable*}{llrrrr}
}{
    \begin{deluxetable}{llrrrr}
}
\tablewidth{0pc}
\tabletypesize{\scriptsize}
\tablecaption{
    Summary of photometric observations
    \label{tab:photobs}
}
\tablehead{
    \multicolumn{1}{c}{Instrument/Field\tablenotemark{a}} &
    \multicolumn{1}{c}{Date(s)} &
    \multicolumn{1}{c}{\# Images} &
    \multicolumn{1}{c}{Cadence\tablenotemark{b}} &
    \multicolumn{1}{c}{Filter} &
    \multicolumn{1}{c}{Precision\tablenotemark{c}} \\
    \multicolumn{1}{c}{} &
    \multicolumn{1}{c}{} &
    \multicolumn{1}{c}{} &
    \multicolumn{1}{c}{(sec)} &
    \multicolumn{1}{c}{} &
    \multicolumn{1}{c}{(mmag)}
}
\startdata
\sidehead{\textbf{\hatcur{65}}}
~~~~HAT-6/G342 & 2009 Sep--2009 Dec & 2738 & 231 & $r$ & 16.7 \\
~~~~HAT-8/G342 & 2009 Sep--2009 Dec & 3174 & 235 & $r$ & 16.6 \\
~~~~FLWO~1.2\,m/KeplerCam & 2011 Jun 10 & 86 & 124 & $i$ & 1.6 \\
~~~~FLWO~1.2\,m/KeplerCam & 2011 Jun 26 & 108 & 135 & $i$ & 1.3 \\
~~~~FLWO~1.2\,m/KeplerCam & 2011 Jul 14 & 73 & 133 & $i$ & 2.3 \\
~~~~FLWO~1.2\,m/KeplerCam & 2011 Sep 20 & 117 & 124 & $i$ & 1.9 \\
~~~~FLWO~1.2\,m/KeplerCam & 2013 Sep 16 & 188 & 60 & $z$ & 3.3 \\
~~~~FLWO~1.2\,m/KeplerCam & 2013 Sep 29 & 295 & 60 & $i$ & 1.4 \\
~~~~FLWO~1.2\,m/KeplerCam & 2013 Oct 04 & 294 & 60 & $i$ & 1.3 \\
\sidehead{\textbf{\hatcur{66}}}
~~~~HAT-10/G101 & 2011 Feb--2012 Mar & 2029 & 212 & $r$ & 16.8 \\
~~~~HAT-5/G101 & 2011 Feb--2012 Apr & 1520 & 214 & $r$ & 18.1 \\
~~~~HAT-6/G101 & 2011 Feb--2012 Mar & 1178 & 214 & $r$ & 16.9 \\
~~~~HAT-7/G101 & 2011 Feb--2012 Mar & 4931 & 212 & $r$ & 14.9 \\
~~~~HAT-8/G101 & 2011 May--2012 Jun & 4157 & 212 & $r$ & 14.5 \\
~~~~HAT-9/G101 & 2011 Oct--2012 Jan & 260 & 212 & $r$ & 13.6 \\
~~~~FLWO~1.2\,m/KeplerCam & 2015 Apr 29 & 204 & 59 & $i$ & 1.9 \\
~~~~FLWO~1.2\,m/KeplerCam & 2015 Nov 26 & 273 & 60 & $z$ & 1.9 \\
~~~~FLWO~1.2\,m/KeplerCam & 2015 Dec 08 & 131 & 60 & $i$ & 1.9 \\
\enddata
\tablenotetext{a}{
    For HATNet data we list the HATNet unit and field name from which
    the observations are taken. HAT-5, -6, -7 and -10 are located at
    Fred Lawrence Whipple Observatory in Arizona. HAT-8 and -9 are
    located on the roof of the Smithsonian Astrophysical Observatory
    Submillimeter Array hangar building at Mauna Kea Observatory in
    Hawaii. Each field corresponds to one of 838 fixed pointings used
    to cover the full 4$\pi$ celestial sphere. All data from a given
    HATNet field are reduced together, while detrending through
    External Parameter Decorrelation (EPD) is done independently for
    each unique unit+field combination.
}
\tablenotetext{b}{
    The median time between consecutive images rounded to the nearest
    second. Due to factors such as weather, the day--night cycle,
    guiding and focus corrections the cadence is only approximately
    uniform over short timescales.
}
\tablenotetext{c}{
    The RMS of the residuals from the best-fit model.
}
\ifthenelse{\boolean{emulateapj}}{
    \end{deluxetable*}
}{
    \end{deluxetable}
}
%% --------------------------------------------------------------------

% =====================================================================
\subsection{Spectroscopic Observations}
\label{sec:obsspec}
% ++++++++++++++++++++++++++++++++++++++++++++++++++++++++++++++++++++++
% ++++++++++++++++++++++++++++++++++++++++++++++++++++++++++++++++++++++

Spectroscopic observations of both \hatcur{65} and \hatcur{66} were
carried out using the Tillinghast Reflector Echelle Spectrograph
\citep[TRES;][]{furesz:2008} on the 1.5\,m Tillinghast Reflector at
FLWO, and HIRES \citep{vogt:1994} on the Keck-I~10\,m at MKO. For
\hatcur{65} we also obtained observations using the FIbre-fed
\'Echelle Spectrograph (FIES) on the 2.5\,m Nordic Optical Telescope
\citep[NOT;][]{djupvik:2010} at the Observatorio del Roque de los
Muchachos on the Spanish island of La Palma. For \hatcur{66} spectroscopic observations were also collected using the SOPHIE spectrograph on the 1.93\,m telescope at the Observatoire de Haute-Provence \citep[OHP;][]{bouchy:2009} in France. The spectroscopic observations
collected for each system are summarized in
\reftabl{specobs}. Phase-folded high-precision RV and spectral line
bisector span (BS) measurements are plotted in \reffigl{rvbis}
together with our best-fit models for the RV orbital wobble of the
host stars (\refsecl{globmod}). The individual RV and BS measurements
are made available in \reftabl{rvs} at the end of the paper.

The TRES observations were reduced to spectra and cross-correlated
against synthetic stellar templates to measure the RVs and to estimate
\teffstar, \loggstar, and \vsini. Here we followed the procedure of
\cite{2010ApJ...720.1118B}, initially making use of a single order
containing the gravity and temperature-sensitive Mg~b lines. Based on
these ``reconnaissance'' observations we quickly ruled out common
false positive scenarios, such as transiting M dwarf stars, or blends
between giant stars and pairs of eclipsing dwarf stars \citep[e.g.,][]{2009ApJ...704.1107L}. For
\hatcur{65} we only obtained a single TRES observation which, in
combination with the FIES observations discussed below, rules out
these false positive scenarios. For \hatcur{66} the initial TRES RVs
showed evidence of an orbital variation consistent with a
planetary-mass companion producing the transits detected by HATNet, so
we continued collecting higher S/N observations of this system with
TRES. High precision RVs and BSs were measured from these spectra via
a multi-order analysis \citep[e.g.,][]{2014AJ....147...84B}.

The FIES spectra of \hatcur{65} were reduced in a similar manner to
the TRES data \citep{2010ApJ...720.1118B}, and were used for
reconnaissance. Two exposures were obtained using the
medium-resolution fiber, while the third was obtained with the
high-resolution fiber. One of the two medium resolution observations
had sufficiently high S/N to be used for characterizing the stellar
atmospheric parameters (\refsecl{stelparam}).

The HIRES observations of \hatcur{65} and \hatcur{66} were reduced to
relative RVs in the Solar System barycenter frame following the method
of \citet{butler:1996}, and to BSs following
\citet{2007ApJ...666L.121T}. We also measured Ca~II HK chromospheric
emission indices (the so-called $S$ and $\log_{10}R^{\prime}_{\rm HK}$
indices) following \citet{isaacson:2010} and \citet{noyes:1984}. The
I$_{2}$-free template observations of each system were also used to
determine the adopted stellar atmospheric parameters
(\refsecl{stelparam}).

The SOPHIE spectra of \hatcur{66} were collected as described in
\citet{2013AAP...558A..86B} and reduced following
\citet{santerne:2014}. One of the observations was obtained during a
planetary transit and is excluded from the analysis.

%% --------------------------------------------------------------------
%% Table summarizing spectroscopy observations
%%
\ifthenelse{\boolean{emulateapj}}{
    \begin{deluxetable*}{llrrrrr}
}{
    \begin{deluxetable}{llrrrrrrrr}
}
\tablewidth{0pc}
\tabletypesize{\scriptsize}
\tablecaption{
    Summary of spectroscopy observations
    \label{tab:specobs}
}
\tablehead{
    \multicolumn{1}{c}{Instrument}          &
    \multicolumn{1}{c}{UT Date(s)}             &
    \multicolumn{1}{c}{\# Spec.}   &
    \multicolumn{1}{c}{Res.}          &
    \multicolumn{1}{c}{S/N Range\tablenotemark{a}}           &
    \multicolumn{1}{c}{$\gamma_{\rm RV}$\tablenotemark{b}} &
    \multicolumn{1}{c}{RV Precision\tablenotemark{c}} \\
    &
    &
    &
    \multicolumn{1}{c}{($\lambda$/$\Delta \lambda$)/1000} &
    &
    \multicolumn{1}{c}{(\kms)}              &
    \multicolumn{1}{c}{(\ms)}
}
\startdata
\sidehead{\textbf{\hatcur{65}}}
NOT~2.5\,m/FIES & 2010 Aug 21--22 & 2 & 46 & 24--28 & $-48.131$ & 100 \\
FLWO~1.5\,m/TRES & 2010 Oct 27 & 1 & 44 & 16.5 & $-47.768$ & 100 \\
NOT~2.5\,m/FIES & 2011 Oct 8 & 1 & 67 & 15 & $-47.799$ & 1000 \\
Keck-I/HIRES & 2010 Dec 14 & 1 & 55 & 80 & $\cdots$ & $\cdots$ \\
Keck-I/HIRES+I$_{2}$ & 2010 Dec--2013 Aug & 12 & 55 & 64--106 & $\cdots$ & 25 \\
\sidehead{\textbf{\hatcur{66}}}
FLWO~1.5\,m/TRES & 2014 Nov--2015 Jun & 10 & 44 & 17--22 & $7.973$ & 43 \\
OHP~1.93\,m/SOPHIE & 2015 Mar--2016 Jan & 14 & 39 & 12--33 & $7.226$ & 20 \\
Keck-I/HIRES+I$_{2}$ & 2015 Dec--2016 Jan & 5 & 55 & 78--119 & $\cdots$ & 12 \\
Keck-I/HIRES & 2016 Feb 3 & 1 & 55 & 148 & $\cdots$ & $\cdots$ \\
\enddata 
\tablenotetext{a}{
    S/N per resolution element near 5180\,\AA.
}
\tablenotetext{b}{
    For high-precision RV observations included in the orbit
    determination this is the zero-point RV from the best-fit
    orbit. For other instruments it is the mean value. We do not
    provide this information for Keck-I/HIRES for which only relative
    velocities are measured.
}
\tablenotetext{c}{
    For high-precision RV observations included in the orbit
    determination this is the scatter in the RV residuals from the
    best-fit orbit (which may include astrophysical jitter), for other
    instruments this is either an estimate of the precision (not
    including jitter), or the measured standard deviation. We do not
    provide this quantity for the I$_{2}$-free templates obtained with
    Keck-I/HIRES.
}
\ifthenelse{\boolean{emulateapj}}{
    \end{deluxetable*}
}{
    \end{deluxetable}
}
%% --------------------------------------------------------------------

%
\setcounter{planetcounter}{1}
%
%% --------------------------------------------------------------------
\ifthenelse{\boolean{emulateapj}}{
    \begin{figure*} [ht]
}{
    \begin{figure}[ht]
}
{
\centering
\plottwo{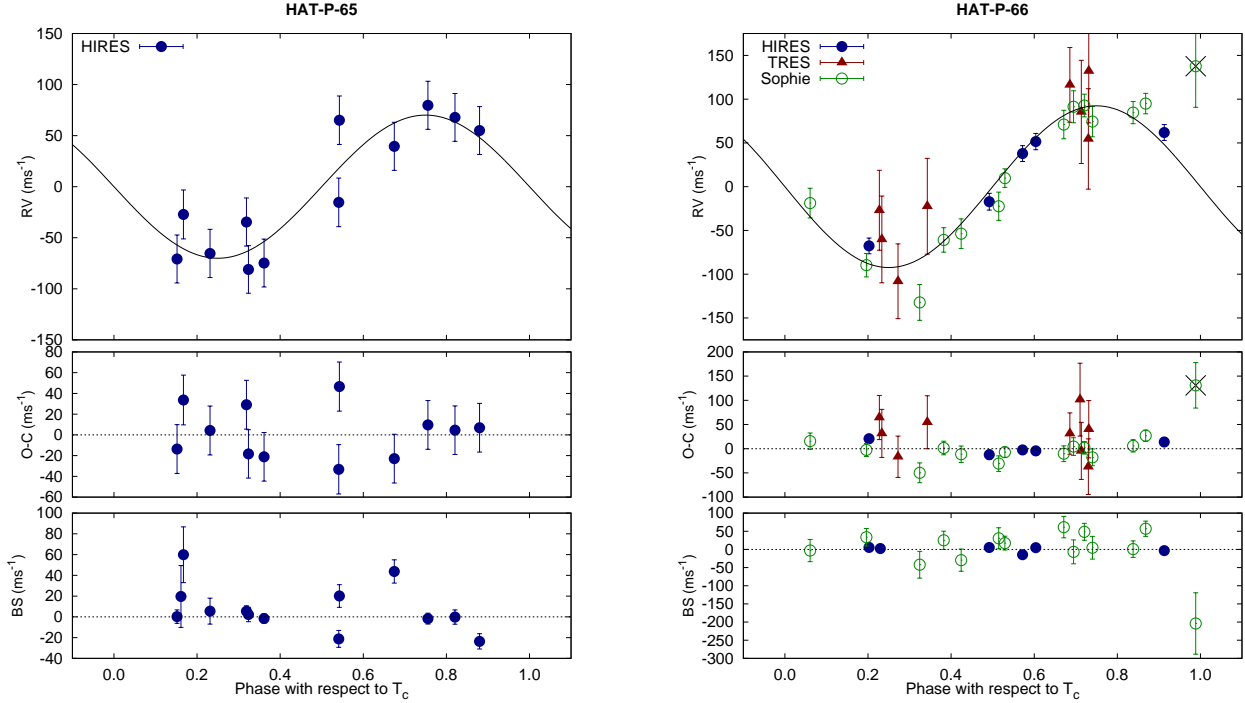}{\hatcurhtr{66}-rv.eps}
}
\caption{
    Phase-folded high-precision RV measurements for \hatcur{65} and
    \hatcur{66}. The instruments used are labelled in the plots. In
    each case we show three panels. The top panel shows the phased
    measurements together with our best-fit circular-orbit model (see
    \reftabl{planetparam}) for each system. Zero-phase corresponds to
    the time of mid-transit. The center-of-mass velocity has been
    subtracted. The second panel shows the velocity $O\!-\!C$
    residuals from the best fit. The error bars include the jitter
    terms listed in Table~\ref{tab:planetparam} added in quadrature to
    the formal errors for each instrument. The third panel shows the
    bisector spans (BS). Note the different vertical scales of the
    panels.  For \hatcur{66} the crossed-out SOPHIE measurement was
    obtained during transit and is excluded from the analysis.
}
\label{fig:rvbis}
\ifthenelse{\boolean{emulateapj}}{
    \end{figure*}
}{
    \end{figure}
}
%% --------------------------------------------------------------------

%% --------------------------------------------------------------------

% =====================================================================
\subsection{Photometric follow-up observations}
\label{sec:phot}
%++++++++++++++++++++++++++++++++++++++++++++++++++++++++++++++++++++++
%++++++++++++++++++++++++++++++++++++++++++++++++++++++++++++++++++++++

%
\setcounter{planetcounter}{1}
%
%% --------------------------------------------------------------------
\begin{figure*}[!ht]
{
\plotone{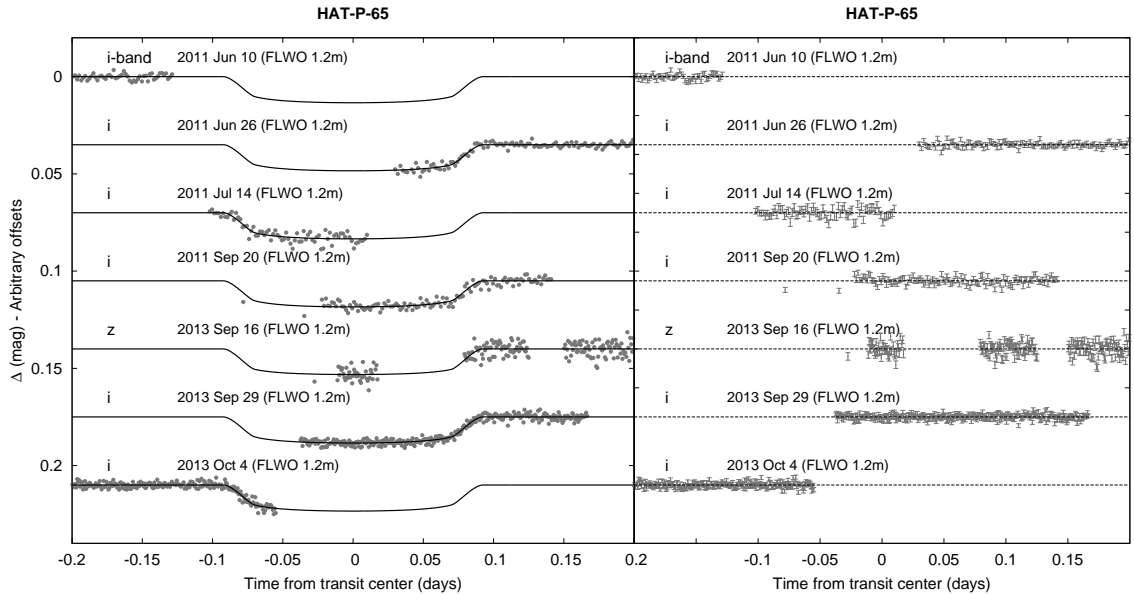}
}
\caption{
    Left: Unbinned transit \lcs{} for \hatcur{65}.  The light curves
    have been filtered of systematic trends, which were fit
    simultaneously with the transit model.
    The dates of the events, filters and instruments used are
    indicated.  Light curves following the first are displaced
    vertically for clarity.  Our best fit from the global modeling
    described in \refsecl{globmod} is shown by the solid lines. Right: The
    residuals from the best-fit model are shown in the same
    order as the original light curves.  The error bars represent the
    photon and background shot noise, plus the readout noise. 
}
\label{fig:lc1}
\end{figure*}

\begin{figure}[!ht]
{
\plotone{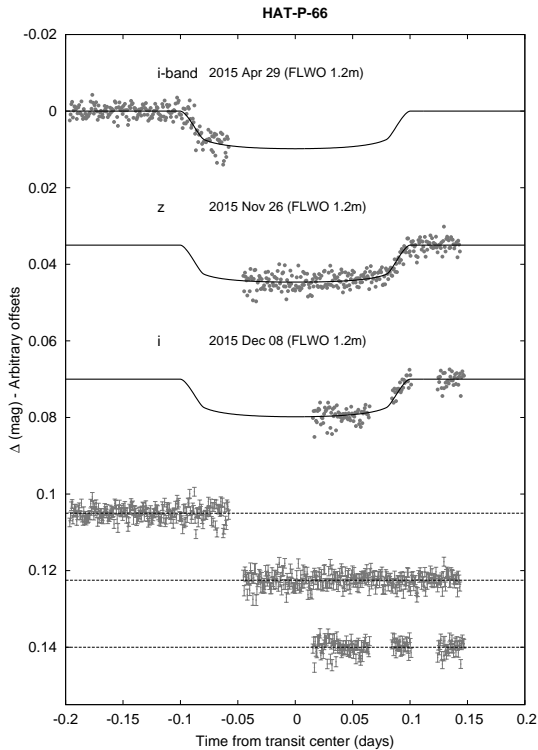}
}
\caption{
    Similar to Figure~\ref{fig:lc1}, here we show unbinned transit \lcs{} for \hatcur{66}. The residuals in this case are shown below in the same order as the original light curves.
}
\label{fig:lc2}
\end{figure}

In order to better determine the physical parameters of each TEP
system, and to aid in excluding blended stellar eclipsing binary false
positive scenarios, we conducted follow-up photometric time-series
observations of each object using KeplerCam on the 1.2\,m telescope at
FLWO. These observations are summarized in \reftabl{photobs}, where we
list the dates of the observed transit events, the number of images
collected for each event, the cadence of the observations, the filters
used, and the per-point photometric precision achieved after
trend-filtering. The images were reduced to light curves via aperture
photometry based on the FITSH package (following
\citet{2010ApJ...710.1724B}), and filtered for trends, which were fit
to the light curves simultaneously with the transit model
(Section~\ref{sec:globmod}). The resulting trend filtered light curves are
plotted together with the best-fit transit model in
Figure~\ref{fig:lc1} for \hatcur{65} and in Figure~\ref{fig:lc2} for
\hatcur{66}. The data are made available in \reftabl{phfu}.

%++++++++++++++++++++++++++++++++++++++++++++++++++++++++++++++++++++++

%
%
%% --------------------------------------------------------------------
\ifthenelse{\boolean{emulateapj}}{
    \begin{deluxetable*}{llrrrrl}
}{
    \begin{deluxetable}{llrrrrl}
}
\tablewidth{0pc}
\tablecaption{
    Light curve data for \hatcur{65} and \hatcur{66}\label{tab:phfu}.
}
\tablehead{
    \colhead{Object\tablenotemark{a}} &
    \colhead{BJD\tablenotemark{b}} & 
    \colhead{Mag\tablenotemark{c}} & 
    \colhead{\ensuremath{\sigma_{\rm Mag}}} &
    \colhead{Mag(orig)\tablenotemark{d}} & 
    \colhead{Filter} &
    \colhead{Instrument} \\
    \colhead{} &
    \colhead{\hbox{~~~~(2,400,000$+$)~~~~}} & 
    \colhead{} & 
    \colhead{} &
    \colhead{} & 
    \colhead{} &
    \colhead{}
}
\startdata
HAT-P-65 & $ 55128.75175 $ & $  -0.00921 $ & $   0.01482 $ & $ \cdots $ & $ r$ &     HATNet\\
HAT-P-65 & $ 55115.72468 $ & $  -0.00946 $ & $   0.01149 $ & $ \cdots $ & $ r$ &     HATNet\\
HAT-P-65 & $ 55120.93574 $ & $  -0.02459 $ & $   0.01224 $ & $ \cdots $ & $ r$ &     HATNet\\
HAT-P-65 & $ 55115.72489 $ & $  -0.01377 $ & $   0.01518 $ & $ \cdots $ & $ r$ &     HATNet\\
HAT-P-65 & $ 55094.88211 $ & $  -0.01227 $ & $   0.01156 $ & $ \cdots $ & $ r$ &     HATNet\\
HAT-P-65 & $ 55128.75370 $ & $  -0.00283 $ & $   0.01249 $ & $ \cdots $ & $ r$ &     HATNet\\
HAT-P-65 & $ 55154.80831 $ & $  -0.01096 $ & $   0.01202 $ & $ \cdots $ & $ r$ &     HATNet\\
HAT-P-65 & $ 55102.69977 $ & $  -0.00937 $ & $   0.01341 $ & $ \cdots $ & $ r$ &     HATNet\\
HAT-P-65 & $ 55128.75445 $ & $  -0.00558 $ & $   0.01427 $ & $ \cdots $ & $ r$ &     HATNet\\
HAT-P-65 & $ 55115.72734 $ & $   0.00413 $ & $   0.01208 $ & $ \cdots $ & $ r$ &     HATNet\\

\enddata
\tablenotetext{a}{
    Either \hatcur{65} or \hatcur{66}.
}
\tablenotetext{b}{
    Barycentric Julian Date is computed directly from the UTC time
    without correction for leap seconds.
}
\tablenotetext{c}{
    The out-of-transit level has been subtracted. For observations
    made with the HATNet instruments (identified by ``HATNet'' in the
    ``Instrument'' column) these magnitudes have been corrected for
    trends using the EPD and TFA procedures applied in signal-reconstruction mode. For
    observations made with follow-up instruments (anything other than
    ``HATNet'' in the ``Instrument'' column), the magnitudes have been
    corrected for a quadratic trend in time, for variations
    correlated with three PSF shape parameters, and with a linear basis of template light curves representing other systematic trends, which are fit simultaneously
    with the transit.
}
\tablenotetext{d}{
    Raw magnitude values without correction for the quadratic trend in
    time, or for trends correlated with the shape of the PSF. These are only
    reported for the follow-up observations.
}
\tablecomments{
    This table is available in a machine-readable form in the online
    journal.  A portion is shown here for guidance regarding its form
    and content.
}
\ifthenelse{\boolean{emulateapj}}{
    \end{deluxetable*}
}{
    \end{deluxetable}
}
%% --------------------------------------------------------------------

% =====================================================================
\subsection{Imaging Constraints on Resolved Neighbors}
\label{sec:image}
%++++++++++++++++++++++++++++++++++++++++++++++++++++++++++++++++++++++
\begin{comment}
\end{comment}
%++++++++++++++++++++++++++++++++++++++++++++++++++++++++++++++++++++++

In order to detect possible neighboring stars which may be diluting
the transit signals we obtained $J$ and $K_{S}$-band snapshot images
of both targets using the WIYN High-Resolution Infrared Camera (WHIRC)
on the WIYN~3.5\,m telescope at Kitt Peak National Observatory (KPNO)
in AZ. Observations were obtained on the nights of 2016 April 24, 27
and 28, with seeing varying between $\sim 0\farcs5$ and $\sim
1\arcsec$. Images were collected at different nod positions. These
were calibrated, background-subtracted, registered and median-combined
using the same tools that we used for reducing the KeplerCam images.

\begin{figure}[!ht]
\plotone{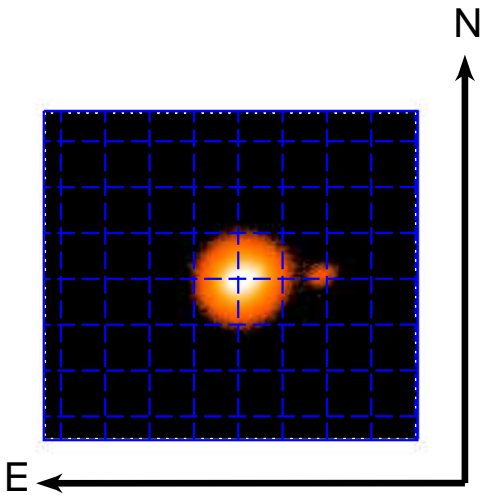}
\caption[]{ $J$-band image of \hatcur{65} from WHIRC on the
  WIYN~3.5\,m showing the $\Delta J = 4.91 \pm 0.01$\,mag neighbor
  located $3\farcs6$ to the west. The grid spacing is $2\arcsec$.
\label{fig:neighborimage}}
\end{figure}

We find that \hatcur{65} has a neighbor located $3\farcs6$ to the west
with a magnitude difference of $\Delta J = 4.91 \pm 0.01$\,mag and
$\Delta K = 4.95 \pm 0.03$\,mag relative to \hatcur{65}
(\reffigl{neighborimage}). The neighbor is too faint and distant to be
responsible for the transits detected in either the HATNet or
KeplerCam observations. The neighbor has a $J-K$ color that is the
same as \hatcur{65} to within the uncertainties, and is thus a
background star with an effective temperature that is similar to that
of \hatcur{65}, and not a physical companion. No neighbor is detected
within 10\arcsec\ of \hatcur{66}.

\reffigl{constrastcurves} shows the $J$ and $K$-band magnitude
contrast curves for \hatcur{65} and \hatcur{66} based on these
observations. These curves are calculated using the method and software
described by \citet{2016arXiv160600023E}. The bands shown in these
images represent the variation in the contrast limits depending on the
position angle of the putative neighbor.

%
%
%% ----------------
\ifthenelse{\boolean{emulateapj}}{
    \begin{figure*}[!ht]
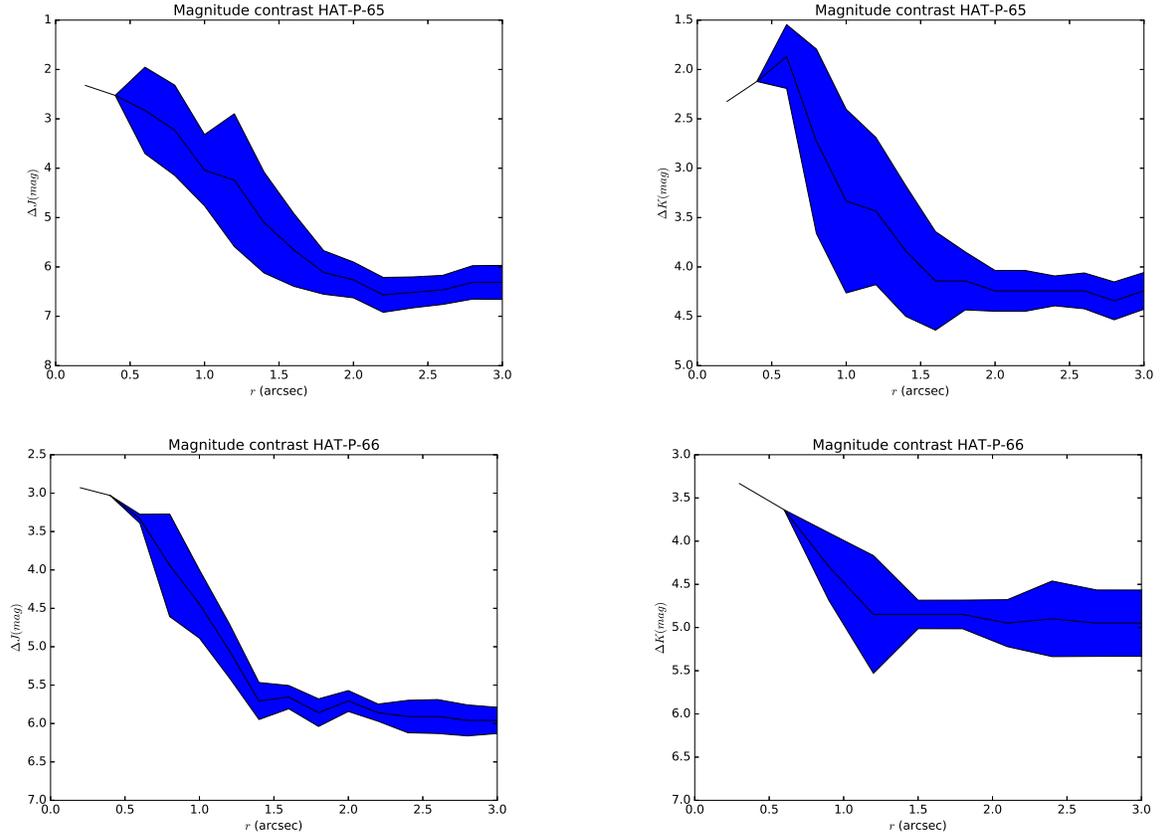

}{
    \begin{figure}[!ht]
}
\plottwo{\hatcurhtr{65}_20160424_whirc_J.eps}{\hatcurhtr{65}_20160428_whirc_K.eps}
\plottwo{\hatcurhtr{66}_20160427_whirc_J.eps}{\hatcurhtr{66}_20160427_whirc_K.eps}
\caption[]{
    Contrast curves for \hatcur{65} (top) and \hatcur{66} (bottom) in
    the $J$-band (left) and $K$-band (right) based on observations
    made with WHIRC on the WIYN~3.5\,m as described in
    Section~\ref{sec:image}. The bands show the variation in the
    contrast limits depending on the position angle of the putative
    neighbor.
\label{fig:constrastcurves}}
\ifthenelse{\boolean{emulateapj}}{
    \end{figure*}
}{
    \end{figure}
}
%% ----------------

% #####################################################################
%% Analysis
\section{Analysis}
\label{sec:analysis}
%++++++++++++++++++++++++++++++++++++++++++++++++++++++++++++++++++++++
%++++++++++++++++++++++++++++++++++++++++++++++++++++++++++++++++++++++

% =====================================================================
\subsection{Properties of the parent star}
\label{sec:stelparam}
%++++++++++++++++++++++++++++++++++++++++++++++++++++++++++++++++++++++
%++++++++++++++++++++++++++++++++++++++++++++++++++++++++++++++++++++++

High-precision atmospheric parameters, including the effective surface
temperature \teffstar, the surface gravity \loggstar, the metallicity
\feh, and the projected rotational velocity \vsini, were determined by
applying the Stellar Parameter Classification \citep[SPC;][]{buchhave:2012:spc} procedure to our high resolution spectra. For \hatcur{65} this
analysis was performed on the highest S/N FIES spectrum and on our
Keck-I/HIRES I$_{2}$-free template spectrum (we adopt the weighted
average of each parameter determined from the two spectra). For
\hatcur{66} this analysis was performed on our Keck-I/HIRES
I$_{2}$-free template spectrum. We assume a minimum uncertainty of 50\,K
on \teffstar, 0.10\,dex on \loggstar, 0.08\,dex on \feh, and 0.5\,\kms\ on
\vsini, which reflects the systematic uncertainty in the method, and
is based on applying the SPC analysis to observations of spectroscopic
standard stars.

Following \citet{2007ApJ...664.1190S} we combine the \teffstar\ and
\feh\ values measured from the spectra with the stellar densities
(\rhostar) determined from the light curves (based on the analysis in
\refsecl{globmod}) to determine the physical parameters of the
host stars (i.e., their masses, radii, surface gravities, ages,
luminosities, and broad-band absolute magnitudes) via interpolation
within the Yonsei-Yale theoretical stellar isochrones
\citep[YY;][]{yi:2001}. Figure~\ref{fig:iso} compares the model
isochrones to the measured \teffstar\ and \rhostar\ values for each
system.

For \hatcur{65} the \loggstar\ value determined from this analysis
differed by 0.19\,dex ($\sim 1.9\sigma$) from the initial value
determined through SPC. A difference of this magnitude is typical and
reflects the difficulty of accurately measuring all four atmospheric
parameters simultaneously via cross-correlation with synthetic
templates \citep[e.g.,][]{torres:2012}. We therefore carried out a second SPC
analysis of \hatcur{65} with \loggstar\ fixed based on this analysis, and
then repeated the light curve analysis and stellar parameter
determination, finding no appreciable change in \loggstar. For \hatcur{66}
the \loggstar\ value determined from the YY isochrones differed by only
$0.007$\,dex from the initial spectroscopically determined value, so
we did not carry out a second SPC iteration in this case.

The adopted stellar parameters for \hatcur{65} and \hatcur{66} are
listed in Table~\ref{tab:stellar}. We also collect in this table a variety of
photometric and kinematic properties for each system from
catalogs. Distances are determined using the listed photometry and
assuming a $R_{V} = 3.1$ \citet{cardelli:1989} extinction law.

The two stars are quite similar, with masses of
\hatcurISOmlong{65}\,\msun\ and \hatcurISOmlong{66}\,\msun\ for
\hatcur{65} and \hatcur{66}, respectively, and with respective radii
of \hatcurISOrlong{65}\,\rsun\ and \hatcurISOrlong{66}\,\rsun. The
stars are moderately evolved, with ages of \hatcurISOage{65}\,Gyr and
\hatcurISOage{66}\,Gyr (these are $84 \pm 10$\% and $83^{+9}_{-20}$\% of each star's full
lifetime, respectively). As we point out in \refsecl{discussion}, there appears to
be a general trend among the host stars of highly inflated planets in
which the largest planets are preferentially found around moderately
evolved stars. \hatcur{65} and \hatcur{66} are in line with this
trend.

%% --------------------------------------------------------------------
\ifthenelse{\boolean{emulateapj}}{
    \begin{figure*}[!ht]
}{
    \begin{figure}[!ht]
}
{
\plottwo{\hatcurhtr{65}-iso-rho.eps}{\hatcurhtr{66}-iso-rho.eps}
}
\caption{
    Model isochrones from \cite{\hatcurisocite{65}} for the measured
    metallicities of \hatcur{65} and \hatcur{66}. We show models for ages of 0.2\,Gyr and 1.0 to 14.0\,Gyr in 1.0\,Gyr increments (ages increasing from left to right). The
    adopted values of $\teffstar$ and \rhostar\ are shown together with
    their 1$\sigma$ and 2$\sigma$ confidence ellipsoids.  The initial
    values of \teffstar\ and \rhostar\ for \hatcur{65} from the first SPC and \lc\
    analyses are represented with a triangle.
}
\label{fig:iso}
\ifthenelse{\boolean{emulateapj}}{
    \end{figure*}
}{
    \end{figure}
}

%
%
%% --------------------------------------------------------------------
%% Table of stellar parameters. 
%%
\ifthenelse{\boolean{emulateapj}}{
    \begin{deluxetable*}{lccl}
}{
    \begin{deluxetable}{lccl}
}
\tablewidth{0pc}
\tabletypesize{\footnotesize}
\tablecaption{
    Stellar parameters for \hatcur{65} and \hatcur{66} \tablenotemark{a}
    \label{tab:stellar}
}
\tablehead{
    \multicolumn{1}{c}{} &
    \multicolumn{1}{c}{\bf HAT-P-65} &
    \multicolumn{1}{c}{\bf HAT-P-66} &
    \multicolumn{1}{c}{} \\
    \multicolumn{1}{c}{~~~~~~~~Parameter~~~~~~~~} &
    \multicolumn{1}{c}{Value}                     &
    \multicolumn{1}{c}{Value}                     &
    \multicolumn{1}{c}{Source}
}
\startdata
\noalign{\vskip -3pt}
\sidehead{Astrometric properties and cross-identifications}
~~~~2MASS-ID\dotfill               & \hatcurtwomassshort{65}  & \hatcurtwomassshort{66} & \\
~~~~GSC-ID\dotfill                 & \hatcurCCgsc{65}      & \hatcurCCgsc{66}     & \\
~~~~R.A. (J2000)\dotfill            & \hatcurCCra{65}       & \hatcurCCra{66}    & 2MASS\\
~~~~Dec. (J2000)\dotfill            & \hatcurCCdec{65}      & \hatcurCCdec{66}   & 2MASS\\
~~~~$\mu_{\rm R.A.}$ (\masy)              & \hatcurCCpmra{65}     & \hatcurCCpmra{66} & UCAC4\\
~~~~$\mu_{\rm Dec.}$ (\masy)              & \hatcurCCpmdec{65}    & \hatcurCCpmdec{66} & UCAC4\\
\sidehead{Spectroscopic properties}
~~~~$\teffstar$ (K)\dotfill         &  \hatcurSMEteff{65}   & \hatcurSMEteff{66} & SPC\tablenotemark{b}\\
~~~~$\loggstar$ (cgs)\dotfill         &  \hatcurSMEilogg{65}   & \hatcurSMEilogg{66} & SPC\tablenotemark{c}\\
~~~~$\feh$\dotfill                  &  \hatcurSMEzfeh{65}   & \hatcurSMEzfeh{66} & SPC               \\
~~~~$\vsini$ (\kms)\dotfill         &  \hatcurSMEvsin{65}   & \hatcurSMEvsin{66} & SPC                \\
~~~~$\vmac$ (\kms)\dotfill          &  $1.0$   & $1.0$ & Assumed              \\
~~~~$\vmic$ (\kms)\dotfill          &  $2.0$   & $2.0$ & Assumed              \\
~~~~$\gamma_{\rm RV}$ (\ms)\dotfill&  \hatcurRVgammaabs{65}  & \hatcurRVgammaabs{66} & TRES\tablenotemark{d}  \\
~~~~$S_{\rm HK}$\dotfill           & \hatcurSindex{65} & $\cdots$ & HIRES \\
~~~~$\log R^{\prime}_{\rm HK}$\dotfill           & \hatcurRHKindex{65} & $\cdots$ & HIRES \\
\sidehead{Photometric properties}
~~~~$B$ (mag)\dotfill               &  \hatcurCCtassmB{65}  & \hatcurCCtassmB{66} & APASS\tablenotemark{e} \\
~~~~$V$ (mag)\dotfill               &  \hatcurCCtassmv{65}  & \hatcurCCtassmv{66} & APASS\tablenotemark{e} \\
~~~~$I$ (mag)\dotfill               &  \hatcurCCtassmI{65}  & \hatcurCCtassmI{66} & TASS Mark IV\tablenotemark{f} \\
~~~~$g$ (mag)\dotfill               &  \hatcurCCtassmg{65}  & \hatcurCCtassmg{66} & APASS\tablenotemark{e} \\
~~~~$r$ (mag)\dotfill               &  \hatcurCCtassmr{65}  & \hatcurCCtassmr{66} & APASS\tablenotemark{e} \\
~~~~$i$ (mag)\dotfill               &  \hatcurCCtassmi{65}  & \hatcurCCtassmi{66} & APASS\tablenotemark{e} \\
~~~~$J$ (mag)\dotfill               &  \hatcurCCtwomassJmag{65} & \hatcurCCtwomassJmag{66} & 2MASS           \\
~~~~$H$ (mag)\dotfill               &  \hatcurCCtwomassHmag{65} & \hatcurCCtwomassHmag{66} & 2MASS           \\
~~~~$K_s$ (mag)\dotfill             &  \hatcurCCtwomassKmag{65} & \hatcurCCtwomassKmag{66} & 2MASS           \\
\sidehead{Derived properties}
~~~~$\mstar$ ($\msun$)\dotfill      &  \hatcurISOmlong{65}   & \hatcurISOmlong{66} & YY+$\rhostar$+SPC \tablenotemark{g}\\
~~~~$\rstar$ ($\rsun$)\dotfill      &  \hatcurISOrlong{65}   & \hatcurISOrlong{66} & YY+$\rhostar$+SPC         \\
~~~~$\loggstar$ (cgs)\dotfill       &  \hatcurISOlogg{65}    & \hatcurISOlogg{66} & YY+$\rhostar$+SPC         \\
~~~~$\rhostar$ (\gcmc)\dotfill       &  \hatcurLCrho{65}    & \hatcurLCrho{66} & Light curves         \\
~~~~$\lstar$ ($\lsun$)\dotfill      &  \hatcurISOlum{65}     & \hatcurISOlum{66} & YY+$\rhostar$+SPC         \\
~~~~$M_V$ (mag)\dotfill             &  \hatcurISOmv{65}      & \hatcurISOmv{66} & YY+$\rhostar$+SPC         \\
~~~~$M_K$ (mag,\hatcurjhkfilset{65})\dotfill &  \hatcurISOMK{65} & \hatcurISOMK{66} & YY+$\rhostar$+SPC         \\
~~~~Age (Gyr)\dotfill               &  \hatcurISOage{65}     & \hatcurISOage{66} & YY+$\rhostar$+SPC         \\
~~~~$A_{V}$ (mag)\dotfill               &  \hatcurXAv{65}     & \hatcurXAv{66} & YY+$\rhostar$+SPC         \\
~~~~Distance (pc)\dotfill           &  \hatcurXdistred{65}\phn  & \hatcurXdistred{66} & YY+$\rhostar$+SPC\\ [-1.5ex]
\enddata
\tablecomments{
For both systems the fixed-circular-orbit model has a higher Bayesian evidence than the eccentric-orbit model. We therefore assume a fixed circular orbit in generating the parameters listed here.
}
\tablenotetext{a}{
    We adopt the IAU 2015 Resolution B3 nominal values for the Solar
    and Jovian parameters \citep{prsa:2016} for all of our
    calculations, taking \rjup\ to be the nominal equatorial radius of
    Jupiter. Where necessary we assume $G = 6.6408 \times
    10^{-11}$\,m$^{3}$kg$^{-1}$s$^{-1}$. Because \citet{yi:2001} do
    not specify the assumed value for $G$ or \msun, we take the
    stellar masses from these isochrones at face value without
    conversion. Any discrepancy results in an error that is less than
    one percent, which is well below the observational uncertainty. We
    note that the standard values assumed in prior HAT planet
    discovery papers are very close to the nominal values adopted
    here. In all cases the conversion results in changes to measured
    parameters that are indetectable at the level of precision to
    which they are listed.
}
\tablenotetext{b}{
    SPC = Stellar Parameter Classification procedure for the analysis
    of high-resolution spectra \citep{buchhave:2012:spc}, applied to
    the TRES spectra of \hatcur{65} and the Keck/HIRES spectra of \hatcur{66}. These parameters
    rely primarily on SPC, but have a small dependence also on the
    iterative analysis incorporating the isochrone search and global
    modeling of the data.
}
\tablenotetext{c}{
    The spectroscopically determined value of $\loggstar$ is from our initial SPC analysis where \teffstar, \loggstar, \feh\ and \vsini\ were all varied. Systematic errors are common when all four parameters are varied.  The adopted values for \teffstar, \feh\ and \vsini\ stem from a second iteration of SPC, where \loggstar\ is fixed to the value determined through the light curve modeling and isochrone comparison. This value is listed under the ``Derived Properties'' section of the table.
}
\tablenotetext{d}{
    In addition to the uncertainty listed here, there is a $\sim
    0.1$\,\kms\ systematic uncertainty in transforming the velocities
    to the IAU standard system.
} \tablenotetext{e}{
    From APASS DR6 for as
    listed in the UCAC 4 catalog \citep{zacharias:2013:ucac4}.  
}
\tablenotetext{f}{
    \citet{droege:2006}.
}
\tablenotetext{g}{
    \hatcurisoshort{65}+\rhostar+SPC = Based on the \hatcurisoshort{65}
    isochrones \citep{\hatcurisocite{65}}, \rhostar\ as a luminosity
    indicator, and the SPC results.
}
\ifthenelse{\boolean{emulateapj}}{
    \end{deluxetable*}
}{
    \end{deluxetable}
}
%% --------------------------------------------------------------------

% =====================================================================
\subsection{Excluding blend scenarios}
\label{sec:blend}
%++++++++++++++++++++++++++++++++++++++++++++++++++++++++++++++++++++++
%++++++++++++++++++++++++++++++++++++++++++++++++++++++++++++++++++++++

In order to exclude blend scenarios we carried out an analysis
following \citet{2012AJ....144..139H}. Here we attempt to model
the available photometric data (including light curves and catalog
broad-band photometric measurements) for each object as a blend
between an eclipsing binary star system and a third star along the
line of sight (either a physical association, or a chance
alignment). The physical properties of the stars are constrained using
the Padova isochrones \citep{girardi:2002}, while we also require that
the brightest of the three stars in the blend have atmospheric
parameters consistent with those measured with SPC. We also simulate
composite cross-correlation functions (CCFs) and use them to predict
RVs and BSs for each blend scenario considered.

Based on this analysis we rule out blended stellar eclipsing binary
scenarios for both \hatcur{65} and \hatcur{66}. For \hatcur{65} we are
able to exclude blend scenarios, based solely on the photometry, with
greater than $3.7\sigma$ confidence, while for \hatcur{66} we are able
to exclude them with greater than $3.9\sigma$ confidence. For both
objects, the blend models which come closest to fitting the
photometric data (those which could not be rejected with $5\sigma$
confidence) can additionally be rejected due to the predicted large
amplitude BS and RV variations which we do not observe.

% =====================================================================
\subsection{Global modeling of the data}
\label{sec:globmod}
%++++++++++++++++++++++++++++++++++++++++++++++++++++++++++++++++++++++
%++++++++++++++++++++++++++++++++++++++++++++++++++++++++++++++++++++++

%%
In order to determine the physical parameters of the TEP systems, we
carried out a global modeling of the HATNet and KeplerCam photometry,
and the high-precision RV measurements following
\citet{2008ApJ...680.1450P,2010ApJ...710.1724B,2012AJ....144..139H}. We
use the \citet{mandel:2002} transit model to fit the light curves,
with limb darkening coefficients fixed to the values tabulated by
\citet{claret:2004} for the atmospheric parameters of the stars and
the broad-band filters used in the observations. For the KeplerCam
follow-up light curves we account for instrumental variations by using
a set of linear basis vectors in the fit. The vectors that we use
include the time of observations, the time squared, three parameters
describing the shape of the PSF, and light curves for the twenty
brightest non-variable stars in the field (TFA templates). For the TFA
templates we use the same linear coefficient (which is varied in the
fit) for all light curves collected for a given transiting planet
system through a given filter, while for the other basis vectors we
use a different coefficient for each light curve. For the HATNet light
curves we use a \citet{mandel:2002} model, and apply the fit to the
signal-reconstruction TFA data (see Section~\ref{sec:detection}). The
RV curves are modeled using a Keplerian orbit, where we allow the
zero-point for each instrument to vary independently in the fit, and
we include an RV jitter term added in quadrature to the formal
uncertainties. The jitter is treated as a free parameter which we fit
for, and is taken to be independent for each instrument.

All observations of an individual system are modeled simultaneously
using a Differential Evolution Markov Chain Monte Carlo procedure
\citep{terbraak:2006}. We visually inspect the Markov Chains and
also apply a \citet{geweke:1992} test to verify convergence and
determine the burn-in period. For both systems we consider two models:
a fixed-circular-orbit model, and an eccentric-orbit model. To
determine which model to use we estimate the Bayesian evidence ratio
from the Markov Chains following \citet{weinberg:2013}, and find that
for both systems the fixed-circular model has a greater evidence, and
therefore adopt the parameters that come from this model. The
resulting parameters for both planetary systems are listed in
\reftabl{planetparam}. We also list the 95\% confidence upper-limit on
the eccentricity for each system. 

We find that \hatcurb{65} has a mass
of \hatcurPPmlong{65}\,\mjup, a radius of \hatcurPPrlong{65}\,\rjup,
an equilibrium temperature (assuming zero albedo, and full
redistribution of heat) of \hatcurPPteff{65}\,K, and is consistent
with a circular orbit, with a 95\% confidence upper limit on the
eccentricity of $e\hatcurRVeccentwosiglimeccen{65}$. \hatcurb{66} has
similar properties, with a mass of \hatcurPPmlong{66}\,\mjup, a radius
of \hatcurPPrlong{66}\,\rjup, an equilibrium temperature (same
assumptions) of \hatcurPPteff{66}\,K, and an eccentricity of
$e\hatcurRVeccentwosiglimeccen{66}$ with 95\% confidence.

%
% ---------------------------------------------------------------------
\ifthenelse{\boolean{emulateapj}}{
    \begin{deluxetable*}{lcc}
}{
    \begin{deluxetable}{lcc}
}
\tabletypesize{\scriptsize}
\tablecaption{Orbital and planetary parameters for \hatcurb{65} and \hatcurb{66} \tablenotemark{a}\label{tab:planetparam}}
\tablehead{
    \multicolumn{1}{c}{} &
    \multicolumn{1}{c}{\bf HAT-P-65b} &
    \multicolumn{1}{c}{\bf HAT-P-66b} \\ 
    \multicolumn{1}{c}{~~~~~~~~~~~~~~~Parameter~~~~~~~~~~~~~~~} &
    \multicolumn{1}{c}{Value} &
    \multicolumn{1}{c}{Value}
}
\startdata
\noalign{\vskip -3pt}
\sidehead{\Lc{} parameters}
~~~$P$ (days)             \dotfill    & $\hatcurLCP{65}$ & $\hatcurLCP{66}$ \\
~~~$T_c$ (${\rm BJD}$)    
      \tablenotemark{b}   \dotfill    & $\hatcurLCT{65}$ & $\hatcurLCT{66}$ \\
~~~$T_{14}$ (days)
      \tablenotemark{b}   \dotfill    & $\hatcurLCdur{65}$ & $\hatcurLCdur{66}$ \\
~~~$T_{12} = T_{34}$ (days)
      \tablenotemark{b}   \dotfill    & $\hatcurLCingdur{65}$ & $\hatcurLCingdur{66}$ \\
~~~$\arstar$              \dotfill    & $\hatcurPPar{65}$ & $\hatcurPPar{66}$ \\
~~~$\zrstar$ \tablenotemark{c}             \dotfill    & $\hatcurLCzeta{65}$\phn & $\hatcurLCzeta{66}$\phn \\
~~~$\rpl/\rstar$          \dotfill    & $\hatcurLCrprstar{65}$ & $\hatcurLCrprstar{66}$ \\
~~~$b^2$                  \dotfill    & $\hatcurLCbsq{65}$ & $\hatcurLCbsq{66}$ \\
~~~$b \equiv a \cos i/\rstar$
                          \dotfill    & $\hatcurLCimp{65}$ & $\hatcurLCimp{66}$ \\
~~~$i$ (deg)              \dotfill    & $\hatcurPPi{65}$\phn & $\hatcurPPi{66}$\phn \\

\sidehead{Limb-darkening coefficients \tablenotemark{d}}
~~~$c_1,r$                  \dotfill    & $\hatcurLBir{65}$ & $\hatcurLBir{66}$ \\
~~~$c_2,r$                  \dotfill    & $\hatcurLBiir{65}$ & $\hatcurLBiir{66}$ \\
~~~$c_1,i$                  \dotfill    & $\hatcurLBii{65}$ & $\hatcurLBii{66}$ \\
~~~$c_2,i$                  \dotfill    & $\hatcurLBiii{65}$ & $\hatcurLBiii{66}$ \\
~~~$c_1,z$               \dotfill    & $\hatcurLBiz{65}$  & $\hatcurLBiz{66}$           \\
~~~$c_2,z$               \dotfill    & $\hatcurLBiiz{65}$ & $\hatcurLBiiz{66}$          \\

\sidehead{RV parameters}
~~~$K$ (\ms)              \dotfill    & $\hatcurRVK{65}$\phn\phn & $\hatcurRVK{66}$ \\
~~~$e$ \tablenotemark{e}               \dotfill    & $\hatcurRVeccentwosiglimeccen{65}$ & $\hatcurRVeccentwosiglimeccen{66}$ \\
~~~RV jitter HIRES (\ms) \tablenotemark{f}       \dotfill    & \hatcurRVjitter{65} & \hatcurRVjittertwosiglimA{66} \\
~~~RV jitter TRES (\ms)        \dotfill    & $\cdots$ & \hatcurRVjittertwosiglimB{66} \\
~~~RV jitter SOPHIE (\ms)        \dotfill    & $\cdots$ & \hatcurRVjittertwosiglimC{66} \\

\sidehead{Planetary parameters}
~~~$\mpl$ ($\mjup$)       \dotfill    & $\hatcurPPmlong{65}$ & $\hatcurPPmlong{66}$ \\
~~~$\rpl$ ($\rjup$)       \dotfill    & $\hatcurPPrlong{65}$ & $\hatcurPPrlong{66}$ \\
~~~$C(\mpl,\rpl)$
    \tablenotemark{g}     \dotfill    & $\hatcurPPmrcorr{65}$ & $\hatcurPPmrcorr{66}$ \\
~~~$\rhopl$ (\gcmc)       \dotfill    & $\hatcurPPrho{65}$ & $\hatcurPPrho{66}$ \\
~~~$\log g_p$ (cgs)       \dotfill    & $\hatcurPPlogg{65}$ & $\hatcurPPlogg{66}$ \\
~~~$a$ (AU)               \dotfill    & $\hatcurPParel{65}$ & $\hatcurPParel{66}$ \\
~~~$T_{\rm eq}$ (K)        \dotfill   & $\hatcurPPteff{65}$ & $\hatcurPPteff{66}$ \\
~~~$\Theta$ \tablenotemark{h} \dotfill & $\hatcurPPtheta{65}$ & $\hatcurPPtheta{66}$ \\
~~~$\log_{10}\langle F \rangle$ (cgs) \tablenotemark{i}
                          \dotfill    & $\hatcurPPfluxavglog{65}$ & $\hatcurPPfluxavglog{66}$ \\ [-1.5ex]
\enddata
\tablecomments{
For both systems the fixed-circular-orbit model has a higher Bayesian evidence than the eccentric-orbit model. We therefore assume a fixed circular orbit in generating the parameters listed here.
}
\tablenotetext{a}{
    We adopt the IAU 2015 Resolution B3 nominal values for the Solar
    and Jovian parameters \citep{prsa:2016} for all of our
    calculations, taking \rjup\ to be the nominal equatorial radius of
    Jupiter. Where necessary we assume $G = 6.6408 \times
    10^{-11}$\,m$^{3}$kg$^{-1}$s$^{-1}$. Because \citet{yi:2001} do
    not specify the assumed value for $G$ or \msun, we take the
    stellar masses from these isochrones at face value without
    conversion. Any discrepancy results in an error that is less than
    one percent, which is well below the observational uncertainty. We
    note that the standard values assumed in prior HAT planet
    discovery papers are very close to the nominal values adopted
    here. In all cases the conversion results in changes to measured
    parameters that are indetectable at the level of precision to
    which they are listed.
}
\tablenotetext{b}{
    Times are in Barycentric Julian Date calculated directly from UTC {\em without} correction for leap seconds.
    \ensuremath{T_c}: Reference epoch of
    mid transit that minimizes the correlation with the orbital
    period.
    \ensuremath{T_{14}}: total transit duration, time
    between first to last contact;
    \ensuremath{T_{12}=T_{34}}: ingress/egress time, time between first
    and second, or third and fourth contact.
}
\tablenotetext{c}{
   Reciprocal of the half duration of the transit used as a jump parameter in our MCMC analysis in place of $\arstar$. It is related to $\arstar$ by the expression $\zrstar = \arstar(2\pi(1+e\sin\omega))/(P\sqrt{1-b^2}\sqrt{1-e^2})$ \citep{2010ApJ...710.1724B}.
}
\tablenotetext{d}{
    Values for a quadratic law, adopted from the tabulations by
    \cite{claret:2004} according to the spectroscopic (SPC) parameters
    listed in \reftabl{stellar}.
}
\tablenotetext{e}{
    The 95\% confidence upper limit on the eccentricity determined
    when $\sqrt{e}\cos\omega$ and $\sqrt{e}\sin\omega$ are allowed to
    vary in the fit.
}
\tablenotetext{f}{
    Term added in quadrature to the formal RV uncertainties for each
    instrument. This is treated as a free parameter in the fitting
    routine. In cases where the jitter is consistent with zero we list the 95\% confidence upper limit.
}
\tablenotetext{g}{
    Correlation coefficient between the planetary mass \mpl\ and radius
    \rpl\ estimated from the posterior parameter distribution.
}
\tablenotetext{h}{
    The Safronov number is given by $\Theta = \frac{1}{2}(V_{\rm
    esc}/V_{\rm orb})^2 = (a/\rpl)(\mpl / \mstar )$
    \citep[see][]{hansen:2007}.
}
\tablenotetext{i}{
    Incoming flux per unit surface area, averaged over the orbit.
}
\ifthenelse{\boolean{emulateapj}}{
    \end{deluxetable*}
}{
    \end{deluxetable}
}

% ---------------------------------------------------------------------

%% EOF Analysis

% #####################################################################
%% Discussion
\section{Discussion}
\label{sec:discussion}
% ++++++++++++++++++++++++++++++++++++++++++++++++++++++++++++++++++++++
% ++++++++++++++++++++++++++++++++++++++++++++++++++++++++++++++++++++++

\subsection{Large Radius Planets More Commonly Found Around More Evolved Stars}

With radii of \hatcurPPr{65}\,\rjup\ and \hatcurPPr{66}\,\rjup,
\hatcurb{65} and \hatcurb{66} are among the largest hot Jupiters
known. Both planets are found around moderately evolved stars
approaching the end of their main sequence lifetimes. With an
estimated age of \hatcurISOage{65}\,Gyr, \hatcur{65} is $84 \pm 10$\% of the
way through its total lifespan, while \hatcur{66}, with an age of
\hatcurISOage{66}\,Gyr, is $83^{+9}_{-20}$\% of the way through its lifespan.
Looking at the broader sample of TEPs that have been discovered to
date, we find that the largest exoplanets are preferentially found
around moderately evolved stars. 

This effect may be a by-product of the more physically important
correlation between planet radius and equilibrium temperature (e.g.,
\citealp{fortney:2007}; \citealp{enoch:2012}; and
\citealp{spiegel:2013}), with the planet equilibrium temperature
increasing in time as its host star evolves and becomes more
luminous. We will address the question of how the correlation between
planet radius and host star fractional age which we demonstrate here
relates to the radius--equilibrium temperature correlation in
Section~\ref{sec:rteqcorr}.  While the planet radius--equilibrium
temperature correlation is well known, whether or not the radii of
planets can actually increase in time as their equilibrium
temperatures increase has not been previously established. As
discussed in Sections~\ref{sec:introduction} and
Sections~\ref{sec:theoreticalsignificance} the answer to this question
has important theoretical implications for understanding the physical
mechanism behind the inflation of close-in giant planets. To address
this question we will first attempt to determine whether or not there
is a statistically signification correlation between planetary radii
and the evolutionary status of their host stars.

%
%
%% ----------------
\ifthenelse{\boolean{emulateapj}}{
    \begin{figure*}[!ht]
}{
    \begin{figure}[!ht]
}
\plotone{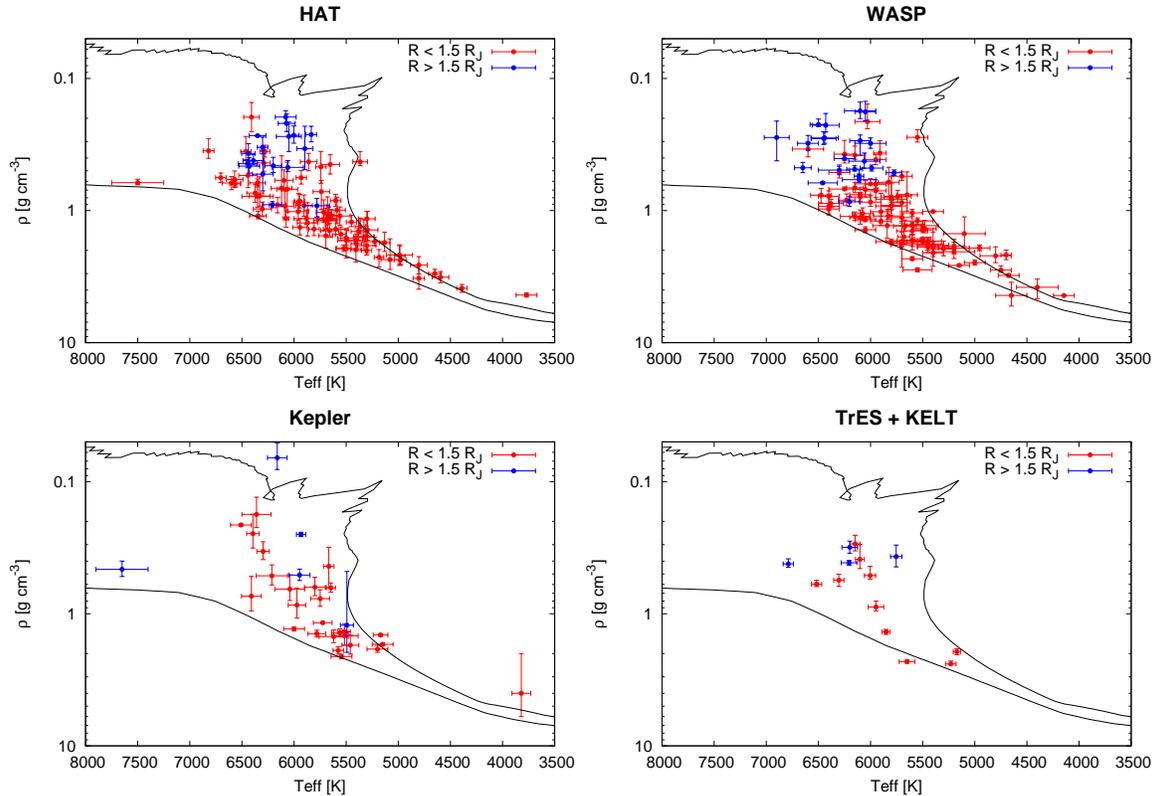}
\caption[]{
    Host stars for TEPs with $R > 0.5$\,\rjup\ and $P < 10$\,days from
    the HAT, WASP, {\em Kepler}, TrES and KELT surveys. The lower line
    in each panel is the 200\,Myr solar-metallicity isochrone from the
    YY stellar evolution models, while the upper line is the locus of
    points for stars having an age that is the lesser of 13.7\,Gyr or
    90\% of their total lifetime, again assuming solar metallicity and
    using the YY models. Note that the maximum stellar age is a smooth
    function of stellar mass according to the models
    (Figure~\ref{fig:maxagevsmass}), but the 90\% lifetime locus in
    the \teffstar--$\rho$ plane is jagged due to the sensitive dependence
    on mass of the late stages of stellar evolution. We distinguish
    here between stars with planets having $R_{P} > 1.5$\,\rjup\ and stars
    with planets having $R_{P} < 1.5$\,\rjup. Large planets have been
    preferentially discovered around more evolved stars than smaller
    planets. This appears to be true for all of the surveys
    considered. Moreover, few, if any, ZAMS stars are known to host
    planets with $R_{P} > 1.5$\,\rjup.
\label{fig:teffrhobyrp}}
\ifthenelse{\boolean{emulateapj}}{
    \end{figure*}
}{
    \end{figure}
}
%% ----------------

Figure~\ref{fig:teffrhobyrp} shows TEP host stars on a
\teffstar--\rhostar\ diagram. These two parameters are directly
measured for TEP systems, and together with the \feh\ of the star, are
the primary parameters used to characterize the stellar hosts. Here we
limit the sample to systems with planets having $\rpl >
0.5$\,\rjup\ and $P < 10$\,days. Because observational selection
effects vary from survey to survey, we show separately the systems
discovered by HAT (both HATNet and HATSouth), WASP, {\em Kepler}, TrES
and KELT, which are the surveys that have discovered
well-characterized planets with $\rpl > 1.5$\,\rjup. The data for the
HAT, WASP, TrES and KELT systems are drawn from a database of TEPs
which we privately maintain, and are listed, together with references,
in Table~\ref{tab:tepparameters} at the end of this paper. These are
planets which have been announced on the arXiv pre-print server as of
2016 June 2, and supplemented by some additional fully confirmed
planets from HAT which had not been announced by that
date. For {\em Kepler} we take the data from the NASA Exoplanet
archive\footnote{\url{http://exoplanetarchive.ipac.caltech.edu},
  accessed 2016 Mar 4}. In Figure~\ref{fig:teffrhobyrp} we distinguish
between hosts with planets having $\rpl > 1.5$\,\rjup, and hosts with
planets having $\rpl < 1.5$\,\rjup. The lower bound in each panel
shows the solar metallicity, 200\,Myr ZAMS isochrone from the YY
models, while the upper bound shows the locus of points for stars having
an age that is the lesser of 13.7\,Gyr or 90\% of their total
lifetime, again assuming solar metallicity and using the YY
models. For all of the surveys considered, planets with $R >
1.5$\,\rjup\ tend to be found around host stars that are more evolved
(closer to the 90\% lifetime locus) than planets with $R <
1.5$\,\rjup. Moreover, very few highly inflated planets have been
discovered around stars close to the ZAMS. The largest planets also
tend to be found around hot/massive stars, and have the highest level
of irradiation.

%
%
%% ----------------
\ifthenelse{\boolean{emulateapj}}{
    \begin{figure*}[!ht]
}{
    \begin{figure}[!ht]
}
\plotone{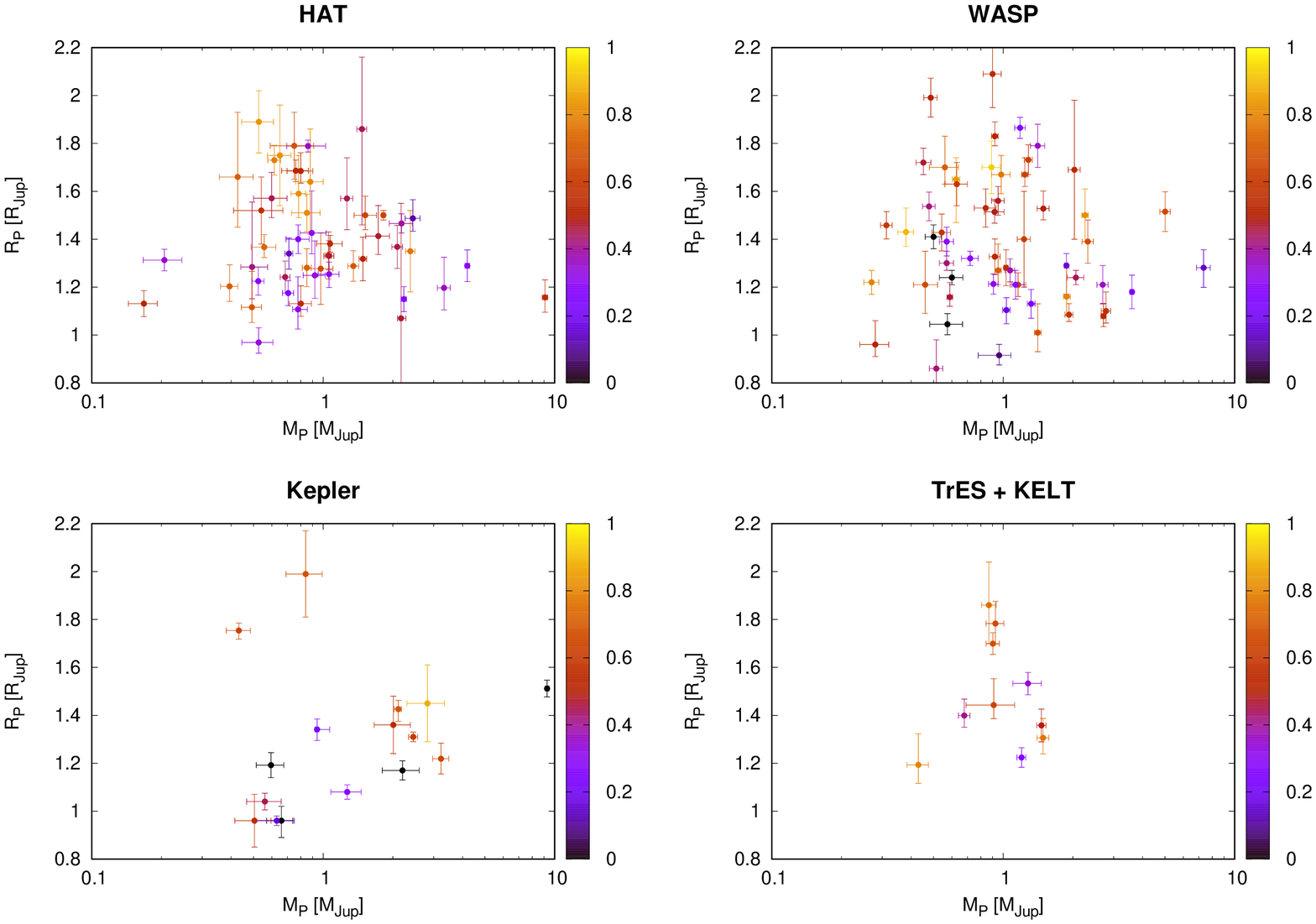}
\caption[]{
    Mass--radius relation for TEPs from HAT (top left), WASP (top
    right), {\em Kepler} (bottom left) and TrES and KELT (bottom
    right) with $R > 0.5$\,\rjup\ and $P < 10$\,days around stars with
    total lifetimes $t_{\rm tot} < 10$\,Gyr. The color-scale for each
    point indicates the fractional age of the system (taken to be
    $\tau = (t - 200\,{\rm Myr})/(t_{\rm tot} - 200\,{\rm Myr})$,
    where $t$ is the age determined from the YY isochrones using
    $\teffstar$, $\rhostar$ and $\feh$ and $t_{\rm tot}$ is the
    maximum age in the YY models for a star with the same mass and
    \feh). A handful of stars with bulk densities indicating very
    young ages show up as black points in the figure. The largest
    planets are found almost exclusively around moderately evolved
    ($\tau \ga 0.5$) stars.
\label{fig:massradage}}
\ifthenelse{\boolean{emulateapj}}{
    \end{figure*}
}{
    \end{figure}
}
%% ----------------

\begin{figure}[!ht]
\plotone{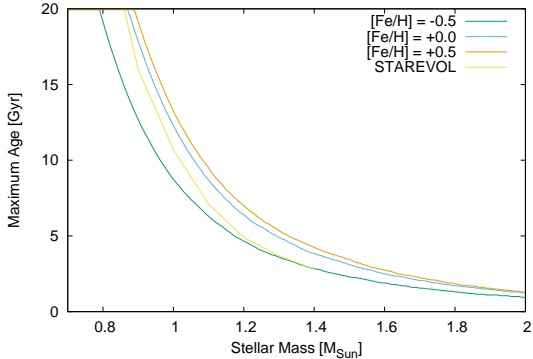}
\caption[]{
    The maximum age of a star as a function of its mass based on
    interpolating within the YY isochrones. These are shown for three
    representative metallicities.  The maximum age is artificially
    capped at 19.95\,Gyr which is the largest age at which the models
    are tabulated. For stars with $M \ga 0.85$\,\msun, which have
    maximum ages below this artificial cap, there is a smooth
    power-law dependence between the maximum age and mass. We use this
    relation to estimate the fractional age $\tau$ of a planetary
    system. For comparison we also show the terminal main sequence age
    as a function of stellar mass from the STAREVOL evolution tracks
    \citep{charbonnel:2004,lagarde:2012}, taken from Table B.6 of
    \citet{santerne:2016}.
\label{fig:maxagevsmass}}
\end{figure}

%% ----------------
\ifthenelse{\boolean{emulateapj}}{
    \begin{figure*}[!ht]
}{
    \begin{figure}[!ht]
}
\plotone{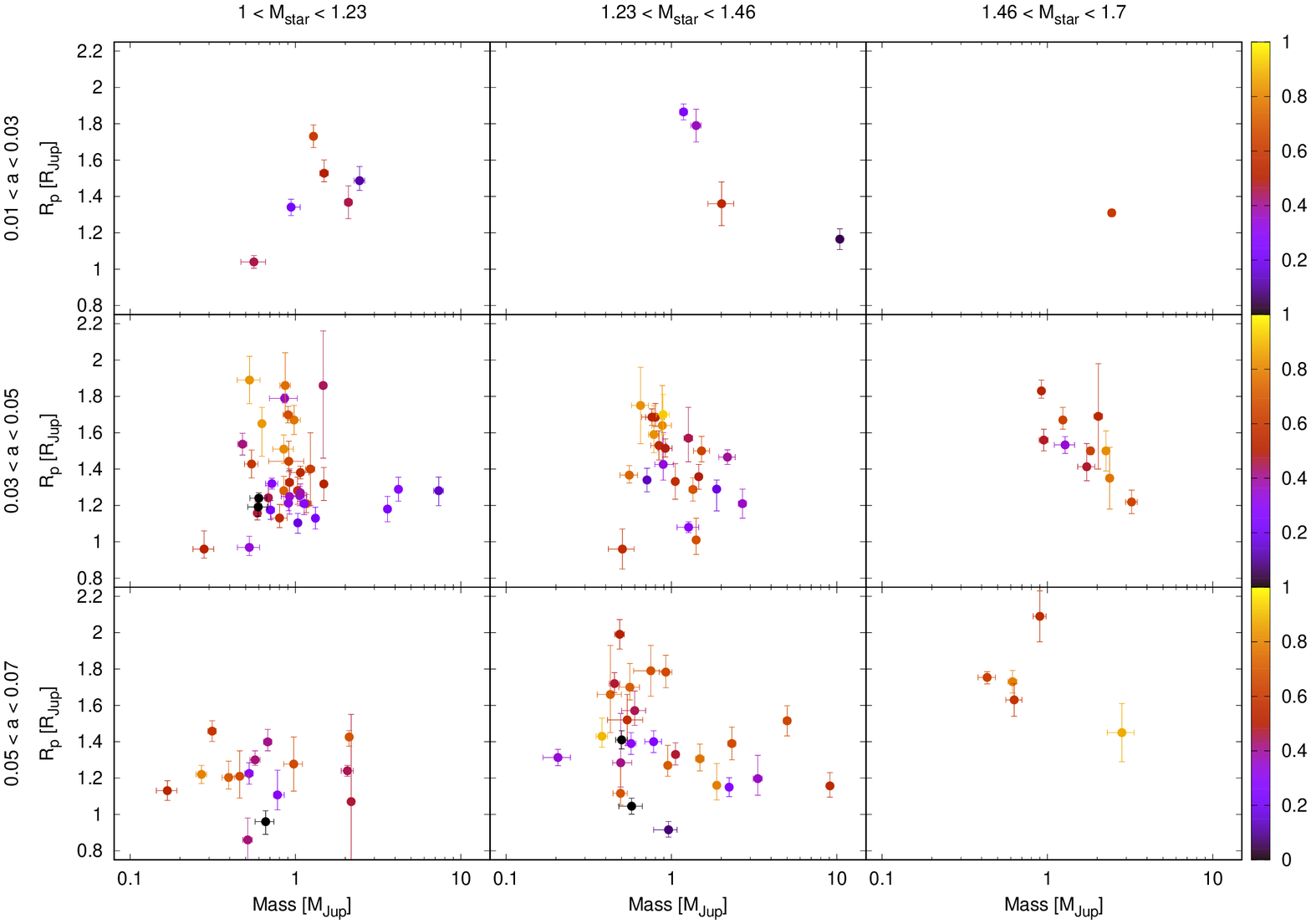}
\caption[]{
    Similar to Figure~\ref{fig:massradage}, here we combine all of the
    data from the different surveys, and show the mass--radius
    relation for different host star mass ranges (the selections are
    shown at the top of each column in solar mass units) and orbital
    semi-major axes (the selections are shown to the left of each row
    in AU). The overall range of semi-major axis and stellar mass
    shown here is chosen to encompass the sample of well-characterized highly
    inflated planets with $R > 1.5$\,\rjup\ around stars with total
    lifetimes $t_{\rm tot} < 10$\,Gyr. Within a given panel the
    largest planets tend to be found around more evolved stars. This
    is the opposite of what one would expect if high irradiation slows
    a planet's contraction, but does not supply energy deep enough into
    the interior of the planet to re-inflate as the luminosity of its host star increases.
\label{fig:massradagegrid}}
\ifthenelse{\boolean{emulateapj}}{
    \end{figure*}
}{
    \end{figure}
}
%% ----------------

For another view of the data, in Figure~\ref{fig:massradage} we plot
the mass--radius relation of close-in TEPs with the color-scale of
each point showing the fractional isochrone-based age of the system
(taken to be equal to $\tau = (t - 200\,{\rm Myr})/(t_{\rm tot} -
200\,{\rm Myr})$). Here $t_{\rm tot}$ for a system is the maximum age
of a star with a given mass and metallicity according to the YY models
(Figure~\ref{fig:maxagevsmass}). We show the fractional age, rather
than the age in Gyr, as the stellar lifetime is a strong function of
stellar mass, and the largest planets also tend to be found around
more massive stars with shorter total lifetimes. Because the star
formation rate in the Galaxy has been approximately constant over the
past $\sim 8$\,Gyr \citep[e.g.,][]{snaith:2015}, for a star of a given
mass we expect $\tau$ to be uniformly distributed between 0 and 1. In
order to perform a consistent analysis, we re-compute ages for all of
the WASP, {\em Kepler}, TrES and KELT systems using the YY models
together with the spectroscopically measured $\teffstar$ and \feh, and
transit-inferred stellar densities listed in
Table~\ref{tab:tepparameters}. In Figure~\ref{fig:massradage} we focus
on systems with $P < 10$\,days and $t_{\rm tot} < 10$\,Gyr. Again it
is apparent that the largest radius planets tend to be around stars
that are relatively old. Note that due to the finite age of the
Galaxy, there has been insufficient time for stars with $t_{\rm tot} >
10$\,Gyr to reach their main sequence lifetimes. The restriction on
$t_{\rm tot}$, which is effectively a cut on host star mass, limits
the sample to stars which could be discovered at any stage in their
evolution. If we do not apply this cut then the apparent correlation
between fractional age and planet radius becomes even more
significant, but this is likely due to observational bias.

The planets shown in Figure~\ref{fig:massradage} have a variety of
orbital separations and host star masses. Because the evolution of a
planet depends on its stellar environment, we expect there to be a
variance in the planet radius at fixed planet mass. In order to better
compare planets likely to have similar histories (but which have
different ages, and thus are at different stages in their history), in
Figure~\ref{fig:massradagegrid} we re-plot the mass--radius relation,
but this time binning by host star mass and orbital semi-major
axis. Note that in comparing planets with the same semi-major axis we
are assuming that orbital evolution can be neglected. Again we use the
color-scale of points to denote the fractional age of the system. We
choose a $3\times 3$ binning to allow a sufficient number of planets
in at least some of the bins to be able to detect a statistical
trend. Unfortunately because we limited by the small sample of
planets, we are forced to use relatively large bins, so there is
likely to still be significant variation in the evolution of different
planets within a bin. Bearing this caveat in mind, we note that the
same trend of larger radius planets, at a given planet mass, being
found around more evolved stars is seen when comparing only planets
with similar host star masses and at similar orbital separations. If
anything the gradient in fractional age with planet radius is more
pronounced in Figure~\ref{fig:massradagegrid} than it is in
Figure~\ref{fig:massradage} (see especially the center row and bottom,
center panel of Figure~\ref{fig:massradagegrid}). If enhanced
irradiation acts to slow a planet's contraction, but does not
re-inflate the planet, then we would expect to see the opposite trend
in Figure~\ref{fig:massradagegrid}. Namely, a planet of a given mass
at a given orbital separation around a star of a given mass should
decrease, or remain constant, in size over time, despite the
increasing irradiation as its host star evolves. This is not what we
see. In Section~\ref{sec:rteqcorr} we follow a more statistically
rigorous method to show that planetary radii increase in time with
increasing irradiation, rather than being set by the initial
irradiation.

%
%
%% ----------------
\ifthenelse{\boolean{emulateapj}}{
    \begin{figure*}[!ht]
}{
    \begin{figure}[!ht]
}
\plotone{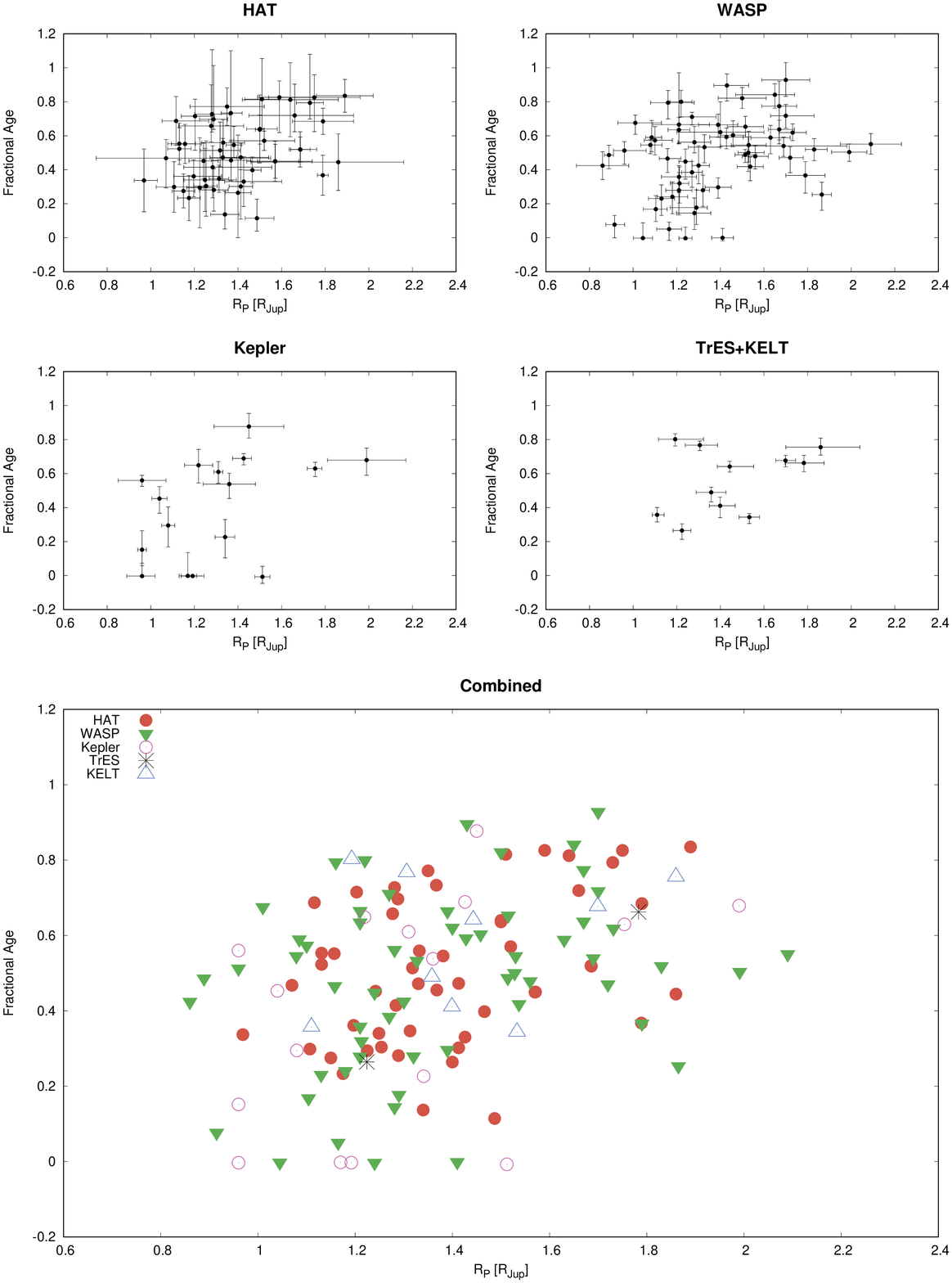}
\caption[]{
   {\em Top:} The fractional isochrone-based age of the system (see
   Figure~\ref{fig:massradage}) vs.\ the planetary radius, shown
   separately for TEP systems discovered by HAT (left) and WASP
   (right). We only show systems with $P < 10$\,days and $t_{\rm tot}
   < 10$\,Gyr. Both the HAT and WASP samples have positive
   correlations between $R_{P}$ and $\tau$. For HAT a Spearman
   non-parametric rank-order correlation test gives a correlation
   coefficient of $\SpearmanrhoHAT$ with a \SpearmanprobHAT\% false alarm probability. For
   the WASP sample we find a correlation coefficient of $\SpearmanrhoWASP$ and a
   false alarm probability of \SpearmanprobWASP\%. 
   {\em Middle:} Same as the top, here we show planets from {\em Kepler}, TrES and KELT, with the same selections applied. 
   {\em Bottom:} Same as top, here we combine data from all of the surveys. The combined data set has a
   correlation coefficient of $\Spearmanrhocomb$ and a false alarm probability
   \Spearmanprobcomb\%. 
\label{fig:fracagerp}}
\ifthenelse{\boolean{emulateapj}}{
    \end{figure*}
}{
    \end{figure}
}
%% ----------------

To establish the statistical significance of the trends seen in
Figures~\ref{fig:teffrhobyrp}--\ref{fig:massradagegrid}, in
Figure~\ref{fig:fracagerp} we plot the fractional isochrone-based age
$\tau$ against planetary radius, restricted to systems with $t_{\rm
  tot} < 10$\,Gyr. Both the HAT and WASP data have positive
correlations between $R_{P}$ and the fractional age. A Spearman
non-parametric rank-order correlation test gives a correlation
coefficient of $\SpearmanrhoHAT$ between $R_{P}$ and the fractional
age for HAT, with a \SpearmanprobHAT\% false alarm probability. For
the WASP sample we find a correlation coefficient of
$\SpearmanrhoWASP$ and a false alarm probability of
\SpearmanprobWASP\%. The {\em Kepler}, TrES and KELT datasets are too
small to perform a robust test for correlation, but they each show a
similar trend. When all of the data are combined, we find a
correlation coefficient of $\Spearmanrhocomb$ and a false alarm
probability of only \Spearmanprobcomb\%. While the correlation is
relatively weak, explaining only a modest amount of the overall
scatter in the data, it has a high statistical significance, and is
extremely unlikely to be due to random chance.

%% ----------------
\ifthenelse{\boolean{emulateapj}}{
    \begin{figure*}[!ht]
}{
    \begin{figure}[!ht]
}
\plotone{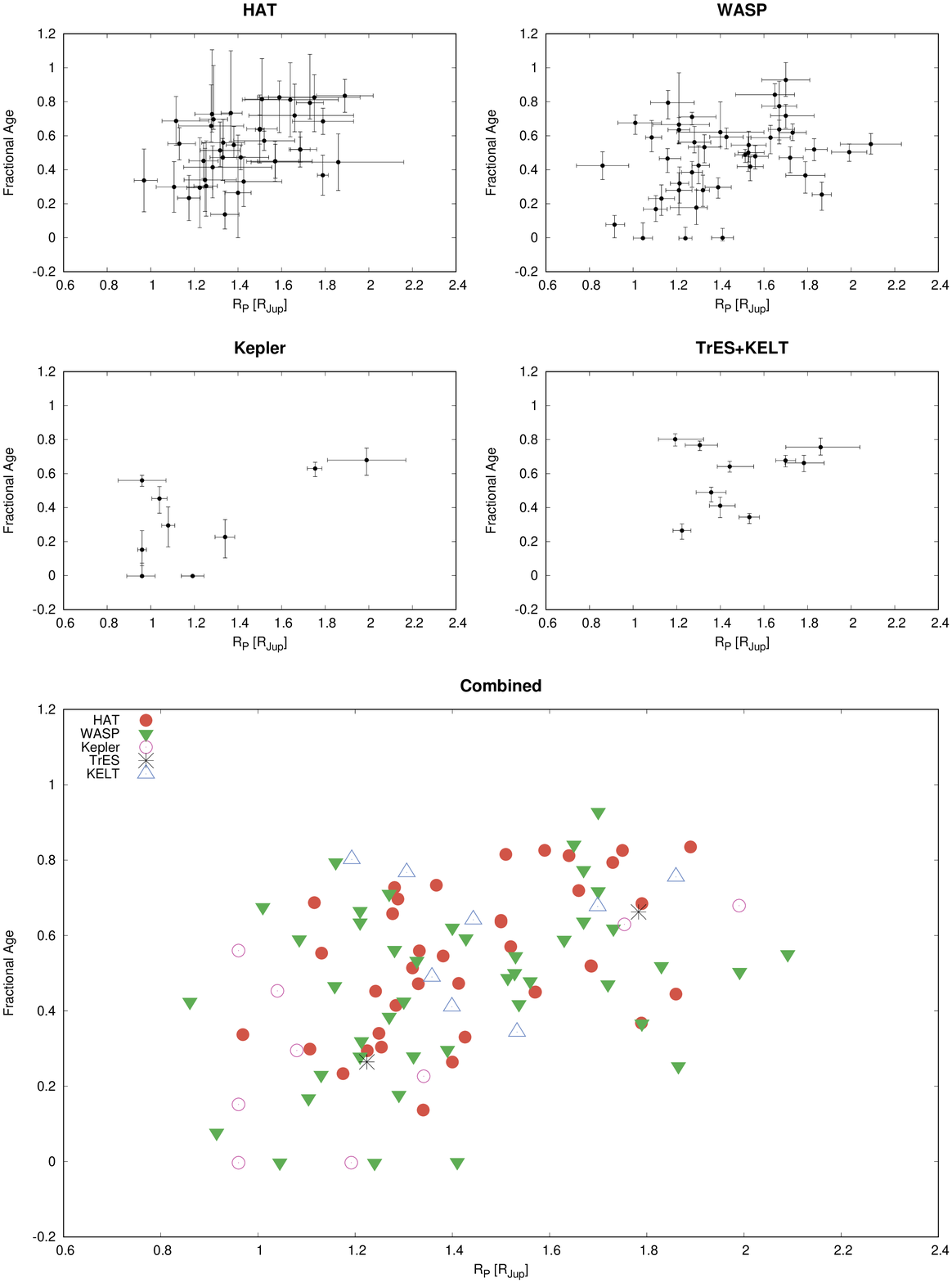}
\caption[]{
   Same as Figure~\ref{fig:fracagerp}, here we only consider systems
   with planets having masses in the range $0.4\,\mjup < \mpl <
   2.0\,\mjup$, which is roughly the radius range over which highly
   inflated planets have been discovered. In this case the Spearman
   non-parametric rank-order correlation test gives a correlation
   coefficient of $\SpearmanMprhoHAT$ with a \SpearmanMpprobHAT\%
   false alarm probability. For the WASP sample we find a correlation
   coefficient of $\SpearmanMprhoWASP$ and a false alarm probability
   of \SpearmanMpprobWASP\%. The combined sample has a correlation
   coefficient of $\SpearmanMprhocomb$ and a false alarm probability
   of $\SpearmanMpprobcomb$\%. 
\label{fig:fracagerpmp}}
\ifthenelse{\boolean{emulateapj}}{
    \end{figure*}
}{
    \end{figure}
}
%% ----------------

Figure~\ref{fig:fracagerpmp} is similar to Figure~\ref{fig:fracagerp},
except that here we restrict the analysis to planets with $0.4\,\mjup
< \mpl < 2.0\,\mjup$, which is roughly the range over which the most
highly inflated planets have been discovered (e.g.,
\reffigl{massradage}). In this case we still find a statistically
significant difference between the fractional ages of stars hosting
large radius planets and those hosting small radius planets, though,
due to the smaller sample size, the overall significance is somewhat
reduced compared to the sample when no restriction is placed on planet
mass (the correlation coefficient itself is somewhat
higher). Quantitatively we find that the HAT sample has a Spearman
correlation coefficient of $\SpearmanMprhoHAT$ and a false alarm
probability of \SpearmanMpprobHAT\%, the WASP sample has a correlation
coefficient of $\SpearmanMprhoWASP$ and a false alarm probability of
\SpearmanMpprobWASP\%, and the combined sample has a correlation
coefficient of $\SpearmanMprhocomb$ and a false alarm probability of
\SpearmanMpprobcomb\%.

%% ----------------
\ifthenelse{\boolean{emulateapj}}{
    \begin{figure*}[!ht]
}{
    \begin{figure}[!ht]
}
\plotone{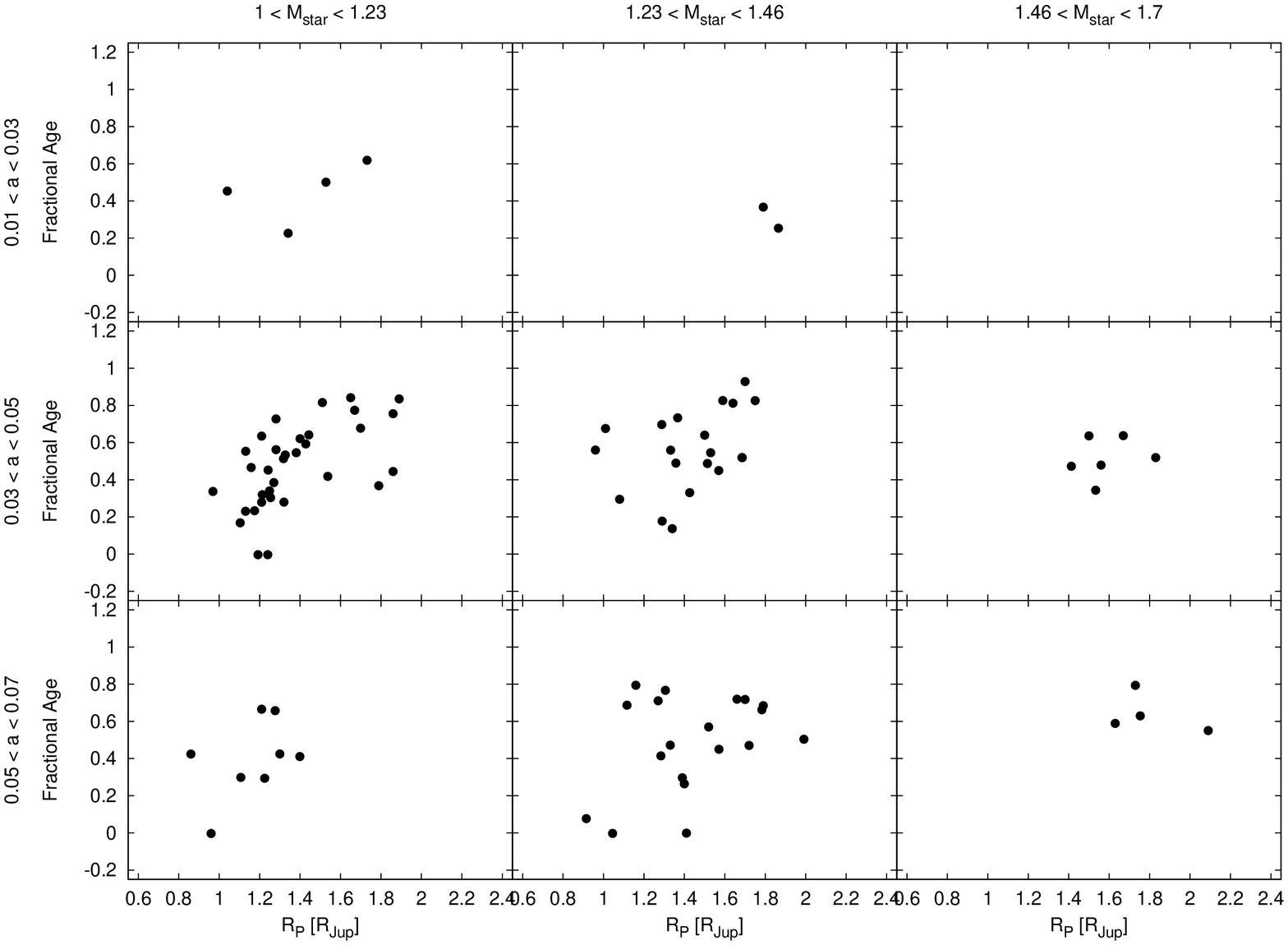}
\caption[]{
    Similar to Figures~\ref{fig:fracagerpmp}
    and~\ref{fig:massradagegrid}, here we combine all of the data from
    the different surveys, and show the fractional isochrone-based age
    vs.\ planet radius for different host star mass ranges (the
    selections are shown at the top of each column in solar mass
    units) and orbital semi-major axes (the selections are shown to
    the left of each row in AU). The overall range of semi-major axis
    and stellar mass shown here is chosen to encompass the sample of
    well-characterized highly inflated planets with $R >
    1.5$\,\rjup\ around stars with total lifetimes $t_{\rm tot} <
    10$\,Gyr. We also restrict the sample to planets with $0.4\,\mjup
    < M_{P} < 2.0\,\mjup$.
\label{fig:fracagerpmpgrid}}
\ifthenelse{\boolean{emulateapj}}{
    \end{figure*}
}{
    \end{figure}
}
%% ----------------

In order to compare planets with similar evolutionary histories, and
in analogy to Figure~\ref{fig:massradagegrid}, in
Figure~\ref{fig:fracagerpmpgrid} we plot the fractional age against
planet radius gridded by host star mass and orbital semimajor
axis. Here we combine all of the data, but restrict the sample to only
planets with $0.4\,\mjup < M_{P} < 2.0\,\mjup$ around stars with
$t_{tot} < 10$\,Gyr. We see the correlation again in several grid
cells, so long as there is a sufficiently large sample.

Of course these correlations are likely biased due to observational
selection effects. We estimate the effect of observational selections
on the measured correlation below in
Section~\ref{sec:selectioneffects}, where we conclude that the
correlation is reduced, but still significant, after accounting for
selections.

We conclude that there is a statistically significant
positive correlation between $\rpl$ and the fractional age of the
system.
This correlation is seen in samples of transiting planets
found by multiple surveys, with strikingly similar results found for
the largest two samples (from WASP and HAT). The largest radius
planets have generally been found around more evolved stars.

\subsubsection{Relation to the Correlation Between Radius and Equilibrium Temperature}
\label{sec:rteqcorr}

It is important to note that the correlation between radius and equilibrium temperature
(or flux) is much stronger than the apparent
correlation between planet radius and the fractional age of the host
star. In fact the data are consistent with the latter correlation
being entirely a by-product of the former correlation. However, the
data also indicate that the radii of planets dynamically increase in time as
their host stars become more luminous and the planetary
equilibrium temperatures increase.

\begin{deluxetable*}{rrrrr}
\tablewidth{0pc}
\tabletypesize{\scriptsize}
\tablecaption{
    Parameter Estimates and Bayesian Evidence For Models of the Form Eq.~\ref{eqn:rplinmodel}
    \label{tab:bayesevmodel}
}
\tablehead{
    \multicolumn{1}{c}{$\ln T_{\rm eq,now}$} &
    \multicolumn{1}{c}{$\ln T_{\rm eq,ZAMS}$} &
    \multicolumn{1}{c}{$\ln a$} &
    \multicolumn{1}{c}{$\tau$} &
    \multicolumn{1}{c}{Bayesian} \\
    \multicolumn{1}{c}{$c_{1}$} &
    \multicolumn{1}{c}{$c_{2}$} &
    \multicolumn{1}{c}{$c_{3}$} &
    \multicolumn{1}{c}{$c_{4}$} &
    \multicolumn{1}{c}{Evidence}
}
\startdata
$0.92 \pm 0.12$ & 0 & $0.273 \pm 0.075$ & 0 & $2.66 \times 10^{9}$ \\
$0.81 \pm 0.14$ & 0 & $0.216 \pm 0.086$ & $0.107 \pm 0.088$ & $9.99 \times 10^{8}$ \\
$0.96 \pm 0.20$ & $-0.07 \pm 0.21$ & $0.263 \pm 0.079$ & 0 & $5.01 \times 10^{8}$ \\
$0.558 \pm 0.099$ & 0 & 0 & $0.223 \pm 0.077$ & $2.53 \times 10^{8}$ \\
$0.77 \pm 0.25$ & $0.03 \pm 0.23$ & $0.216 \pm 0.086$ & $0.111 \pm 0.096$ & $2.09 \times 10^{8}$ \\
$0.57 \pm 0.25$ & $-0.01 \pm 0.24$ & 0 & $0.217 \pm 0.088$ & $4.75 \times 10^{7}$ \\
$0.658 \pm 0.099$ & 0 & 0 & 0 & $2.93 \times 10^{7}$ \\
0 & $0.492 \pm 0.098$ & 0 & $0.332 \pm 0.077$ & $1.73 \times 10^{7}$ \\
$0.90 \pm 0.22$ & $-0.29 \pm 0.21$ & 0 & 0 & $1.17 \times 10^{7}$ \\
0 & $0.62 \pm 0.14$ & $0.123 \pm 0.084$ & $0.291 \pm 0.080$ & $9.95 \times 10^{6}$ \\
0 & $0.76 \pm 0.14$ & $0.224 \pm 0.086$ & 0 & $7.84 \times 10^{4}$ \\
0 & $0.53 \pm 0.11$ & 0 & 0 & $1.44 \times 10^{4}$ \\
0 & 0 & $-0.153 \pm 0.067$ & $0.392 \pm 0.086$ & $2.37 \times 10^{3}$ \\
0 & 0 & 0 & $0.364 \pm 0.087$ & $9.04 \times 10^{2}$ \\
0 & 0 & $-0.094 \pm 0.074$ & 0 & $4.87 \times 10^{-1}$ \\
\enddata
\tablecomments{
    The models tested are sorted from highest to lowest evidence. The Bayesian evidence is reported relative to that for a model with only a freely varying mean for the $\ln \rpl$ values.  For each parameter we report the mean and standard deviation of its posterior probability distribution.
}
\end{deluxetable*}
%% --------------------------------------------------------------------

To demonstrate this we perform a Bayesian linear regression model comparison
using the {\bf BayesFactor} package in R\footnote{\url{http://bayesfactorpcl.r-forge.r-project.org/}} which follows the approach of \citet{liang:2008} and \citet{rouder:2012}. We test models of the form:
\begin{equation}
\label{eqn:rplinmodel}
\ln R_{p} = c_{0} + c_{1}\ln T_{\rm eq,now} + c_{2}\ln T_{\rm eq,ZAMS} + c_{3}\ln a + c_{4}\tau
\end{equation}
where $c_{0}$--$c_{4}$ are varied linear parameters, and we compare
all combinations of models where parameters other than $c_{0}$ are
fixed to $0$. This particular parameterization is motivated by
\citet{enoch:2012} who found that the radii of close-in Jupiter-mass
planets are best modelled by a function of the form given above with
$c_{2} \equiv c_{4} \equiv 0$, and we now include the $T_{\rm
  eq,ZAMS}$ and $\tau$ parameters to test whether age is an important
additional variable, and/or whether the data could be equally well
described if we used the initial equilibrium temperature of the planet
(which does not change in time) rather than the present-day
equilibrium temperature (which increases in time due to the evolution of the
host). Here we consider the full sample of well-characterized planets
with $P < 10$\,days, $\rpl > 0.5$\,\rjup, $0.4\,\mjup < \mpl <
2.0\,\mjup$, and $t_{\rm tot} < 10$\,Gyr. Table~\ref{tab:bayesevmodel}
lists the linear coefficient estimates and the Bayesian evidences,
sorting from highest to lowest, for each of the 15 models under
comparison.

We find that the model with the highest Bayesian evidence has the form:
\begin{equation}
\ln R_{p} = c_{0} + c_{1}\ln T_{\rm eq,now} + c_{3}\ln a
\end{equation}
with an evidence that is $2.7 \times 10^9$ times higher than the
evidence for a model where $c_{1} \equiv c_{2} \equiv c_{3} \equiv
c_{4} \equiv 0$. This finding is consistent with that of
\citet{enoch:2012}. The next highest evidence model has the form:
\begin{equation}
\ln R_{p} = c_{0} + c_{1}\ln T_{\rm eq,now} + c_{3}\ln a + c_{4}\tau
\end{equation}
with an evidence that $0.37$ times that of the highest evidence
model. So indeed including $\tau$ provides no additional explanatory
power beyond what is already provided by $T_{\rm eq,now}$ and
$a$. 

At the same time, we also find that models with $\ln T_{\rm eq,now}$
have substantially higher evidence than models using $\ln T_{\rm
  eq,ZAMS}$ in place of $\ln T_{\rm eq,now}$, while the model using
both $\ln T_{\rm eq,ZAMS}$ and $\tau$ has higher evidence than the
model using $\ln T_{\rm eq,ZAMS}$ alone. Moreover, we find that the
maximum posterior value for $c_{4}$ is greater than zero in all cases
where it is allowed to vary.  In other words, planet radii are more
strongly correlated with the present day equilibrium temperature than
they are with the ZAMS equilibrium temperature, and if the latter is
used in place of the former then a significant positive correlation
between radius and host star fractional age remains.

Based on this we conclude that the radii of close-in Jupiter-mass
giant planets are determined by their present-day equilibrium
temperature and semi-major axis, and that the radii of planets increase
over time as their equilibrium temperatures increase.

\subsubsection{Selection Effects}
\label{sec:selectioneffects}

\paragraph{The Effect of Stellar Evolution on the Detectability of Planets:}
The sample of known TEPs suffers from a broad range of observational
selection effects which in principle might explain a preference for
finding large planets around evolved stars. As stars evolve their
radii increase, which, for fixed \rpl, $a$ and \mstar, reduces the
transit depth by a factor of $\rstar^{-2}$ (reducing their
detectability), but increases the duration of the transits by a factor
of $\rstar$ (increasing their detectability). It also increases the
geometrical probability of a planet being seen to transit by a factor
of $\rstar$ (increasing planet detectability). As stars evolve they
also become more luminous, meaning that at fixed distance they may be
monitored with greater photometric precision (further increasing
planet detectability).

To determine the relative balance of these competing factors for the
HAT surveys, for each TEP system discovered by HAT we estimate the
relative number of ZAMS stars with the same stellar mass and
metallicity around which one could expect to find a planet with the
same radius and orbital period and with the same signal-to-noise
ratio. To do this we use the following expression:
\begin{equation}
N_{\rm ZAMS}/N_{t} = \frac{V_{\rm ZAMS}}{V_{t}}\frac{{\rm Prob}_{\rm ZAMS}}{{\rm Prob}_{t}}
\label{eqn:nzams}
\end{equation}
where $N_{\rm ZAMS}/N_{t}$ is the relative number of ZAMS planet hosts expected compared to those with age $t$, $V_{\rm ZAMS}/V_{t}$ is the relative volume surveyed for
ZAMS-equivalent versions of the TEP system (for simplicity we assume a
uniform space density of stars), and ${\rm Prob}_{\rm ZAMS}/{\rm
  Prob}_{t} = R_{\star,ZAMS}/R_{\star,t}$ is the relative transit
probability, which is equal to the ratio of the stellar radii. To
estimate $V_{\rm ZAMS}/V_{t}$ we note that for fixed photometric
precision, and assuming white noise-dominated observations, the
transit S/N of a given TEP scales as the transit depth times the
square root of the number of points in transit, or as
$(R_{\star,ZAMS}/R_{\star,t})^{-3/2}$. If the data are red-noise
dominated, then the S/N scales simply as the transit depth, which
would increase $V_{\rm ZAMS}/V_{t}$. We then determine the $r$
magnitude of stars in the HAT field containing the TEP in question for
which the per-point RMS is larger than the RMS of the observed TEP
light curve by $(R_{\star,ZAMS}/R_{\star,t})^{-3/2}$. Accounting for the change in absolute
$r$ magnitude between the ZAMS and the present day for the system,
this gives us the relative distance of a ZAMS-equivalent system for
which the transits would be detected with the same S/N. The cube of
this distance is equal to $V_{\rm ZAMS}/V_{t}$.

\begin{figure}[!ht]
\plotone{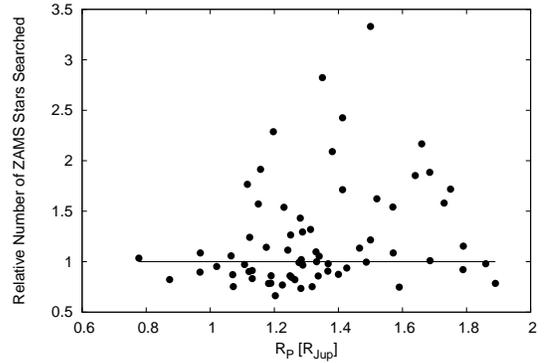}
\caption[]{
    The relative number of ZAMS-equivalent stars searched for a given
    TEP discovered by HAT to the same transit S/N as the observed TEP
    system (equation~\ref{eqn:nzams}) computed as described in
    Section~\ref{sec:selectioneffects}. This is shown as a function of
    planet radius. For the largest planets with $R_{P} >
    1.5$\,\rjup\ the HAT survey is more sensitive to finding the same
    planet around a ZAMS star than it is to finding the planet around
    the moderately evolved star where it was discovered.
\label{fig:NZAMS}}
\end{figure}

In Figure~\ref{fig:NZAMS} we show $N_{\rm ZAMS}/N_{t}$ vs \rpl\ for HAT
planets. For most TEP systems discovered by HAT, including most of the
systems with $\rpl > 1.5$\,\rjup, we have $N_{\rm ZAMS}/N_{t} > 1$. In
other words, the greater transit depths expected for ZAMS systems more
than compensates for the lower luminosities, shorter duration
transits, and lower transit probabilities. Thus, from a pure
transit-detection point-of-view, we should expect to be {\em more} sensitive
to TEPs around ZAMS stars than to TEPs around stars with
the measured host star ages.

Put another way, while selection effects may
lead to fewer small planets being found around older stars (missing planets in the
upper left corner of Figure~\ref{fig:fracagerp}), based on the
estimate in Figure~\ref{fig:NZAMS}, selection effects due to transit
detectability do not explain why we find fewer large planets around
unevolved stars (missing planets in the lower right corner of
Figure~\ref{fig:fracagerp}). If the occurrence rate of large radius
planets is independent of host star age, or if it is larger for
unevolved stars than for evolved stars, we would expect to have found {\em
  more} large planets around unevolved stars than evolved stars. 

%
%
%% ----------------
\ifthenelse{\boolean{emulateapj}}{
    \begin{figure*}[!ht]
}{
    \begin{figure}[!ht]
}
\plotone{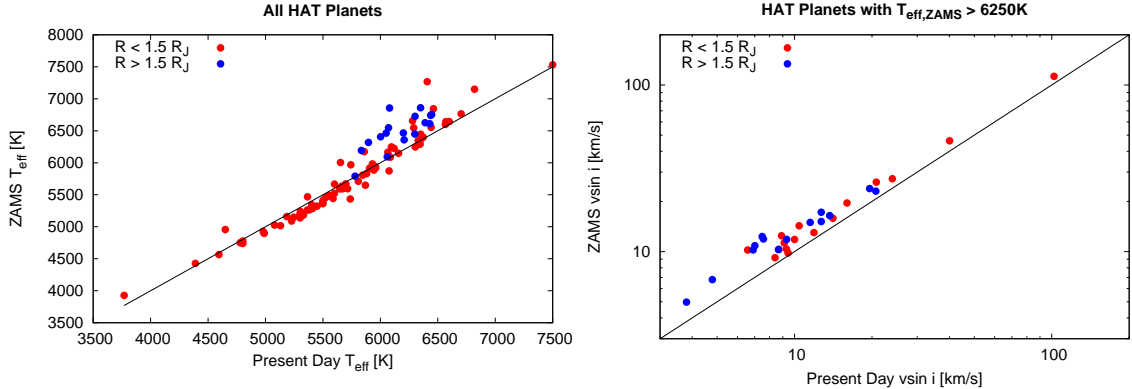}
\caption[]{
    Left: The estimated TEP host-star effective temperature on the
    ZAMS vs.\ its present day measured effective temperature for all
    TEP systems discovered to date by HAT. While stars hosting planets
    with $R_{P} > 1.5$\,\rjup\ had higher effective temperatures on
    the ZAMS, none of them would have been too hot for us to proceed
    with confirmation follow-up observations. Right: The estimated
    projected rotation velocity on the ZAMS vs.\ present day measured
    rotation velocity for HAT TEP hosts with $T_{\rm eff,ZAMS} >
    6250$\,K. The estimated ZAMS $v \sin i$ is calculated by scaling
    the measured $\vsini$ by $R_{\star}/R_{\star, ZAMS}$ assuming the
    spin-down for these radiative-envelope stars is due entirely to
    changes in the moment of inertia, and assuming the latter scales
    as $R_{\star}^2$. While stars hosting planets with $R_{P} >
    1.5$\,\rjup\ would have been rotating more rapidly on the ZAMS
    than at the present day, none of them would have been rotating too
    rapidly for us to proceed with confirmation follow-up
    observations.
\label{fig:zamsteffvsini}}
\ifthenelse{\boolean{emulateapj}}{
    \end{figure*}
}{
    \end{figure}
}
%% ----------------

\paragraph{The Effect of Stellar Evolution on the Ability to Confirm Planets:}
Other observational selection effects may still be at play. If the
orbits of these planets shrink over time due to tidal evolution, then
the transit probability and the fraction of points in transit both
increase in time by more that what we estimated. Beyond simply
detecting the transits, further selections are imposed in the
follow-up program carried out to confirm the
planets. Figure~\ref{fig:zamsteffvsini} compares the present day
effective temperature and $v\sin i$ for HAT TEPs to the expected
values on the ZAMS (estimated as discussed below). For all of the $\rpl > 1.5$\,\rjup\ planets found
by HAT the host star had a higher $\teffstar$ on the ZAMS than at the
present day. The most extreme case is HAT-P-7 which had an
estimated ZAMS effective temperature of $6860$\,K compared to its
present-day temperature of $6350 \pm 100$\,K. While precision RVs are
more challenging for early F dwarfs than for later F dwarfs, the ZAMS
temperatures of the hosts of the largest TEPs found by HAT are still
within the range where we carry out follow-up observations (we do not
follow-up hosts of spectral type A or earlier if they are faint stars
with $V \ga $13). 

An additional potential selection effect relates to
the stellar rotation. Neglecting tidal interactions between the stars
and planets, the host stars would have had higher projected
rotation velocities at ZAMS, primarily resulting from their lower
moments of inertia (most of the hosts of the largest planets found by
HAT have radiative envelopes, or would have had them for much of their
main sequence lifetimes, and thus would not lose substantial angular
momentum from magnetized stellar winds). Roughly speaking we expect
$v\sin i \propto \rstar^{-1}$ (assuming $I \propto \rstar^{2}$). The
most rapidly rotating ZAMS host is HAT-P-41 for which we estimate a
ZAMS rotation velocity of $24$\,\kms, which is still well within the range
where we continue follow-up (we do not follow-up hosts with $\vsini
\ga 50$\,\kms\ if they are around faint stars with $V \ga $13). Thus
while precise RVs would be somewhat more challenging for ZAMS planet hosts than
for moderately evolved hosts, these factors are unlikely to be
responsible for the lack of highly inflated planets discovered to date
around stars close to ZAMS.

\paragraph{Correcting the Correlation Coefficient for Observational Selections:}
In order to determine quantitatively how selection effects impact
the correlation measured between $\rpl$ and $\tau$, we follow
\citet{efron:1999} in calculating a modified Kendall correlation
coefficient that is applicable to data suffering a non-trivial
truncation. The procedure is as follows. We will call the observed
data points $(R_{P,i},\tau_{i})$ and $(R_{P,j},\tau_{j})$, with $i
\neq j$, comparable if each point falls within the other point's
selection range. Here point $j$ is within the selection range for
point $i$ if, holding everything else constant, we could still have
discovered the planet around star $i$ if the system had values of
$(R_{P,i},\tau_{j})$, $(R_{P,j},\tau_{i})$ or $(R_{P,j},\tau_{j})$
instead of $(R_{P,i},\tau_{i})$. Letting $\cal{J}$ be the set of all
comparable pairs, and $N_{p}$ be the total number of such pairs, the
modified Kendall correlation coefficient is then given by
\begin{equation}
\label{eqn:kendall}
r_{K} = \frac{1}{N_{p}}\sum_{(i,j) \in \cal{J}}  {\rm sign}\left( (R_{P,i} - R_{P,j})(\tau_{i} - \tau_{j})\right).
\end{equation}
For uncorrelated data $r_{K}$ has an expected value of $0$, whereas
perfectly correlated data has $r_{K}=1$ and perfectly anti-correlated data
has $r_{K}=-1$. To determine the probability of finding $|r_{K}| > |r_{K,{\rm
  observed}}|$ it is necessary to carry out bootstrap simulations. To
do this we calculate $r_{K}$ for $N_{\rm sim}$ simulated data sets, and
for each simulated data set we randomly select $N$ values of $i$, with
replacement, from the observed samples, adopt $R_{P,i}$ for each
simulated point, and associate with it a value of $\tau$ drawn at
random from the set of points that are comparable to $i$ (including
$i$ itself in this case).

The primary challenge in calculating equation~\ref{eqn:kendall} for
the observed sample of close-in giant planets is to determine the set
of comparable pairs. We do this for the HAT planets by
subtracting the observed transit signal from the survey light curve,
rescaling the scatter to match the expected change in r.m.s.\ due to
the change in the stellar luminosity with age, adding the expected
transit signal given the new trial planetary and stellar radii, but
assuming the original ephemeris and orbital inclination, and using BLS
to determine whether or not the transit could be
recovered. 
Using the set of comparable pairs determined in this fashion, we
find $r_{K} = \Kendallrtrunc$, with a \Kendallprobtrunc\% false alarm probability, based on
the bootstrap simulations (or $r_{K} = \KendallrtruncMp$ with a \KendallprobtruncMp\% false alarm
probability when restricted to planets with $0.4\,\mjup < \mpl <
2.0\,\mjup$). For comparison, if we ignore the selection effects and
assume all points are comparable, we find $r_{K} = \Kendallrnotrunc$ with a false
alarm probability of \Kendallprobnotrunc\% (or $r_{K} = \KendallrnotruncMp$ with a \KendallprobnotruncMp\% false
alarm probability when restricted to planets with $0.4\,\mjup < \mpl <
2.0\,\mjup$). The false alarm probabilities in the latter case are
essentially the same as what was reported above using the Spearman
rank-order correlation test instead of the Kendall test, demonstrating
the consistency of the two methods.

We conclude that while observational selections do slightly bias the
measured correlation between $\rpl$ and $\tau$ for HAT, the effect is
small. While we cannot determine the set of comparable pairs for WASP,
the selections are likely very similar to HAT, and we expect the
effect on the measured correlation of accounting for observational
selections to be similarly small. We therefore expect that the full
combined set of planets would still exhibit a highly significant
correlation between $\rpl$ and $\tau$, even after accounting for
observational selections.

\subsubsection{Systematic Errors in Stellar Parameters}
\label{sec:syserrstel}

The radii of TEPs are not measured directly, but rather are measured
relative to the stellar radii, which in turn are determined by
matching the effective temperatures, stellar densities, and stellar
metallicities to models (either theoretical stellar evolution models,
as done for example for most HAT systems, or by utilizing empirical
models calibrated with stellar eclipsing binary systems, as has been
done for many WASP systems). Any systematic error in the stellar
radius would lead to a proportional error in the planet radius, and
the fact that the largest planets are more commonly found around the
most evolved (and largest) stars is what one would expect to see if
there were significant unaccounted-for systematic errors. Here we
consider a variety of potential systematic errors, and argue that none
of these are responsible for the observed correlation.

\paragraph{Eccentricity:}
One potentially important source of systematic errors in this respect
is the planetary eccentricity, which is constrained primarily by the
RV data, and which is needed to determine the stellar density from the
measured transit duration, impact parameter, and radius ratio. The
host stars of the largest radius planets are among the hottest,
fastest rotating, and highest jitter stars around which transiting
planets have been found (e.g., Figure~\ref{fig:zamsteffvsini}, and
\citealp{2011ApJ...742...59H}). For these systems the eccentricity is
typically poorly constrained, and circular orbits have often been
adopted.\footnote{If circular orbits are not adopted, then there is a bias toward overestimating the eccentricity as shown by \citet{lucy:1971}. This bias may affect some of the earliest discovered planets especially.} If the systems were actually highly eccentric, with transits
near apastron, then the stellar densities would be higher than what
was inferred assuming circular orbits, and the stellar and planetary
radii would be smaller than what has been estimated. There are,
however, several large planets transiting moderately evolved stars for
which secondary eclipses have been observed, providing tight
constraints on the eccentricity (e.g., TrES-4b
\citealp{knutson:2009}, WASP-12b \citealp{campo:2011}, HAT-P-32b
\citealp{zhao:2014}, and WASP-48b \citealp{orourke:2014}). Moreover, the
most inflated planets are on short period orbits, where we expect
circularization. This expectation has been observationally verified in
cases where sufficiently high precision RVs have been possible, or
when secondary eclipse follow-up observations have been made. We also
note that at least for the majority of the very large radius HAT planets, when
the eccentricity is allowed to vary, the planet and stellar radii
determined from the median of the posterior distributions are found to
be larger than when the eccentricity is fixed to zero.

\begin{figure}[!ht]
\plotone{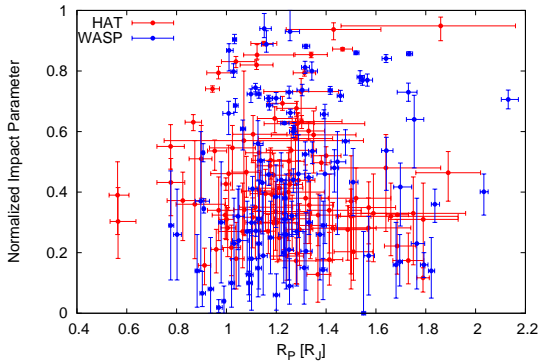}
\caption[]{
   Normalized impact parameter vs.\ planetary radius for TEPs found by
   HAT and WASP. The impact parameters of WASP TEPs appear to be
   uniformly distributed between 0 and 1, as expected for random
   orientations in space. The largest HAT planets have, if anything, a
   bias toward low impact parameters. If these suffer from a
   systematic error, the stellar and planetary radii would be
   underestimated.
\label{fig:impactparameterdist}}
\end{figure}
%% ----------------

\paragraph{Impact Parameter:}
Another potential source of systematic error is if the impact
parameter is in error, perhaps due to an incorrect treatment of limb
darkening \citep[e.g.,][]{espinoza:2015}. Errors in the impact
parameter will translate into concomitant errors in the stellar
density, and in the stellar radius and age. In order to over-estimate
the size of the planets, the impact parameter would need to have been
overestimated. Looking at the distribution of measured planetary
impact parameters, however, shows no evidence for this being the case
(\reffigl{impactparameterdist}). The impact parameter for the WASP
planets appears to be uniformly distributed between 0 and 1, as
expected for random orbital orientations, whereas the HAT planets are,
if anything, biased toward low impact parameters (if these are in
error, the stars and planets would be even larger than currently
estimated).

\begin{figure}[!ht]
\plotone{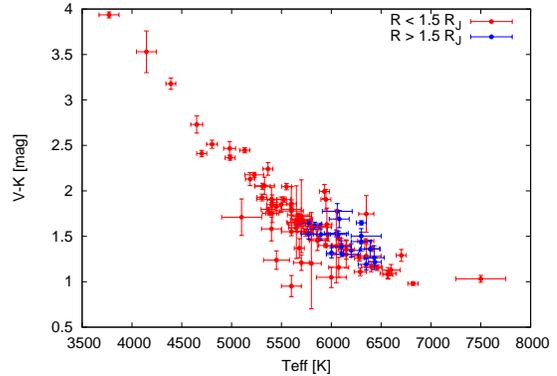}
\caption[]{
   Photometric $V-K$ color (not corrected for reddening)
   vs.\ effective temperature for TEP host stars from HAT and WASP
   with $R_{P} > 0.5$\,\rjup\ and $P < 10$\,days. The blue and red
   colors are used to distinguish between stars hosting planets with
   $R_{P} > 1.5$\,\rjup\ and $R_{P} < 1.5$\,\rjup, respectively. No
   systematic difference in the color-temperature relation is seen
   between these two classes of planets. Such a difference might have
   indicated a systematic error in the stellar effective temperature
   measurements of the stars hosting large planets.
\label{fig:colortemp}}
\end{figure}
%% ----------------

\begin{figure}[!ht]
\plotone{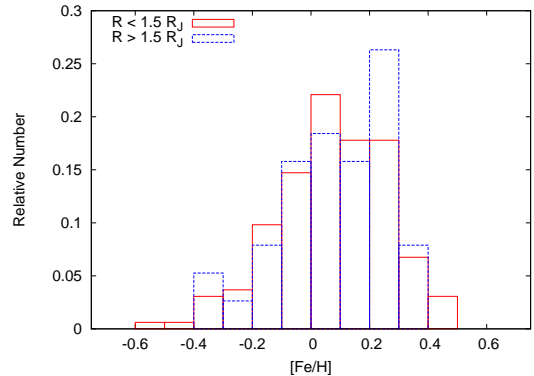}
\caption[]{
   Normalized histograms of \feh\ for transiting planet host stars from HAT and WASP, separated by planetary radius. No significant difference is seen between metallicities of large planet radius host stars, and small planet radius host stars.
\label{fig:metallicitydist}}
\end{figure}
%% ----------------

\paragraph{Stellar Atmospheric Parameters:}
Other potential sources of systematic errors include errors in the
stellar effective temperatures (if the stars are hotter than measured,
they would be closer to the ZAMS) or metallicities, or an error in the
assumed stellar abundance pattern (generally stars are modelled
assuming solar-scaled abundances). A check on the spectroscopic
temperature estimates can be performed by comparing the broad-band
photometric colors to the spectroscopically determined
temperatures. We show this comparison in \reffigl{colortemp} where we
use the color of the points to show the planet radius. While there is
perhaps a slight systematic difference in the $V-K$
vs.\ \teffstar\ relation between large and small radius planets, with
large radius planets being found, on average, around slightly redder
stars at fixed \teffstar\ than small radius planets, the difference is
too small to be responsible for the detected trend between planet
radius and fractional host age. Moreover, the difference is also
consistent with more evolved/luminous stars generally being more
distant from the Solar System than less evolved stars, and thus
exhibiting greater reddening. No systematic difference is seen in the
host star metallicity distributions of small and large radius planets
(\reffigl{metallicitydist}).

\paragraph{Priors Used In Stellar Modelling:}
Stars evolve faster as they age such that a large area on the
Hertzsprung-Russell diagram is covered by stellar models spanning a
small range of ages near the end of a star's life. As a result, when
observed stellar properties are compared to models there can be a bias
toward matching to late-ages. This well-known effect, dubbed the
stellar terminal age bias by \citet{pont:2004}, can be corrected by
adopting appropriate priors on the model parameters (e.g., adopting a
uniform prior on the age). Similarly, failing to account for the
greater prevalence of low-mass stars in the Galaxy relative to
high-mass stars can lead to overpredicting stellar masses
\citep[e.g.,][]{lloyd:2011}. For most transiting planets in the literature,
these possible biases have not been accounted for in performing this
comparison (i.e., generally the analyses have adopted uniform priors
on the relevant observables, namely the effective temperature, density
and metallicity). While these effects could lead to
over-estimated stellar radii, potentially explaining the preference
for large radius planets around evolved stars, in practice we only
expect such biases to be significant if the observed
parameters are not well constrained relative to the scale over which
the astrophysically-motivated priors change substantially. To estimate
the importance of this effect, we calculate new stellar parameters for
each of the HAT TEP systems with $R > 1.5$\,\rjup. Here we place a
uniform prior on the stellar age between the minimum age of the
isochrones and 14\,Gyr (this amounts to assuming a constant star
formation rate over this period), we use the \citet{chabrier:2003}
initial mass function to place a prior on the stellar mass, and we
assume a Gaussian prior on the metallicity with a mean of \feh$=
0$\,dex, and a standard deviation of 0.5\,dex. The details of how we
implement this are described in
Appendix~\ref{appendix:stelparamprior}.

We find that for all 17 systems the changes to the parameters are much
smaller than the uncertainties (the changes are all well below 0.1\%,
and in most cases below 0.01\%). We also find that the prior on the
stellar mass, which increases toward smaller mass stars, generally has
a larger impact than the priors on the age or metallicity. The result
is that in most cases the stellar masses and radii are very slightly
lower when priors are placed on the stellar properties, while the ages
are very slightly higher. The latter is due to the prior on stellar
mass pulling the solution toward lower effective temperatures, which
at the measured stellar densities requires higher ages. We conclude
that since the changes in the stellar parameters are insignificant,
the correlation between the planetary radii and host star fractional
age is not due to biases in the stellar parameters stemming from using
incorrect priors.

\subsubsection{Theoretical Significance}
\label{sec:theoreticalsignificance}

As we have shown, there is a significant correlation between the radii
of close-in giant planets and the fractional ages of their host
stars. This correlation is apparently a by-product of the more
fundamental correlation between planet radius and equilibrium
temperature, but the data also indicate that planetary radii increase
over time as their host stars evolve and become more luminous.

Such an effect is contrary to models of planet evolution where excess
energy associated with a planet's proximity to its host star does not
penetrate deep into the planet interior, but only acts to slow the
planet's contraction. \citet{burrows:2000} and \citet{baraffe:2008}
are examples of such ``default'' models.  Other examples of such
models include \citet{burrows:2007}, who showed, among other things,
how additional opacity which further slows the contraction could
explain the radii of inflated planets known at that time, and
\citet{ibgui:2010} who showed how extended tidal heating of the planet
atmosphere can increase the final ``equilibrium'' radius of a
planet. More generally, \citet{spiegel:2013} explored a variety of
effects related to planet inflation, including the effect on planetary
evolution of varying the depth at which additional energy is deposited
in the interior (see also \citealp{lopez:2016} for a recent
discussion). If the inflation mechanism only slows, but does not
reverse, the contraction, one would expect that at fixed semimajor
axis, older planets should be smaller than younger planets, despite
the increase in equilibrium temperature as the host stars evolve. This
is not what we see (Figure~\ref{fig:massradagegrid}, also
Section~\ref{sec:rteqcorr}).

On the other hand theories in which the energy is deposited deep in
the core of the planet may allow planets to become more inflated as
the energy source increases over time \citep{spiegel:2013}.
Examples of such models include tidal heating of an eccentric planet's
core as considered by \citet{bodenheimer:2001}, \citet{liu:2008}
and \citet{ibgui:2011}, or the Ohmic heating model proposed by
\citet{batygin:2010,batygin:2011} (though see \citealp{huang:2012}
and \citealp{wu:2013} who argue that this mechanism cannot heat the
deep interior). Our finding that planets apparently re-inflate over
time is evidence that some mechanism of this type is in operation.

\section{Summary}
\label{sec:summary}

The existence of highly inflated close-in giant planets is one of the
long-standing mysteries that has emerged in the field of exoplanets.
By continuing to build up the sample of inflated planets we are
beginning to see patterns in their properties, allowing us to narrow
down on the physical processes responsible for the inflation. Here we
presented the discovery of two transiting highly inflated planets
\hatcurb{65} and \hatcurb{66}. The planets are both around moderately
evolved stars, which we find to be a general trend---highly inflated
planets with $R \ga 1.5$\,\rjup\ have been preferentially found around
moderately evolved stars compared to smaller radius planets. This
effect is independently seen in the samples of planets found by HAT,
WASP, {\em Kepler}, TrES and KELT. We argue that this is not due to
observational selection effects, which tend to favor the discovery of
large planets around younger stars, nor is it likely to be the result
of systematic errors in the planetary or stellar parameters. We find
that the correlation can be explained as a by-product of the more
fundamental, and well known, correlation between planet radius and
equilibrium temperature, and that the present day equilibrium
temperature of close-in giant planets, which increases with time as
host stars evolve, provides a significantly better predictor of planet
radii than does the initial equilibrium temperature at the zero age
main sequence.

We conclude that, after contracting during the pre-main-sequence,
close-in giant planets are re-inflated over time as their host stars
evolve. This provides evidence that the mechanism responsible for this
inflation deposits energy deep within the planetary interiors.

The result presented in this paper motivates further observational
work to discover and characterize highly inflated planets. In
particular more work is needed to determine the time-scale for planet
re-inflation. The expected release of accurate parallaxes for these
systems from Gaia should enable more precise ages for all of these
systems. Many more systems are needed to trace the evolution of planet
radius with age as a function of planetary mass, host star mass,
orbital separation, and other potentially important
parameters. Furthermore, the evidence for planetary re-inflation
presented here provides additional motivation to search for highly
inflated long-period planets transiting giant stars.

%% EOF Discussion

% #####################################################################
%% Acknowledgements
\acknowledgements 

HATNet operations have been
funded by NASA grants NNG04GN74G and NNX13AJ15G. Follow-up of HATNet
targets has been partially supported through NSF grant
AST-1108686. G.\'A.B., Z.C., and K.P.\ acknowledge partial support
from NASA grant NNX09AB29G. J.H.\ acknowledges support from NASA grant
NNX14AE87G. K.P.\ acknowledges support from NASA grant NNX13AQ62G. We
acknowledge partial support also from the {\em Kepler} Mission under
NASA Cooperative Agreement NCC2-1390 (D.W.L., PI). A.S.\ is supported
by the European Union under a Marie Curie Intra-European Fellowship
for Career Development with reference FP7-PEOPLE-2013-IEF, number
627202. Part of this work was supported by Funda\c{c}\~ao para a
Ci\^encia e a Tecnologia (FCT, Portugal, ref. UID/FIS/04434/2013)
through national funds and by FEDER through COMPETE2020
(ref. POCI-01-0145-FEDER-007672). Data presented in this paper are
based on observations obtained at the HAT station at the Submillimeter
Array of SAO, and the HAT station at the Fred Lawrence Whipple
Observatory of SAO. This research has made use of the NASA Exoplanet
Archive, which is operated by the California Institute of Technology,
under contract with the National Aeronautics and Space Administration
under the Exoplanet Exploration Program. Data presented herein were
obtained at the WIYN Observatory from telescope time allocated to
NN-EXPLORE through the scientific partnership of the National
Aeronautics and Space Administration, the National Science Foundation,
and the National Optical Astronomy Observatory. This work was
supported by a NASA WIYN PI Data Award, administered by the NASA
Exoplanet Science Institute. We gratefully acknowledge R.~W.~Noyes for
his many contributions to the HATNet transit survey, and we also
gratefully acknowledge contributions from J.~Johnson, and from
G.~Marcy to the collection and reduction of the Keck/HIRES
observations presented here.  The authors wish to recognize and
acknowledge the very significant cultural role and reverence that the
summit of Mauna Kea has always had within the indigenous Hawaiian
community. We are most fortunate to have the opportunity to conduct
observations from this mountain.

%% EOF Acknowledgements

% #####################################################################
%% Bibliography
\clearpage
\bibliographystyle{apj}
\bibliography{htrbib}

\appendix

\section{Estimating Transiting Planet Host Star Parameters With Priors on the Stellar Mass, Age, and Metallicity}
\label{appendix:stelparamprior}

The physical parameters of transiting planet host stars are determined
by comparing the observed parameters \teffstar, \rhostar\ and \feh\ to
theoretical stellar evolution models. In practice the light curve
analysis produces a Markov Chain of \rhostar\ values, which we combine
with simulated chains of \teffstar\ and \feh\ values (we assume the
three parameters are uncorrelated, and that \teffstar\ and \feh\ have
Gaussian uncertainties). For a given $(\rhostar, \teffstar, \feh)$
link in the chain, we perform a trilinear interpolation within a grid
of isochrones from the YY models to get the corresponding stellar
mass, age, radius, luminosity, and absolute magnitudes in various
pass-bands. Our interpolation routine can use any combination of three
parameters as the independent variables, below we make use of this
feature using the mass, age and metallicity as the independent
variables. The resulting chain of stellar physical parameters is then
used to provide best estimates and uncertainties for each of these
parameters.

As discussed in Section~\ref{sec:syserrstel} this process may lead to
systematic errors in the stellar parameters if priors are not adopted
to account for the intrinsic distribution of stars in the Galaxy. The
prior is applied as a multiplicative weight that is associated with
each link in the Markov Chain. The weights are calculated as follows.

Let $P_{m}(m)$, $P_{t}(t)$, and $P_{\feh}(\feh)$ be prior probability
densities to be placed on the stellar mass, age and metallicity,
respectively. Here we use the \citet{chabrier:2003} initial mass
function for the prior on the stellar mass, a uniform distribution for
the prior on the stellar age, and a Gaussian distribution with mean
$0$\,dex and standard deviation $0.5$\,dex for the prior on the
metallicity. Further, let $C_{m}(m)$, $C_{t}(t)$, and $C_{\feh}(\feh)$
be the corresponding cumulative distributions of these prior
probability densities.

For a given $(m, t, \feh)$ link generated from an input set of
$(\rhostar, \teffstar, \feh)$, we find $m^{+} = C^{-1}_{m}(C_{m}(m) +
\Delta u_{m})$, $t^{+} = C^{-1}_{t}(C_{t}(t) + \Delta u_{t})$ and $\feh^{+} =
C^{-1}_{\feh}(C_{\feh}(\feh) + \Delta u_{\feh})$ for some small probability
steps $\Delta u_m \ll 1$, $\Delta u_t \ll 1$ and $\Delta u_{\feh} \ll 1$.  Likewise we calculate $m^{-}$, $t^{-}$ and
$\feh^{-}$ for a negative $\Delta u$. We then perform trilinear
interpolation within the isochrones to find $(\rho_{m+},
T_{{\rm eff},m+}, \feh_{m+})$ associated with the point $(m^{+},
t, \feh)$, and similarly for $m^{-}$, $t^{+}$, etc. 

Letting
\begin{equation}
{\bf v_{m}} = (\rho_{m+} - \rho_{m-}, T_{{\rm eff},m+} - T_{{\rm eff},m-}, \feh_{m+} - \feh_{m-})
\end{equation}
be the vector running from the $m^{-}$ point to the $m^{+}$ point, and similarly for ${\bf v_{t}}$ and ${\bf v_{\feh}}$, the weight $w$ is then calculated as
\begin{equation}
w = \frac{\Delta u_{m} \Delta u_{t} \Delta u_{\feh}}{{\bf v_{m}} \cdot ({\bf v_{t}} \times {\bf v_{\feh}})}
\end{equation}
where the denominator is the volume of the parallelepiped spanned by
the three vectors. We use these weights in calculating the weighted
median and $1\sigma$ confidence regions of each parameter chain.

\clearpage
\LongTables

\tabletypesize{\scriptsize}
\ifthenelse{\boolean{emulateapj}}{

}{
    \end{deluxetable}
}

\end{document}